\documentclass{ptephy}

\usepackage{graphics} 

\begin{document}

\begin{center}
\begin{tabular}{cc}
\begin{minipage}[b]{0.3\linewidth}
\vspace*{1em}
\includegraphics[scale=0.8]{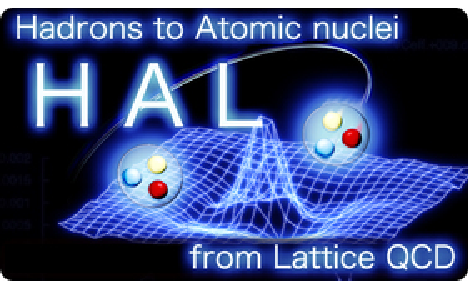}
\end{minipage}
&
\begin{minipage}[b]{100pt}
YITP-14-95 \\
RIKEN-QHP-173 \\
UTCCS-P-78 \\
UTHEP-666
\end{minipage}
\end{tabular}
\end{center}

\title{Coupled channel approach to strangeness $S=-2$ baryon-bayron interactions in Lattice QCD}

\author{
\name{Kenji~Sasaki}{1},
\name{Sinya~Aoki}{1,2},  
\name{Takumi~Doi}{3}, 
\name{Tetsuo~Hatsuda}{3},  
\name{Yoichi~Ikeda}{3},
\name{Takashi~Inoue}{4}, 
\name{Noriyoshi~Ishii}{5},  
\name{Keiko~Murano}{5} 
\name{(HAL~QCD~Collaboration)}{}%
}

\address{
\affil{1}{Center for Computational Sciences, University of Tsukuba, Tsukuba, 305-8577, Japan}
\affil{2}{Yukawa Institute for Theoretical Physics, Kyoto University, Kyoto, 606-8502, Japan}
\affil{3}{Theoretical Research Division, Nishina Center, RIKEN, Wako, 351-0198, Japan}
\affil{4}{Nihon University,  College of Bioresource Sciences, Fujisawa, 252-0880, Japan}
\affil{5}{Research Center for Nuclear Physics (RCNP), Osaka University, Ibaraki, Osaka, 567-0047, Japan}
\email{kenjis@het.ph.tsukuba.ac.jp}}

\begin{abstract}%
The baryon-baryon interactions with strangeness $S=-2$ with the flavor SU(3) breaking 
are calculated for the first time  by using the HAL~QCD method  extended
to coupled channel system in lattice QCD.
The potential matrices are extracted from the Nambu-Bethe-Salpeter wave functions obtained by 
the  $2+1$ flavor gauge configurations of CP-PACS/JLQCD Collaborations with a physical volume of $(1.93~{\rm{fm}})^3$
and with $m_{\pi}/m_K=0.96, 0.90, 0.86$.
The spatial structure and the quark mass dependence of the potential matrix in the baryon basis 
and in the SU(3) basis are investigated.
\end{abstract}

\subjectindex{B38, B64, D32, D34}
\maketitle

\section{Introduction}

Studying the baryon-baryon ($BB$) interactions in strangeness $S=-2$ channel  is an important 
 step to understand hypernuclei such as double-$\Lambda$ hypernuclei 
 and $\Xi$ hypernuclei (see e.g. \cite{Dover:1982ng,Gibson:1995an,Hiyama:2009zz}) as well as 
 exotic hadrons such as the $H$-dibaryon \cite{Jaffe:1976yi,Sakai:1999qm}.
  Moreover, the hyperon superfluidity in the core of the
  neutron stars is intimately related to the hyperon-hyperon interaction in the  $S=-2$ channel
  \cite{Takatsuka:2005bp}.
  Due to the limited experimental data, however, 
  the $BB$ interactions in the $S=-2$ channel are far from being realistic
  even under the constraints from the approximate flavor SU(3) symmetry.
  In addition, there are nearby two-baryon states in the $S=-2$ channel (e.g.
  $\Lambda \Lambda$ and $N \Xi$) so that the coupled-channel treatment is
  essential for studying the $S=-2$ system.

 Recently, the $BB$ interactions in the flavor SU(3) limit
have been studied systematically  in full QCD simulations on the lattice by the HAL QCD method (reviewed in 
 \cite{Aoki:2012tk}) for 
 several different masses of the pseudo-scalar meson $m_{\rm PS}= 470 \sim 1170$ MeV
  (see \cite{Inoue:2011ai} and references therein). 
 In this case, all the two-baryon thresholds are degenerate and the classification of the 
 $BB$ interactions in the flavor basis is applicable. 
 An extension of the HAL QCD method beyond  the inelastic threshold, which is relevant for the 
 $BB$ interactions with the flavor SU(3) breaking, has been also proposed to treat coupled channel
 systems\cite{Aoki:2011gt,Aoki:2012bb}. 
  The main purpose of this paper is to study the $BB$ interactions in 
  the $S=-2$ channel with the explicit SU(3) breaking on the basis of the coupled channel HAL QCD
  method developed in our previous works\cite{Aoki:2011gt,Aoki:2012bb}.

We note here that
the hyperon-nucleon scattering length away from the SU(3) symmetric limit was first evaluated by quench QCD simulation in \cite{Muroya:2004fz} and later by full QCD simulations in \cite{Beane:2006gf,Beane:2012ey}, where the L\"uscher's finite volume method was utilized.
 On the other hand, the hyperon-nucleon potentials, which provide much more information than the scattering lengths,  were derived through the equal-time NBS wave functions with the HAL QCD method in~\cite{Nemura:2008sp,Nemura:2009kc,Nemura:2010nh}. The present study can be regarded as  a coupled channel
  generalization of our previous works below the inelastic threshold. 

This paper is organized as follows.
In Sect.~2, we review the coupled channel approach to the the $BB$ interactions by the HAL QCD method in lattice QCD.
In Sect.~3, we define baryon operators and the  baryon states.
In Sect.~4, numerical setup on the lattice  is summarized.
In Sect.~5, we  present our numerical results of the $BB$ potentials.
Summary and conclusions are given in Sect.~6.

\section{Coupled channel $BB$ potentials}
In this section, we briefly review the coupled channel HAL QCD method \cite{Aoki:2011gt,Aoki:2012bb}
applicable to the inelastic scattering that $a_1+a_2\rightarrow b_1+b_2$,  where $ (a_1,a_2) \not= (b_1, b_2)$.
 
\subsection{Formalism}
We first define equal-time Nambu-Bethe-Salpeter (NBS) wave functions 
with the total energy $W_i$ as 
\begin{eqnarray}
\label{EQ:NBSdef}
\begin{array}{l}
 \psi^{a}_{W_i}(\vec{r}) e^{-{W_i} t} = \dfrac{1}{\sqrt{Z_{a_1}}\sqrt{ Z_{a_2}}}\sum_{\vec{x}} \langle 0 \mid B_{a_1}(\vec{x} +\vec{r},t) B_{a_2}(\vec{x}, t) \vert B=2, W_i \rangle, \\
 \psi^{b}_{W_i}(\vec{r}) e^{-{W_i} t}= \dfrac{1}{\sqrt{Z_{b_1}}\sqrt{ Z_{b_2}}}\sum_{\vec{x}} \langle 0 \mid B_{b_1}(\vec{x} +\vec{r},t) B_{b_2}(\vec{x}, t) \vert B=2, W_i \rangle,
\end{array} 
\end{eqnarray}
where $B_{c_j}(\vec{x},t)$ with $c=a,b$ and $j=1,2$ denotes a local composite operator
 for baryon $B_{c_j}$ with its wave-function renormalization factor $\sqrt{Z_{c_j}}$. The state 
$\vert B=2, W_i \rangle$ stands for a QCD asymptotic in-state with baryon number $2$ and energy $W_i$.
In the present  exploratory studies, we assume 
$\sqrt{Z_{a1}}\sqrt{Z_{a2}}= \sqrt{Z_{b1}}\sqrt{Z_{b2}}$ which implies that the flavor SU(3) breaking in the wave-function renormalization factor is not sizable in the present setup.  The validity of this assumption is left for future studies.

In the asymptotic region at long distance, these NBS wave functions satisfy free Schr\"odinger-type  equations as
\begin{eqnarray}
 \left( \frac{(k^c_i)^2}{2\mu^c} + \frac{\nabla^2}{2\mu^c} \right)\psi^{c}_{W_i}(\vec{r}) = 0, \quad r\equiv\vert\vec r\vert \to \infty, 
\end{eqnarray}
where
the corresponding asymptotic momentum $k_i^{c}$ in the center-of-mass (CM) frame is defined through the relation,
\begin{eqnarray}
 W_i = \sqrt{m_{c_1}^2+(k_i^c)^2} + \sqrt{m_{c_2}^2+(k_i^c)^2},  
\end{eqnarray}
with $m_{c_j}$ being the mass of the baryon $B_{c_j}$, and the reduced mass $\mu^c$ is given 
by $1/\mu^c=1/m_{c_1} + 1/m_{c_2}$.
On the other hand, in the interaction region at short distance, 
we have 
\begin{eqnarray}
\label{EQ.defK}
 K^{c}(\vec{r}, W_i) = \left(\frac{(k_i^c)^2}{2 \mu^c} + \frac{\nabla^2}{2 \mu^c}\right)\psi^{c}_{W_i}(\vec{r}) \not= 0 ,
\end{eqnarray}
from which we define the energy-independent non-local potential matrix as 
\begin{eqnarray}
\label{EQ.fctK}
  K^{c}(\vec{r}, W_i)
 = \sum_{c^\prime=a,b}  \int\!\!d^3 r^\prime \, {U^{c}}_{c'}(\vec r, \vec r^\prime) \, \psi^{c^\prime}_{W_i}(\vec r^\prime).
\end{eqnarray}
This is an extension of the HAL QCD definition for the potential to the coupled channel case~\cite{Aoki:2011gt}. To handle the non-locality of the potential, we introduce the derivative expansion as
  $U(\vec r,\vec r') 
  = (V_{\rm{LO}}(\vec r)+ V_{\rm{NLO}}(\vec r)  + \cdots )\delta(\vec r-\vec r^\prime) $,
where N${^n}$LO term is of $O(\vec{\nabla}^n)$.
At low energies, a good convergence of derivative expansion has been confirmed for the $NN$ case~\cite{Murano:2011nz}.

\subsection{Extraction of potential matrix}
In the leading order of the derivative expansion of the non-local potential,
eqs.~(\ref{EQ.defK}) and (\ref{EQ.fctK}) can be written as a coupled channel form of the Schr\"odinger equation for two independent channels $a$ and $b$,
\begin{eqnarray}
  \label{EQ:CoupledSE}
\left( \begin{array}{l}
 \left( E_i^a  - H_0^{a} \right)
  \psi^{a}_{W_i}(\vec{r}) \\
 \left( E_i^b - H_0^{b} \right)
  \psi^{b}_{W_i}(\vec{r})
\end{array} \right)
=
\left( \begin{array}{cc}
{V^{a}}_{a}(\vec{r}) & {V^{a}}_{b}(\vec{r}) \\
{V^{b}}_{a}(\vec{r}) & {V^{b}}_{b}(\vec{r}) 
\end{array} \right)
\left( \begin{array}{c}
\psi^{a}_{W_i}(\vec{r}) \\
\psi^{b}_{W_i}(\vec{r})
\end{array} \right)
\end{eqnarray}
where the kinetic energy and the free Hamiltonian for channels $c=a,b$ are given by $E_i^c = \frac{(k_i^c)^2}{2\mu^c}$ and  $ {H_0}^c = - \frac{\nabla^2}{2\mu^c} $, respectively.

Two pairs of NBS wave functions, $\{\psi_{W_i}^{a}, \psi_{W_i}^{b}\}_{i=1,2}$ are necessary to  extract the local potential matrix from the above coupled channel equation.
In the infinite volume,
we can have two states, $\vert a, W\rangle$ and $\vert b, W\rangle$ with a given energy $W$, which are connected to the asymptotic scattering states if $W$ is larger than $m_{a_1}+m_{a_2}$ and $m_{b_1}+m_{b_2}$.
This implies that two nearby eigenstates, $|B=2,W_1 \rangle$ and $|B=2, W_2 \rangle $ with $W_1-W_2 = O(L^{-2})$, exist even for finite volume. 
Suppose that $W_1 < W_2$ are two lowest energies of two baryons in the finite volume. 
By using the wall-source operators 
${\mathcal{I}}_a(t)= \overline{\left(B_{a_2} B_{a_1} \right)}(t)$
 and ${\mathcal{I}}_b(t) = \overline{\left(B_{b_2} B_{b_1} \right)}(t)$ 
 \footnote{A detailed definition of the wall source operators will be given in Sect.~\ref{SECT:SETUP}.}, the states
  $\vert B=2, W_1\rangle$ and $\vert B=2, W_2\rangle$ are created as 
\begin{eqnarray}
&&
{\mathcal{I}}_c (0) \vert 0 \rangle  = C_{c1} \vert B=2, W_1 \rangle + C_{c2} \vert B=2, W_2 \rangle + \cdots ,
\end{eqnarray}
where the coefficient matrix $C_{cj}$ can be determined from two-baryon correlation functions.
We then define optimized source operators as 
\begin{eqnarray}
 \left( \begin{array}{c} {\mathcal{I}}_{W_1}(t) \\ {\mathcal{I}}_{W_2} (t) \end{array} \right)
 = 
 \left( \begin{array}{cc}
 C_{a1} & C_{a2} \\
 C_{b1} & C_{b1} \\
 \end{array} \right)^{-1}
 \left( \begin{array}{c} {\mathcal{I}}_a (t) \\ {\mathcal{I}}_b (t) \end{array} \right),
\end{eqnarray}
so that four point (4-pt) function $F_{\mathcal{I}_{W_i}}^c(\vec r,t)$  at large $t$ behaves as
\begin{eqnarray}
F_{\mathcal{I}_{W_i}}^c(\vec r,t) &\equiv & \langle 0 \vert B_{c_1}(\vec x+\vec r,t) B_{c_2}(\vec x,t)  \mathcal{I}_{W_i}(0)\vert 0 \rangle \simeq \psi_{W_i}^c(\vec r) e^{-W_i t} + O\left(e^{-W_3 t} \right)
 \label{eq:4-pt}
\end{eqnarray}
for $i=1,2$ and $c=a,b$, where $W_3$ corresponds to the 3rd state satisfying  $W_1 < W_2 < W_3 < W_{j\ge 4}$.
By using these 4-pt functions, the coupled channel potential matrix can be determined as
\begin{eqnarray}
\label{EQ:CCWR}
&&
 \left( \begin{array}{ll} 
 {V^a}_a (\vec{r}) & {V^a}_b (\vec{r}) \\ 
 {V^b}_a (\vec{r}) & {V^b}_{b} (\vec{r})
 \end{array} \right) 
 \nonumber \\
&& \hspace*{1.5em} 
\simeq
 \left( \begin{array}{cc} 
 (E_1^a-H_0^a ) F^{a}_{ \mathcal{I}_{W_1}}(\vec{r}, t) & 
 ( E_2^a-H_0^a) F^{a}_{\mathcal{I}_{W_2}}(\vec{r}, t) \\
 (E_1^b-H_0^b )F^{b}_{\mathcal{I}_{W_1}}(\vec{r}, t) & 
 ( E_2^b -H_0^b) F^{b}_{\mathcal{I}_{W_2}}(\vec{r}, t) \\
 \end{array} \hspace*{-0.3em} \right)
  \left( \begin{array}{cc}
F^{a}_{\mathcal{I}_{W_1}}(\vec{r}, t) & F^{a}_{\mathcal{I}_{W_2}}(\vec{r}, t) \\
F^{b}_{\mathcal{I}_{W_1}}(\vec{r}, t) & F^{b}_{\mathcal{I}_{W_2}}(\vec{r}, t) \\
 \end{array} \right)^{\hspace*{-0.2em}-1} \hspace*{1.5em}
\end{eqnarray}
for sufficiently large $t$, where the states with $ W (> W_1,W_2)$ can be neglected in the 
above 4-pt functions.
As the volume increases, however, the spectrum becomes denser and  two lowlying 
states $W_1$ and $W_2$ cannot be isolated unless extremely large $t$ is achieved.    
This is why  we need an improved method in practice, as explained in the next subsection.
\subsection{Time-dependent method}
The improved method to extract the potentials without using the ground state saturation has been proposed in
 Ref.~\cite{HALQCD:2012aa} in the case of the single channel. In this subsection, we extend this method to the 
coupled channel case. 

We first introduce the normalized $4$-pt correlation function $R$ defined as
\begin{eqnarray}
 R^c_{\mathcal{I}_d}(\vec r, t) 
 &\equiv& \frac{F^c_{\mathcal{I}_d}(\vec r,t)}{\exp[-(m_{c_1}+m_{c_2}) t]}
 =\sum_{j}   \psi^{c}_{W_j} (\vec{r}) e^{-\Delta W_j^c t} A_d^{W_j} +\cdots, 
\label{EQ.Rdefine}
\end{eqnarray}
where $ \Delta W_j^c =W_j - m_{c_1}-m_{c_2}$ and 
 $A_d^{W_j} = \langle W_j \vert \mathcal{I}_{d}(0) \vert 0 \rangle$.
 The 4-pt function $F^c_{\mathcal{I}_d}(\vec r,t)$ here is defined through 
the original wall-source operator $\mathcal{I}_d (0)$ instead of $\mathcal{I}_{W_i} (0)$.
The ellipses in Eq.(\ref{EQ.Rdefine})
denote inelastic contributions from channels other than $a$ and $b$. 

In the non-relativistic approximation valid at low energies, $ \Delta W_j^c \simeq  E_j^c$,  
we can  replace the kinetic energy term in the equation with the time derivative  as
\begin{eqnarray}
-\frac{\partial}{\partial t} {R^{c}}_{\mathcal{I}_d}(\vec r, t) 
\simeq \sum_j E_j^c  \psi^{c}_{W_j}(\vec r) e^{-\Delta W_j t} A_d^{W_j},
\end{eqnarray}
with which we obtain the Schr\"{o}dinger type equation,
\begin{eqnarray}
\left(-\frac{\partial}{\partial t} - H_0^c \right) {R^{c}}_{\mathcal{I}_d}(\vec r, t)  &=& \int d^3 r^\prime {U^c}_e(\vec r, \vec r^\prime) {\Delta^c}_e
{R^e}_{\mathcal{I}_d}(\vec r^\prime, t),
\end{eqnarray}
where ${\Delta^c}_e = \exp[-(m_{e_1}+m_{e_2})t] /  \exp[-(m_{c_1}+m_{c_2})t] $.
If we go beyond the non-relativistic approximation, higher-order time-derivatives appear, which we will not
 consider in this paper. 
Expanding $U$ in terms of derivatives again, the leading order coupled channel potentials can be obtained as  
\begin{eqnarray}
\label{EQ:CCWR2x2}
&&
 \left( \begin{array}{ll} 
 {{V^a}_{a}} (\vec{r}) & {{V^a}_{b }} (\vec{r}) {\Delta^{a}}_b \\ 
 {{V^b}_{a }} (\vec{r}) {\Delta^b}_a & {{V^b}_{b}} (\vec{r})
 \end{array} \right) 
 \nonumber \\
 && \hspace*{1.5em} 
\simeq
 \left( \begin{array}{cc} 
 ( -\frac{\partial}{\partial t}-H_0^a ) R^{a}_{{\mathcal I}_a}(\vec{r},t) & 
 ( -\frac{\partial}{\partial t}-H_0^a ) R^{a}_{{\mathcal I}_b}(\vec{r},t) \\
 ( -\frac{\partial}{\partial t}-H_0^b ) R^{b}_{{\mathcal I}_a}(\vec{r},t) & 
 ( -\frac{\partial}{\partial t}-H_0^b ) R^{b}_{{\mathcal I}_b}(\vec{r},t) \\
 \end{array} \hspace*{-0.3em} \right)
  \left( \begin{array}{cc}
 R^{a}_{{\mathcal I}_a}(\vec{r},t) & R^{a}_{{\mathcal I}_b}(\vec{r},t) \\
 R^{b}_{{\mathcal I}_a}(\vec{r},t) & R^{b}_{{\mathcal I}_b}(\vec{r},t)
 \end{array} \right)^{\hspace*{-0.2em}-1} .  \hspace*{1.5em} 
\end{eqnarray}
Extension of this formula to three channels is straightforward.
For eq.(\ref{EQ:CCWR2x2}) to work,  two independent source operators $\mathcal{I}_a$ and $\mathcal{I}_b$ are 
needed, while no optimization is required. Note that isolation of each eigenstates is not necessary in this method ~\cite{HALQCD:2012aa}.
  Only the constraint is to keep moderately large $t$ so that
 other channels having larger threshold energies than $a$ and $b$ can be suppressed.
 In the following, we  employ this improved method in our numerical calculations. 

\section{Strangeness $S=-2$ two-baryon system}
We emply the following interpolating operator for octet baryons,
\begin{eqnarray}
B_\alpha(\vec{x}) = \epsilon_{abc} (q^{T}_a(\vec x) C \gamma_5 q_b(\vec x)) q_{c \alpha}(\vec x)
\end{eqnarray}
with the Dirac index $\alpha$, which represents the spin of the octet baryons.
Denoting quark-flavors as $q=u,d,s$ for "up", "down" and "strange", respectively,
  the flavor structures of baryons are given in terms of the isospin multiplets as 
\begin{eqnarray}
\begin{array}{lllcccccc}
S=0 & I=1/2 & : & \multicolumn{3}{l}{p = [ud]u}, & \multicolumn{3}{c}{n = [ud]d}\\
S=-1 & I=1 & : & \multicolumn{2}{l}{\Sigma^+ =-[us]u}, & \multicolumn{2}{c}{\Sigma^0 = -([ds]u+[us]d) / \sqrt{2}}, & \multicolumn{2}{l}{\Sigma^- = -[ds]d } \\
S=-1 & I=0 & : & \multicolumn{6}{l}{\Lambda = ([sd]u+[us]d-2[du]s)/\sqrt{6}} \\
S=-2 & I=1/2 & : & \multicolumn{3}{l}{\Xi^0 = [su]s}, & \multicolumn{3}{c}{\Xi^- = [sd]s}  
\end{array} .
\end{eqnarray}

Considering the Fermi-Dirac statistics of two baryons, the allowed combinations for the $S=-2$ system are given in 
Table~\ref{TAB:S2channel},
where $I_z=0$ components are given as
\begin{eqnarray}
(\Sigma \Sigma)_I &=& 
\left\{ \begin{array}{ll}
\sqrt{\frac{1}{3}} \left( 
\Sigma^+ \Sigma^- + \Sigma^- \Sigma^+ - \Sigma^0 \Sigma^0 \right), & I=0 \\
\sqrt{\frac{1}{2}} \left(
\Sigma^+ \Sigma^- - \Sigma^- \Sigma^+ \right), & I=1 \\
\sqrt{\frac{1}{6}} \left( 
\Sigma^+ \Sigma^- + \Sigma^- \Sigma^+ + 2 \Sigma^0 \Sigma^0 \right), & I=2
\end{array} \right.
 \\
(N \Xi)_I &=& 
\left\{ \begin{array}{ll}
\sqrt{\frac{1}{2}} \left( p \Xi^- - n \Xi^0 \right), & I=0 \\
\sqrt{\frac{1}{2}} \left( p \Xi^- + n \Xi^0 \right), & I=1 \\
\end{array} \right. .
\end{eqnarray}
\begin{table}
\caption{Summary of channels with $S=-2$.}
\label{TAB:S2channel}
\begin{center}
\begin{tabular*}{\linewidth}{@{\extracolsep{\fill}}ccccccccc}
\hline
\multicolumn{2}{c}{Channel} & \multicolumn{4}{c}{Baryon-pairs} & \multicolumn{3}{c}{SU(3) multiplets} \\
\hline
$I=0$ 
 & ${^1S_0}$ & \multicolumn{4}{c}{$\Lambda \Lambda$, ~ $(N \Xi)_0$, ~ $(\Sigma \Sigma)_0$}
 & \multicolumn{3}{c}{$1$, ~ $8_s$, ~ $27$} \\ 
 & ${^3S_1}-{^3D_1}$ 
 & \multicolumn{4}{c}{$(N \Xi)_0$}
 & \multicolumn{3}{c}{$8_a$} \\ 
$I=1$ 
 & ${^1S_0}$ 
 & \multicolumn{4}{c}{$(N \Xi)_1$, ~ $(\Lambda \Sigma)_1$}
 & \multicolumn{3}{c}{$8_s$, ~ $27$} \\ 
 & ${^3S_1}-{^3D_1}$ 
 & \multicolumn{4}{c}{$(N \Xi)_1$, ~ $\Lambda \Sigma$, ~ $(\Sigma \Sigma)_1$}
 & \multicolumn{3}{c}{$8_a$, ~ $10$, ~ $\overline{10}$} \\ 
$I=2$ 
 & ${^1S_0}$ 
 & \multicolumn{4}{c}{$(\Sigma \Sigma)_2$}
 & \multicolumn{3}{c}{$27$} \\ 
\hline
\end{tabular*}
\end{center}
\end{table}

\section{Numerical simulations}
\label{SECT:SETUP}
\begin{table}
\begin{center}
\caption{Lattice parameters and hadron masses in unit of [MeV] are listed.}
\label{TAB:HadronM}
 \begin{tabular}{cccccc}
\hline \hline
 \multicolumn{6}{c}{Lattice parameters} \\
 $\beta$ & $\kappa_s$ & $c_{SW}$ & lattice size & $a$ [fm] & $L$ [fm] \\
\hline
 1.83 & 0.13710 & 1.7610 & $16^3 \times 32$ & $0.1209$ & $1.93$ \\
\hline \hline
 \end{tabular} \\[2mm]
  \begin{tabular}{ccccccccc}
  \hline \hline
   & $N_{conf}$ & $\kappa_{ud}$ & $m_\pi$ & $m_K$ & $m_N$ & $m_\Lambda$ & $m_\Sigma$ & $m_\Xi$ \\
  \hline
  Set 1 & $700$ & $0.13760$ & $875(1)$ & $916(1)$ & $1810(2)$ & $1839(2)$ & $18466(2)$ & $1872(2)$ \\
  Set 2 & $800$ & $0.13800$ & $749(1)$ & $828(1)$ & $1619(2)$ & $1675(2)$ & $1689(2)$ & $1737(2)$ \\
  Set 3 & $800$ & $0.13825$ & $660(1)$ & $768(1)$ & $1482(3)$ & $1556(3)$ & $1575(3)$ & $1640(2)$ \\
  \hline \hline
  \end{tabular}
 \end{center}
\end{table}

We employ $2+1$-flavor full QCD gauge configurations from Japan Lattice Data Grid(JLDG)/International Lattice Data Grid(ILDG)~\cite{JLDGILDG}.
They are generated by the CP-PACS and JLQCD Collaborations~\cite{CPJLCOLLAB} with the renormalization-group improved gauge action and the non-perturbatively $O(a)$ improved Wilson quark action at $\beta = 6/g^2=1.83$ (corresponding lattice spacing in the physical unit, $a = 0.1209~{\rm{ fm}}$~\cite{Ishikawa:2007nn})  
on a $L^3 \times T = 16^3 \times 32$ lattice (corresponding lattice size in the physical unit, 
$(1.93~{\rm{fm}})^3\times 3.87~{\rm fm} $). 
In our calculation, 
the hopping parameter for the $s$-quark is kept as
 $\kappa_{s} = 0.13710$, while the 
 three gauge ensembles, $\kappa_{u,d} = 0.13760$ (Set 1), 0.13800 (Set 2) and 0.13825 (Set 3), are taken for
 $u,d$-quarks. 

Wall source operators which generate positive parity two-baryon states with flavor structures $h_1$ and $h_2$ are given by
\begin{eqnarray}
{\mathcal{I}}^h_{\alpha \beta} = 
\left[ \epsilon_{abc} (\bar{Q}_a C \gamma_5 \bar{Q}^T_b) \bar{Q}_{c \alpha} \right]_{h_2}
\left[ \epsilon_{def} (\bar{Q}_d C \gamma_5 \bar{Q}^T_e) \bar{Q}_{f \beta} \right]_{h_1},
\end{eqnarray}
where $\bar{Q} = \sum_{\vec{x}} \bar{q}(\vec{x})$ is the quark wall-source. 
Projection operators for spin-singlet and spin-triplet states are given by
\begin{eqnarray}
P^{S=0}_{\alpha \beta} \equiv \frac{1 - \vec{\sigma}_1 \cdot \vec{\sigma}_2}{4} 
~~{\rm{and}}~~
P^{S=1}_{\alpha \beta} \equiv \frac{3 + \vec{\sigma}_1 \cdot \vec{\sigma}_2}{4}.
\end{eqnarray}

\begin{figure}
\begin{center}
\begin{tabular}{cc}
\includegraphics[scale=0.55]{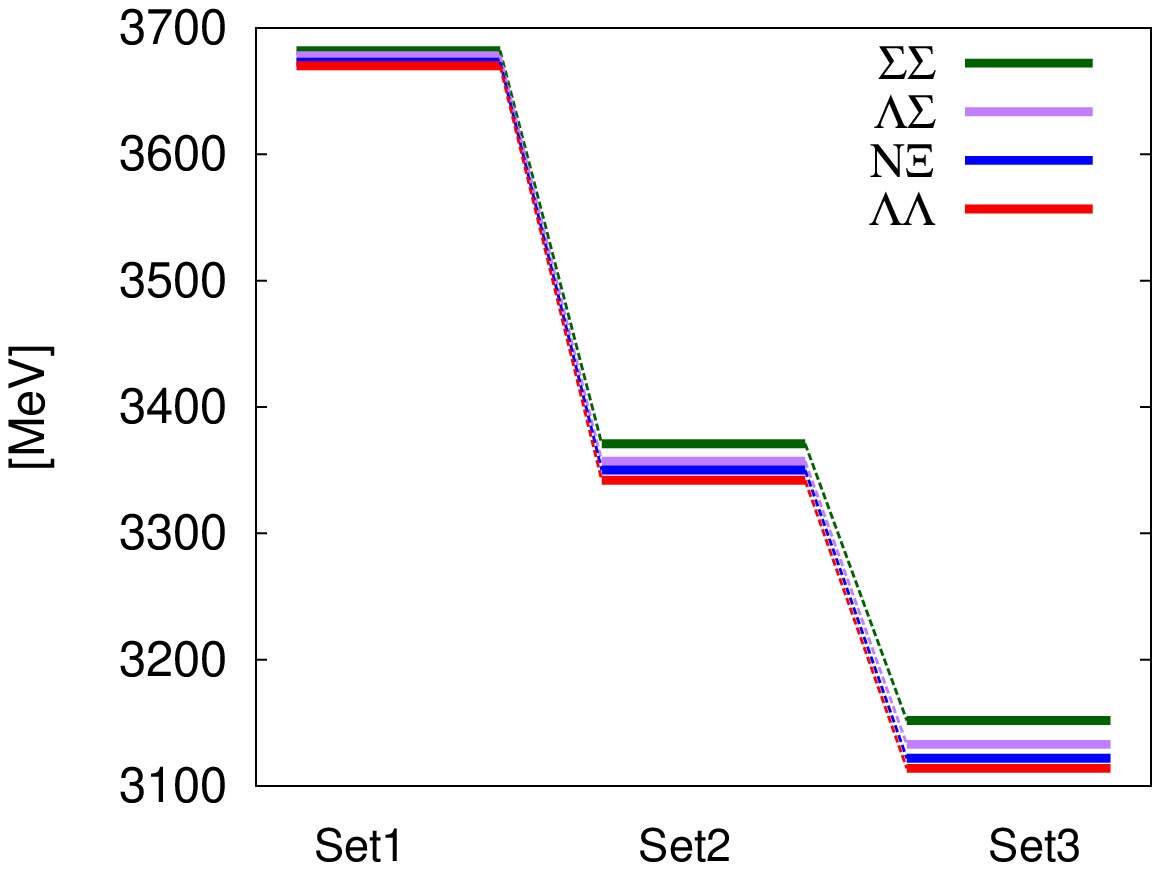}
&
\includegraphics[scale=0.55]{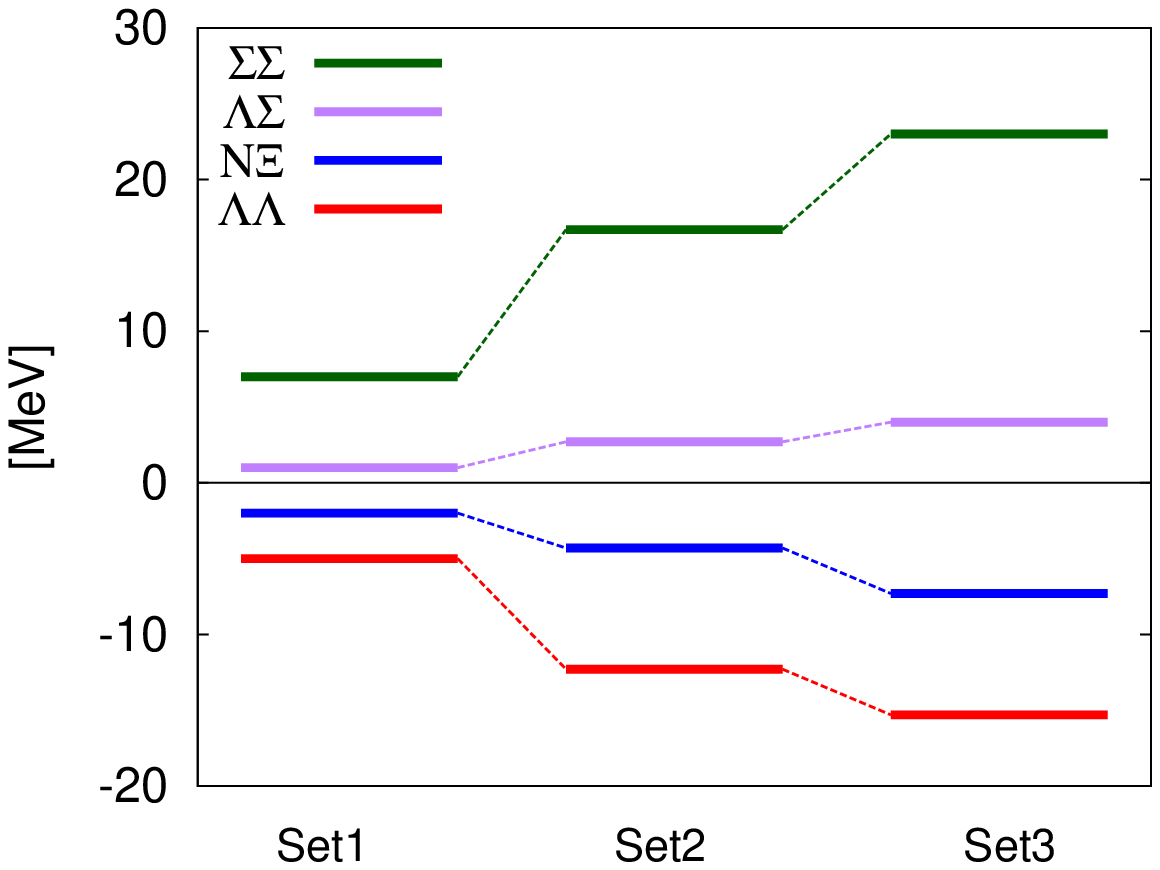}
\end{tabular}
\end{center}
\caption{
Thresholds of two-baryons with the strangeness $S=-2$ for each gauge ensemble.
(Left)  A sum of masses for each channel in units of MeV. (Right) A sum of  two-baryon masses in each channel minus the average of three channels, $(m_N+2m_\Lambda +m_\Xi + 2 m_\Sigma)/3$ .}
\label{FIG.Thresholds}
\end{figure}

Quark propagators are calculated for the wall source at $t_0$ with the Dirichlet boundary condition in the temporal direction at $t = 16+t_0 $. 
The wall source is placed at $32$ different values of $t_0$  on each gauge configuration, in order to increase the statistics, in addition to the average over forward and backward propagations in time. 
The $A_1^+$ projection of the cubic group is taken for the sink operator to obtain the relative $S$-wave in the $BB$ wave function\footnote{In this paper, relative $D$-waves in spin-triplet channels are not explicitly considered but their effect is included implicitly in the effective central potentials for spin-triplet channels.
}. 
Numerical computations have been carried out using the  KEK supercomputer system, Blue Gene/L, and the kaon and jpsi clusters at Fermilab.
Hadron masses obtained in our calculation are given in Table~\ref{TAB:HadronM}.
Thresholds of two-baryons with the strangeness $S=-2$ for each set of gauge configurations are plotted in Fig.~\ref{FIG.Thresholds}.

\section{Numerical Results}
We now present our results of coupled channel $BB$ potentials in strangeness $S=-2$ sector.

\subsection{Time dependence}
\begin{figure}
\begin{tabular}{cc}
  \includegraphics[scale=0.55]{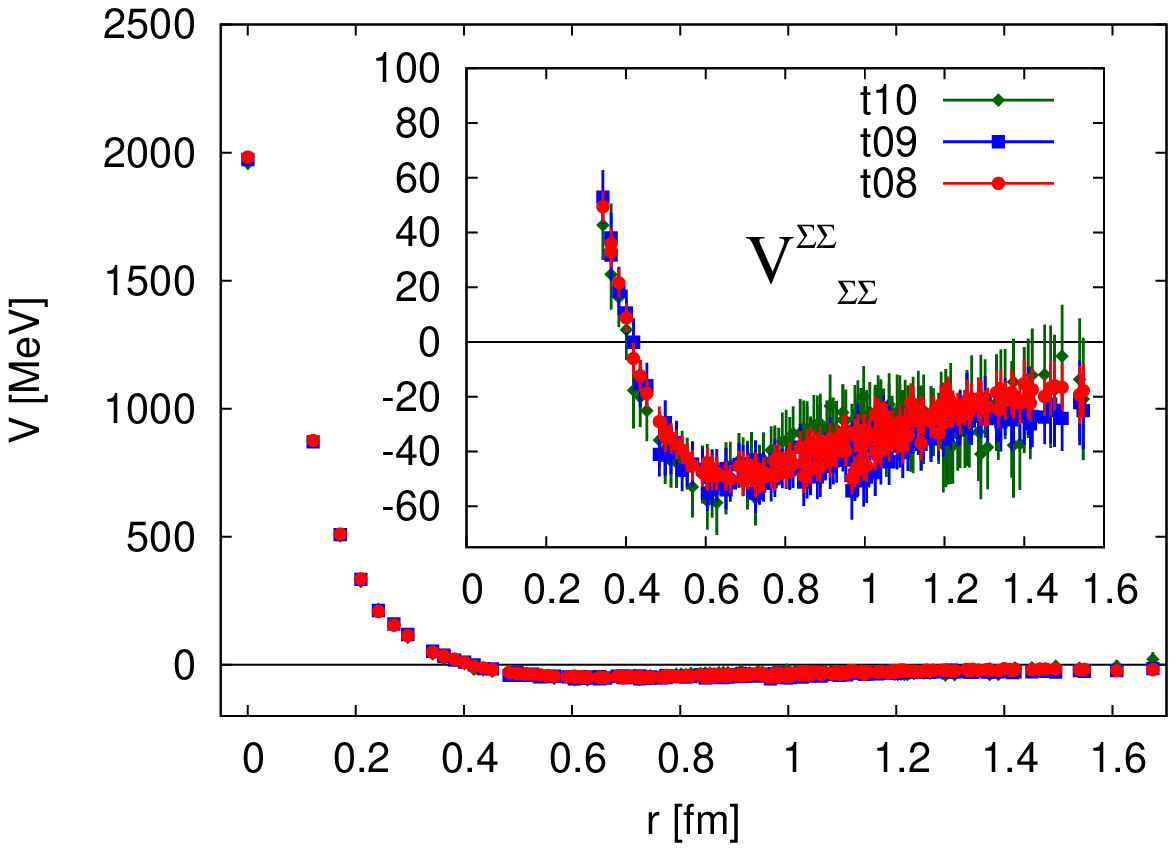} &  
  \includegraphics[scale=0.55]{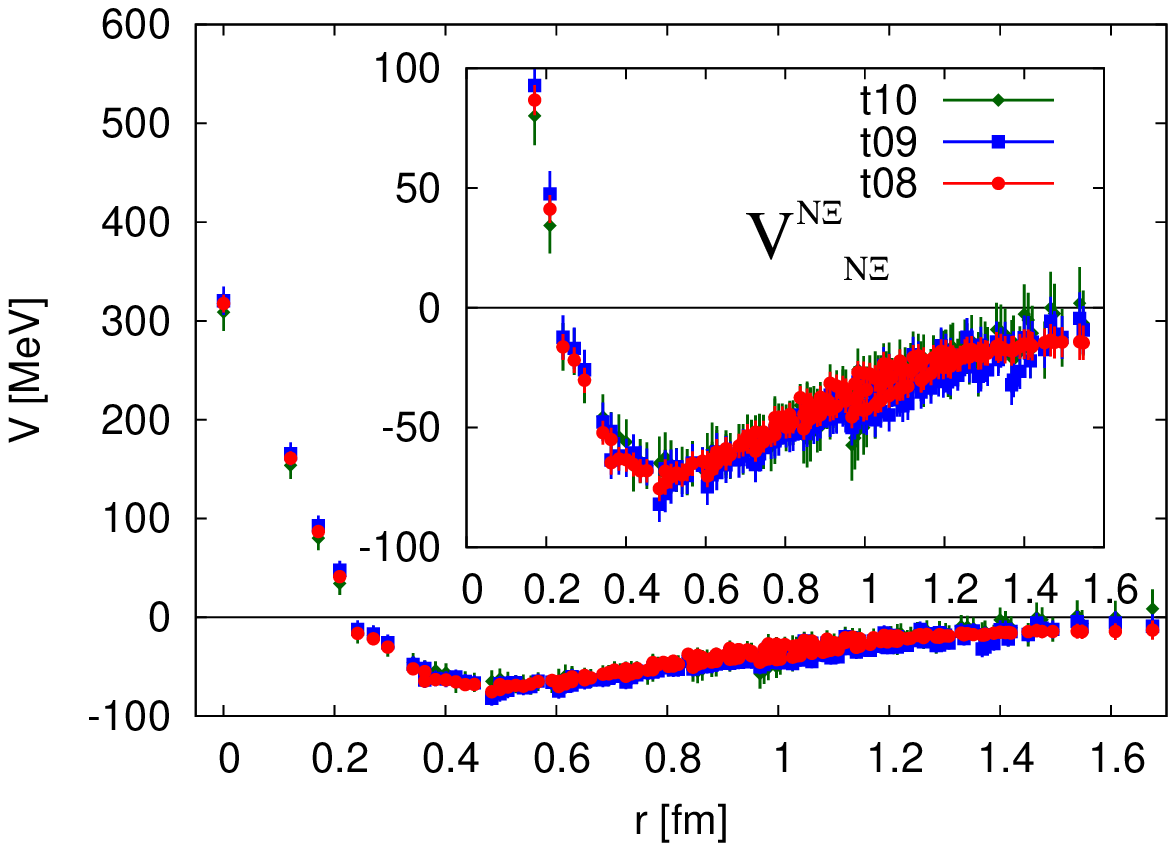} 
\end{tabular}
\caption{ $\Sigma \Sigma$ potential in the ${^1S_0}$ ($I=2$) channel (left) and 
$N\Xi$ potential in the ${^3S_1}$ ($I=0$) channel (right) as a function of $r$ 
at $t-t_0 = 8$(red), $9$ (blue) and $10$ (green) calculated with Set 3.}
\label{FIG:t-dep-single}
\end{figure}

We first show how the  time-dependent method extended to the coupled channel system works in our calculation.
For this purpose, we investigate time dependences of the diagonal potentials.
Fig.~\ref{FIG:t-dep-single} shows ${V^{\Sigma\Sigma}}_{\Sigma\Sigma}$ in $^1S_0$ ($I=2$) channel (5th line in Table~\ref{TAB:S2channel}) and ${V^{N\Xi}}_{N\Xi}$ in  $^3S_1$ ($I=0$) channel (2nd line in Table~\ref{TAB:S2channel}) at three values of $t-t_0$ ($=8,9,10$) with Set 3, which has the lightest pion mass in our calculation.  Within statistical errors, no significant $t-t_0$ dependence is observed for these single channel potentials with Set 3, showing that $t-t_0=8$ is large enough to suppress inelastic contributions and that higher order contributions in the derivative expansion are negligible. 

\begin{figure}
\begin{tabular}{cc}
 \includegraphics[scale=0.55]{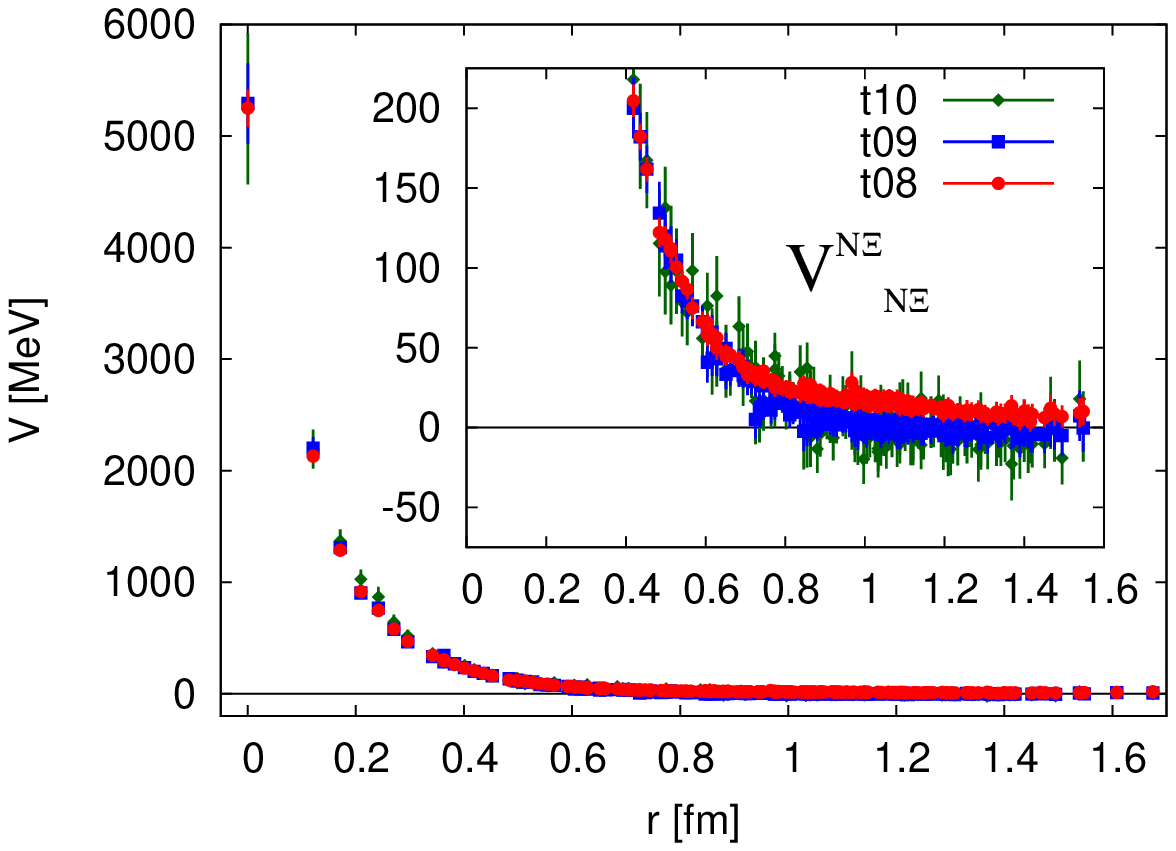}  & 
 \includegraphics[scale=0.55]{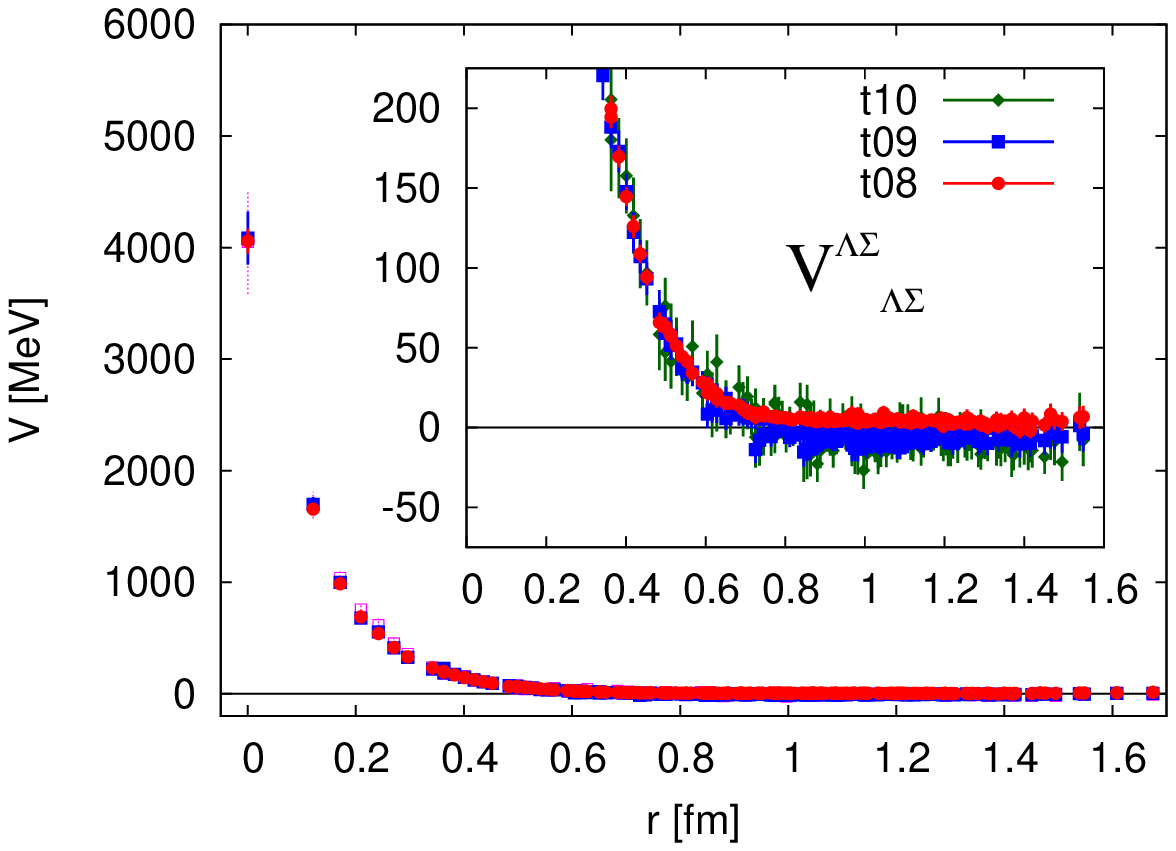}
\end{tabular}
\caption{ Diagonal parts of  potential matrix  in the ${^1S_0}$ ($I=1$) channel, ${V^{N\Xi}}_{N\Xi}$ (left) and 
${V^{\Lambda\Sigma}}_{\Lambda\Sigma}$ (right),  at $t-t_0 = 8$(red), $9$ (blue) and $10$ (green) calculated with Set 1.}
\label{FIG:t-dep-double}
\end{figure}

\begin{figure}
\begin{tabular}{ccc}
\hspace*{-1em}
  \includegraphics[scale=0.40]{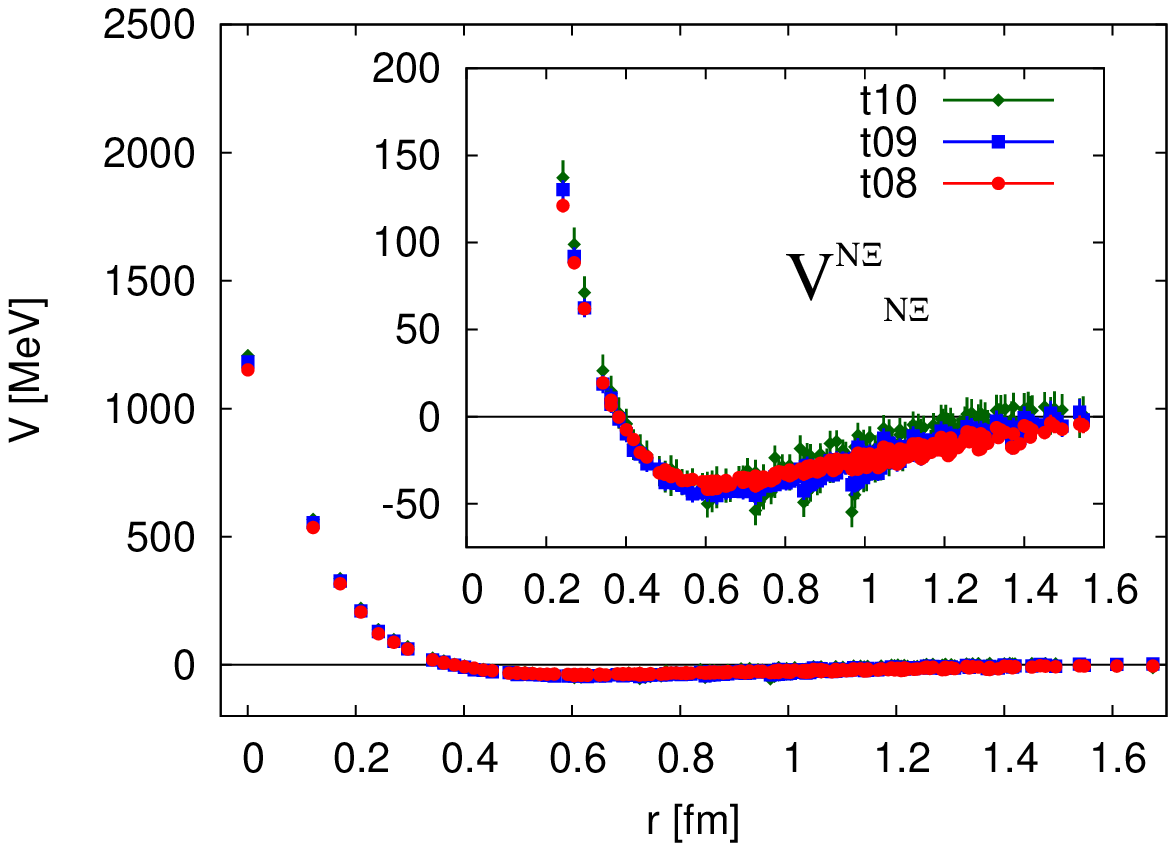} \hspace*{-1.5em}
& \includegraphics[scale=0.40]{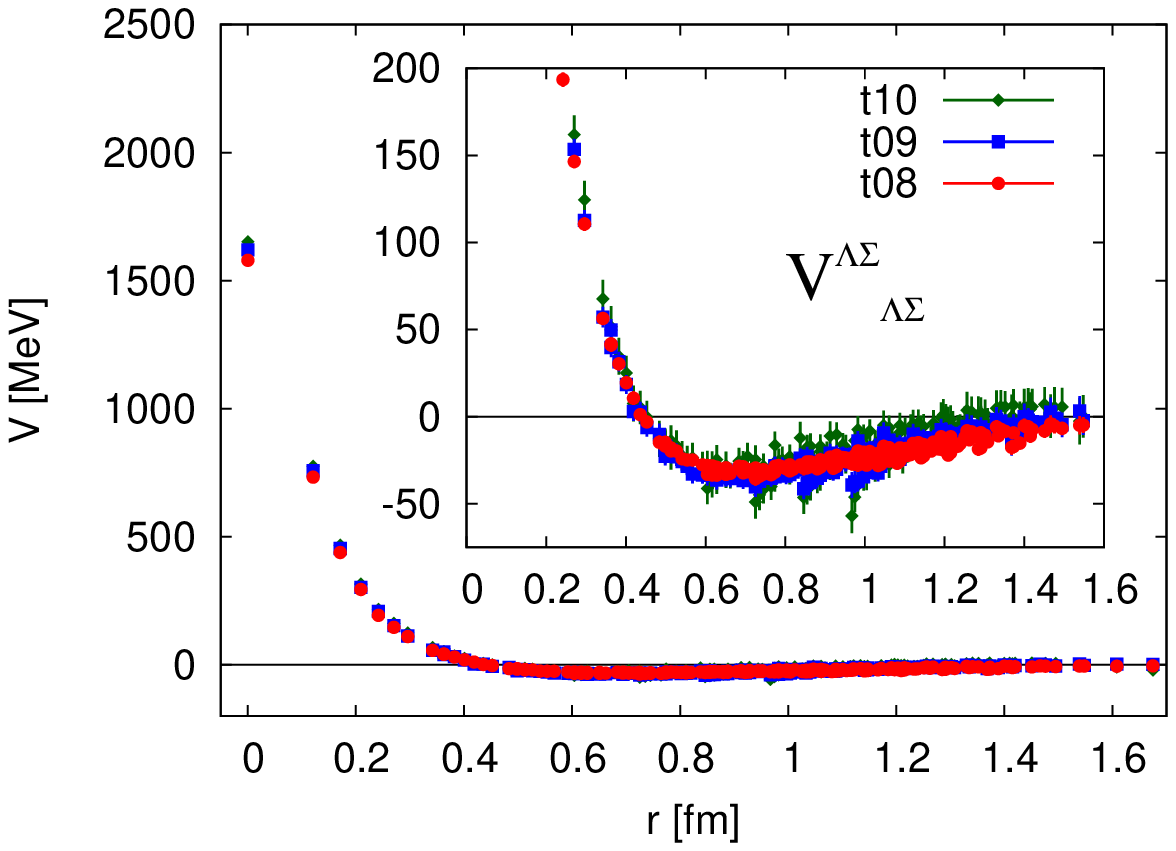} \hspace*{-1.5em}
& \includegraphics[scale=0.40]{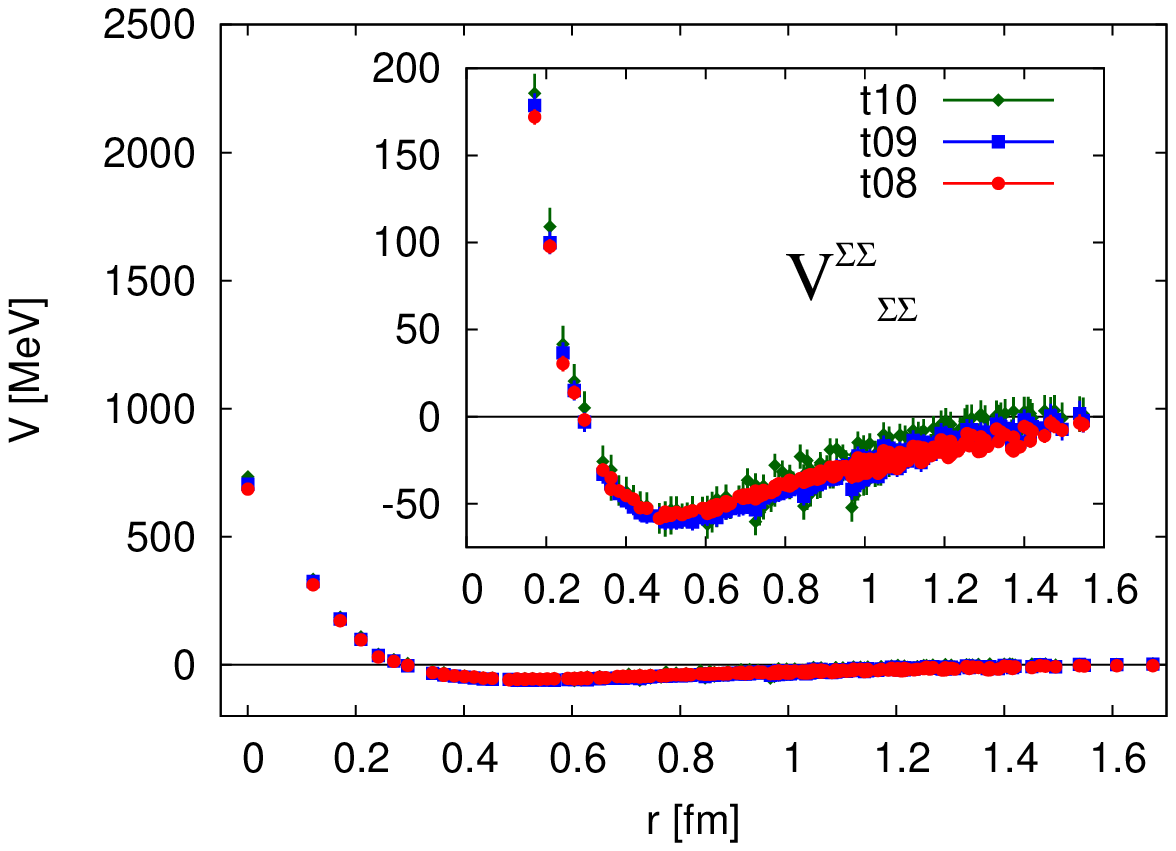} \hspace*{-1em}
\\
\hspace*{-1em}
  \includegraphics[scale=0.40]{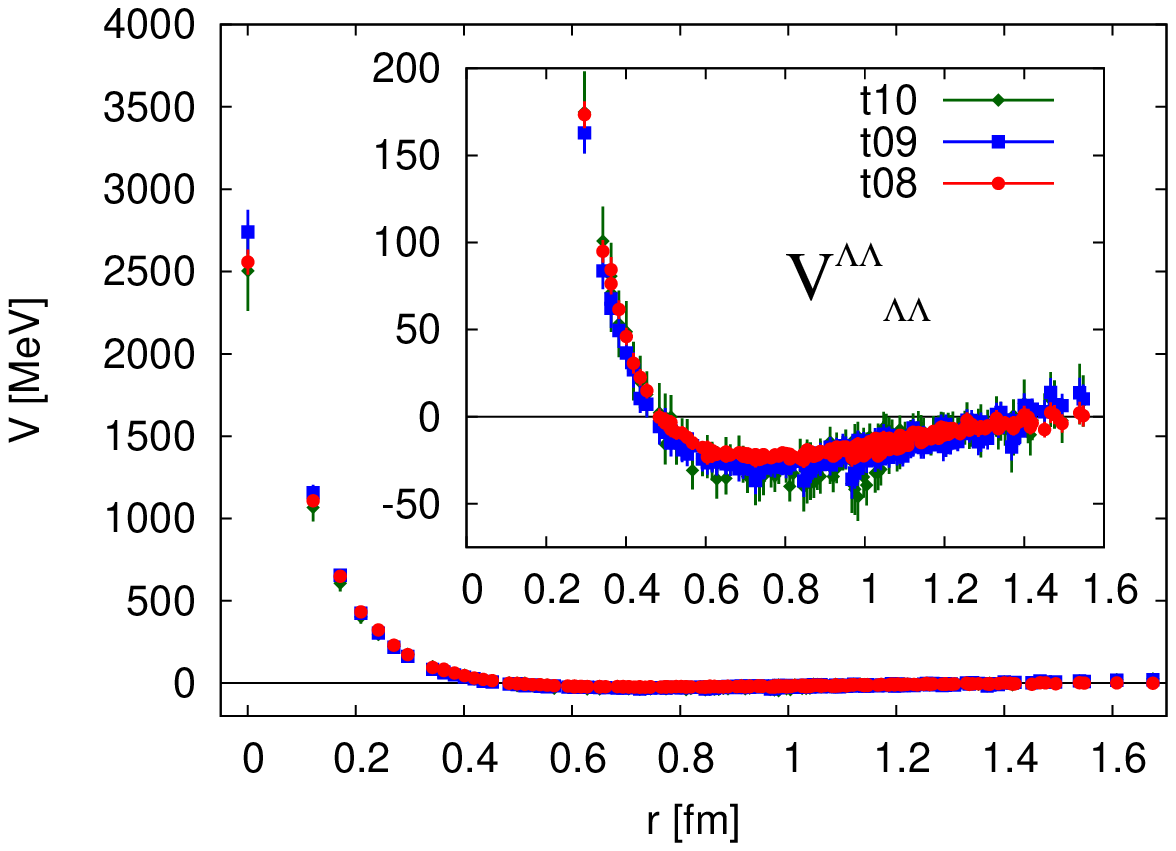} \hspace*{-1.5em}
& \includegraphics[scale=0.40]{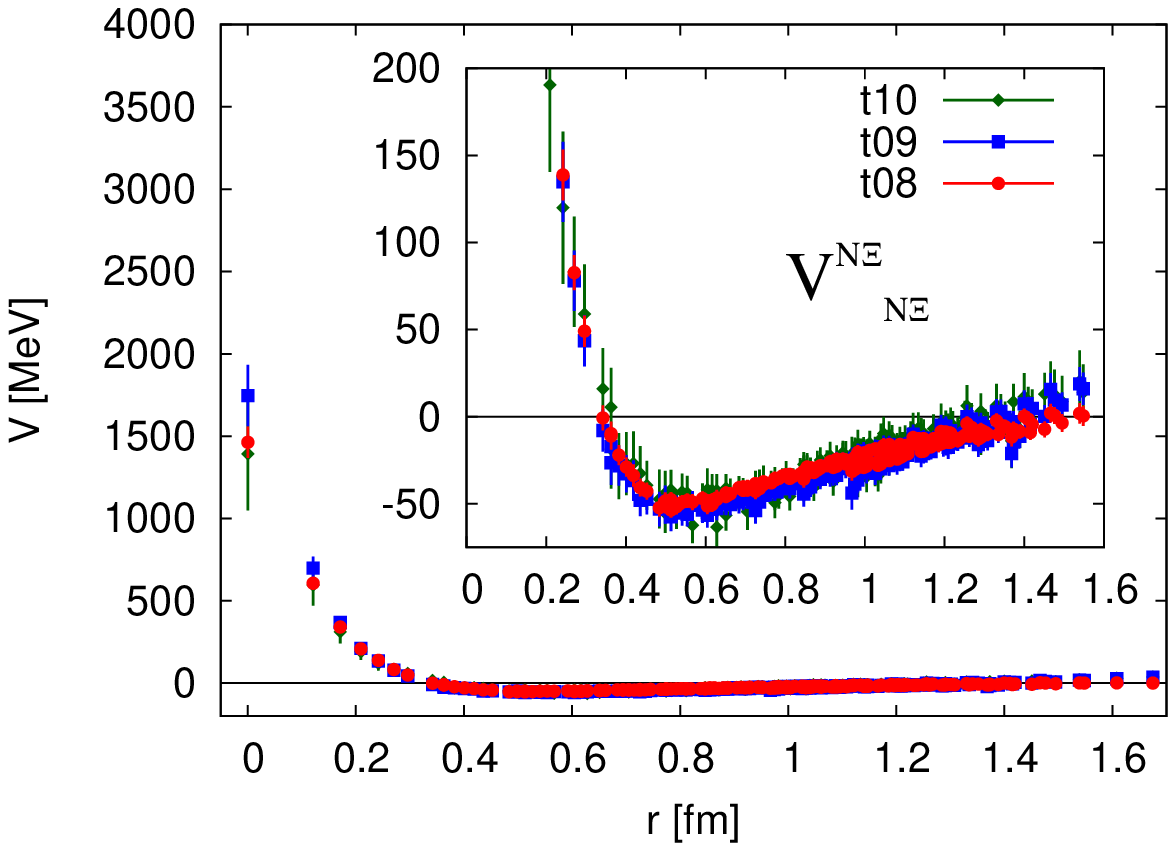} \hspace*{-1.5em}
& \includegraphics[scale=0.40]{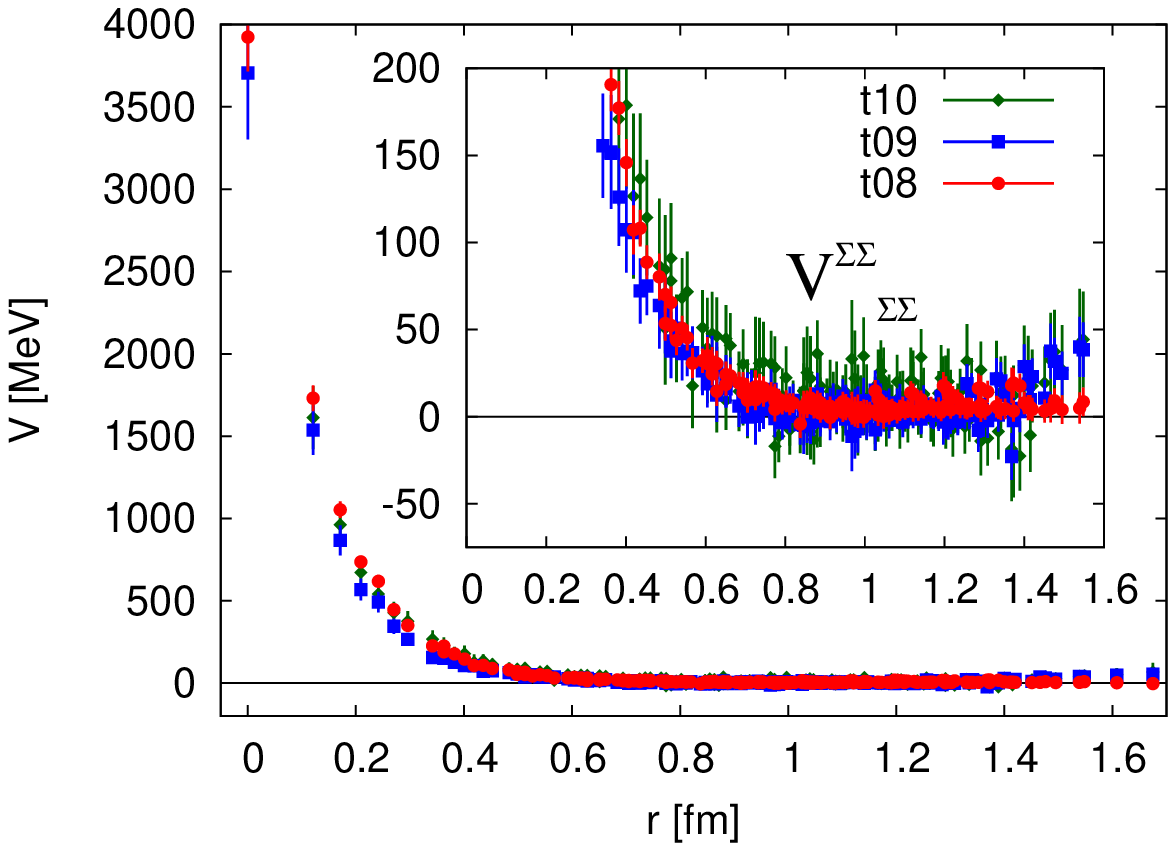} \hspace*{-1em}
\end{tabular}
\caption{(Upper) Diagonal parts of  potential matrix in the ${^3S_1}$ ($I=1$) channel, ${V^{N\Xi}}_{N\Xi}$ (left),
${V^{\Lambda\Sigma}}_{\Lambda\Sigma}$ (center), and ${V^{\Sigma\Sigma}}_{\Sigma\Sigma}$ (right),  at $t-t_0 = 8$(red), $9$ (blue) and $10$ (green) calculated with Set 2.
(Lower) Same as above but in the ${^1S_0}$ ($I=0$) channel, ${V^{\Lambda\Lambda}}_{\Lambda\Lambda}$ (left),
${V^{N\Xi}}_{N\Xi}$ (center), and ${V^{\Sigma\Sigma}}_{\Sigma\Sigma}$ (right).
}
\label{FIG:t-dep-triple}
\end{figure}

In Fig.~\ref{FIG:t-dep-double}, two diagonal potentials in $^1S_0$ ($I=1$) channel (3rd line in Table~\ref{TAB:S2channel}) calculated with Set 1 are shown at $t-t_0=8 \sim 10$. Again no significant $t-t_0$ dependence is observed at this quark mass and this is true at other quark masses. Similarly,  three diagonal potentials in $^3S_1$ ($I=1$) channel (4th line in 
Table~\ref{TAB:S2channel}) and those in $^1S_0$ ($I=0$) channel (1st line in Table~\ref{TAB:S2channel})
 show no significant $t-t_0$ dependence at all quark masses, as seen in Fig.~\ref{FIG:t-dep-triple}
for the Set 2.

Since no significant $t-t_0$ dependences are observed for all diagonal potentials at $t-t_0=8,9,10$, we hereafter consider results at $t-t_0=8$, where statistical errors are smallest. 

\subsection{Hermiticity}
Hermiticity of the potential matrix is a sufficient condition for the probability 
conservation, though it is not a necessary condition. 
In this subsection, we investigate the Hermiticity of the potential matrix, ${V^a}_b = {V^b}_a$, since it
 is not automatically guaranteed in the definition of the coupled channel potential matrix in eq.~(\ref{EQ.fctK}).
As in the case of the diagonal parts, we confirm that off-diagonal parts of potential matrix show no significant $t-t_0$ dependence, so we take results at $t-t_0=8$ in our analysis.

\begin{figure}
\begin{center}
\begin{tabular}{c}
\includegraphics[scale=0.55]{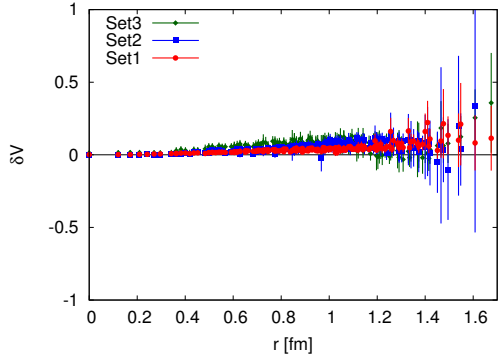} 
\end{tabular}
\end{center}
\caption{Hermiticity measure for off-diagonal elements of potential matrix, $\delta V_{N\Xi-\Lambda\Sigma}$ in the ${^1S_0}$ ($I=1$) channel on Set 1 (red), Set 2 (blue) and Set 3 (green).
}
\label{FIG:1S0I1-h}
\end{figure}

We introduce an Hermiticity measure$\delta V_{a-b} \equiv 2({V^a}_b - {V^b}_a)/({V^a}_b + {V^b}_a)$ to see the relative magnitude of the hermiticiy  violation of the potential matrix.
Fig.~\ref{FIG:1S0I1-h} presents $\delta V_{N\Xi-\Lambda\Sigma}$ in the $^1S_0$ ($I=1$) channel  
with Set 1 (red), Set 2 (blue) and Set 3 (green).
It satisfies the Hermiticity well within the statistical errors.

\begin{figure}
\begin{tabular}{ccc}
\hspace*{-1em}
  \includegraphics[scale=0.40]{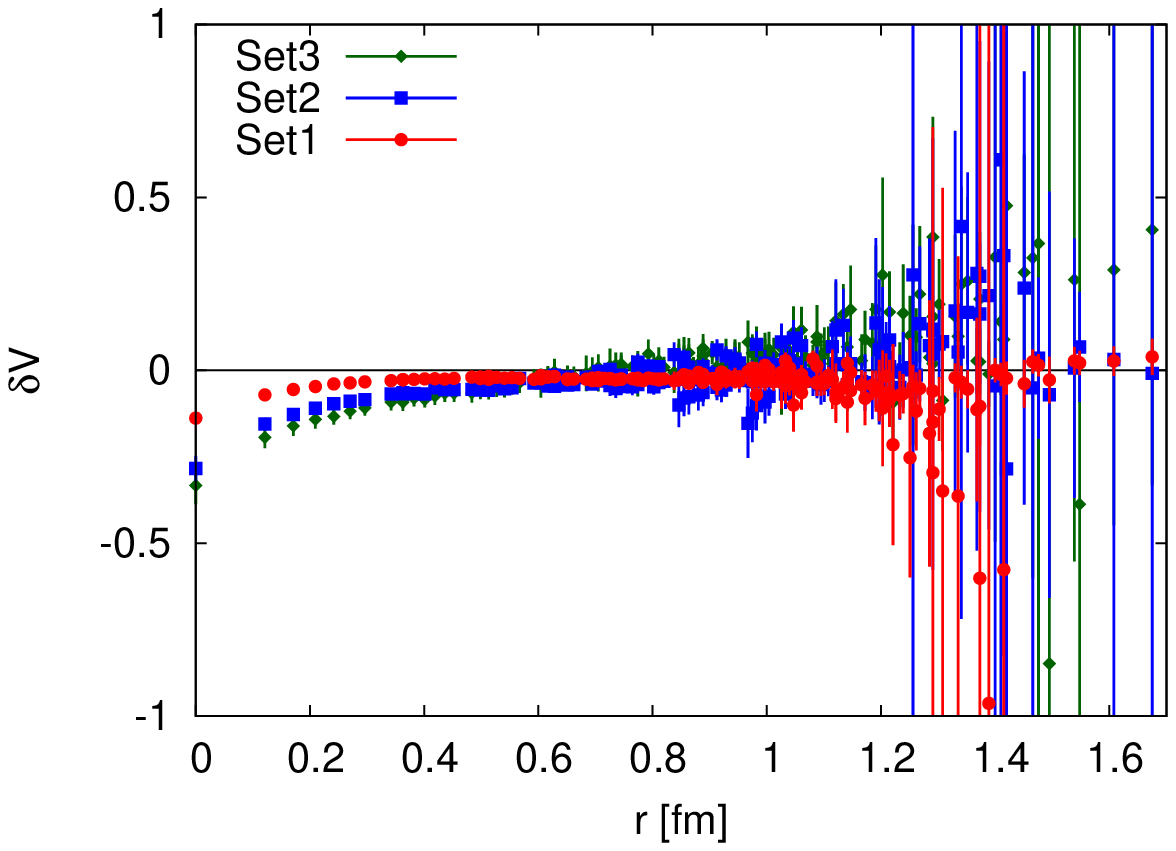} \hspace*{-1.5em}
& \includegraphics[scale=0.40]{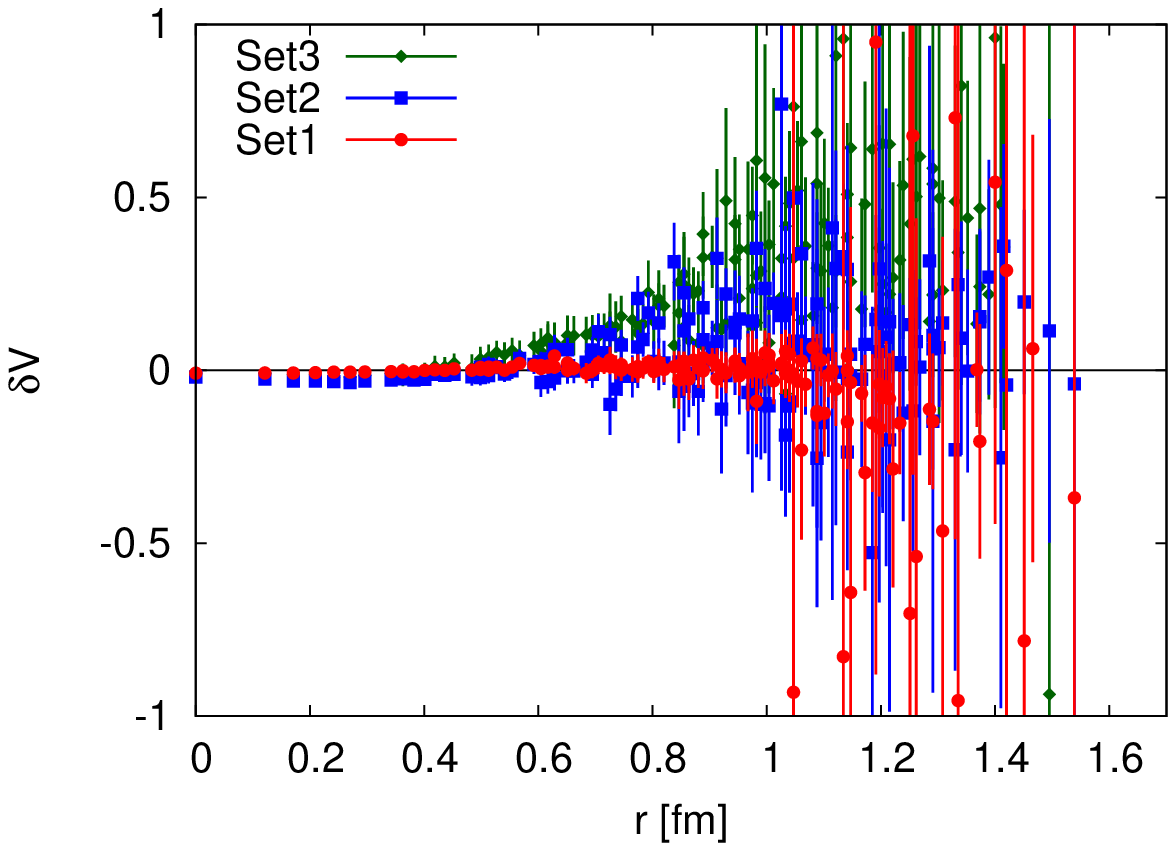} \hspace*{-1.5em}
& \includegraphics[scale=0.40]{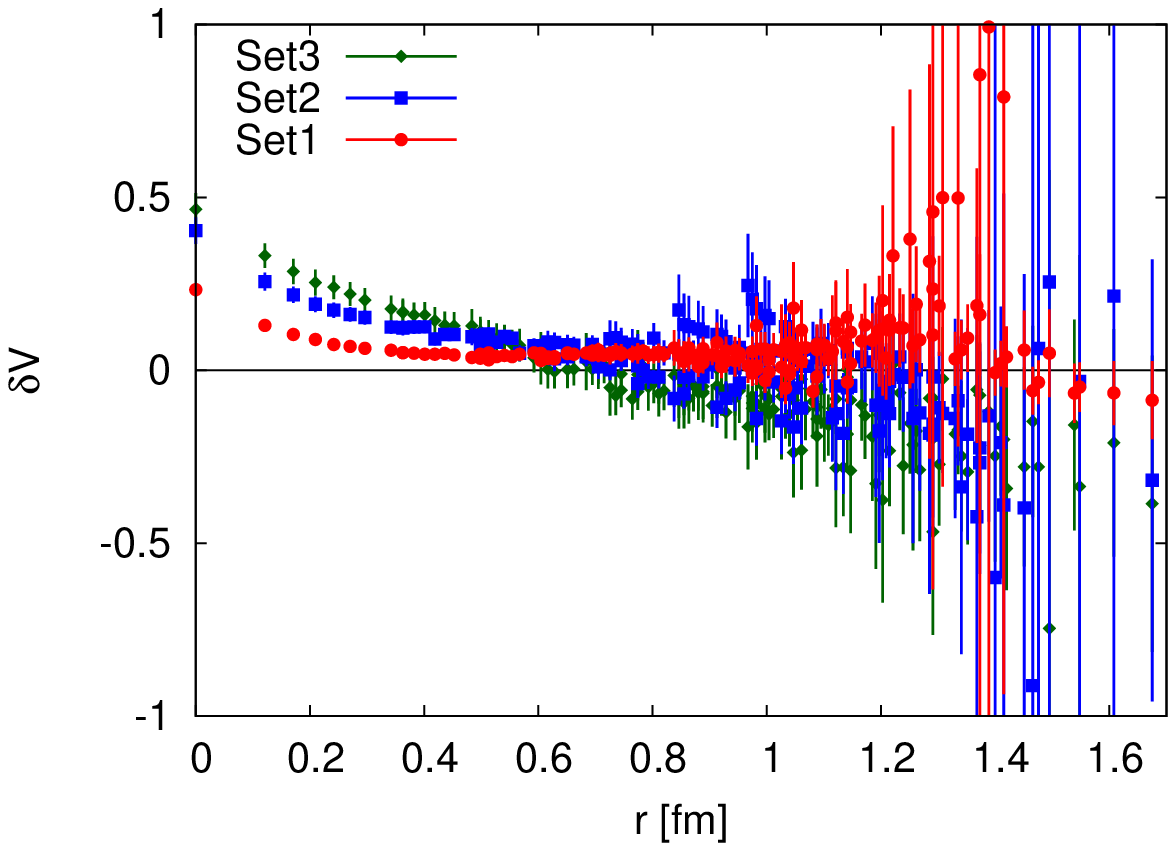} \hspace*{-1em}
\end{tabular}
\caption{Violation of Hermiticity  in the channel  ${^3S_1}$ ($I=1$): 
(left) $\delta V_{N\Xi-\Lambda\Sigma}$ (center) $\delta V_{N\Xi-\Sigma\Sigma}$
(right) $\delta V_{\Lambda\Sigma-\Sigma\Sigma}$ 
on Set 1 (red), Set 2 (blue) and Set 3 (green).
}
\label{FIG:3S1I1-h}
\end{figure}

\begin{figure}
\begin{tabular}{ccc}
\hspace*{-1em}
  \includegraphics[scale=0.40]{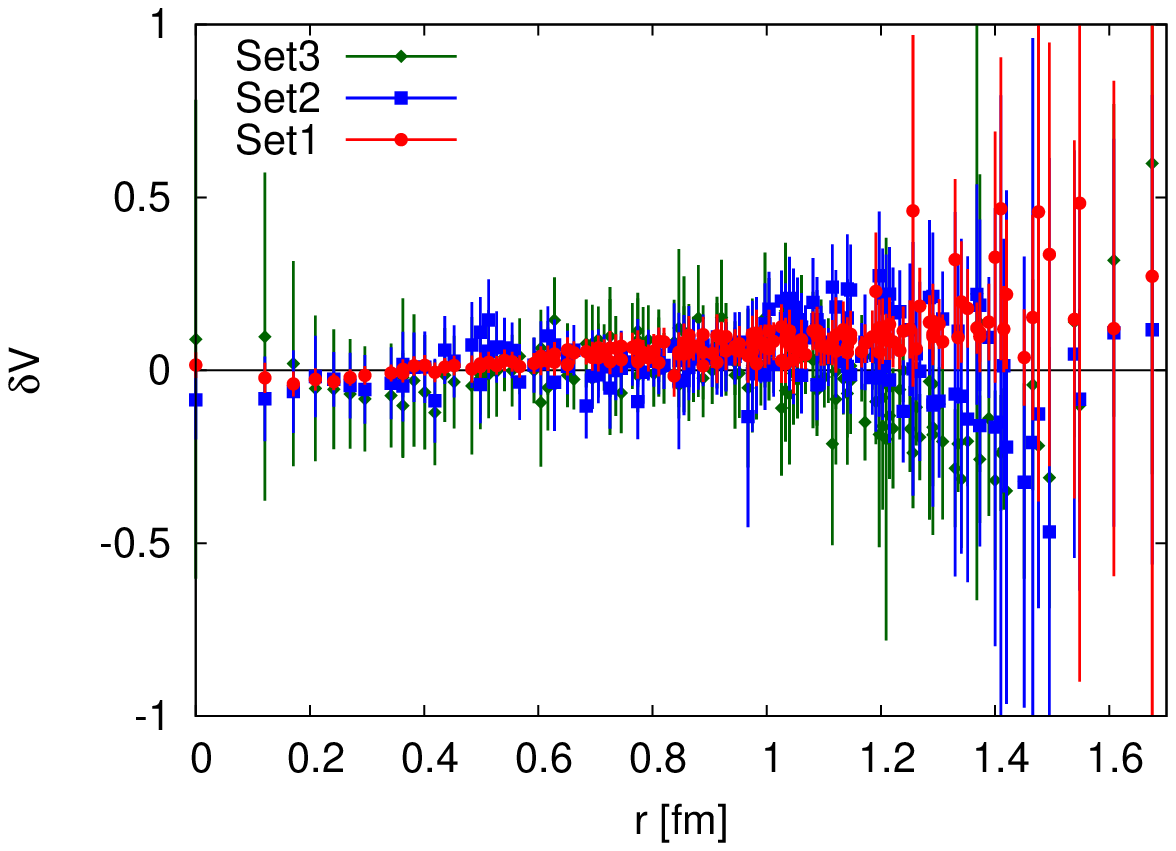} \hspace*{-1.5em}
& \includegraphics[scale=0.40]{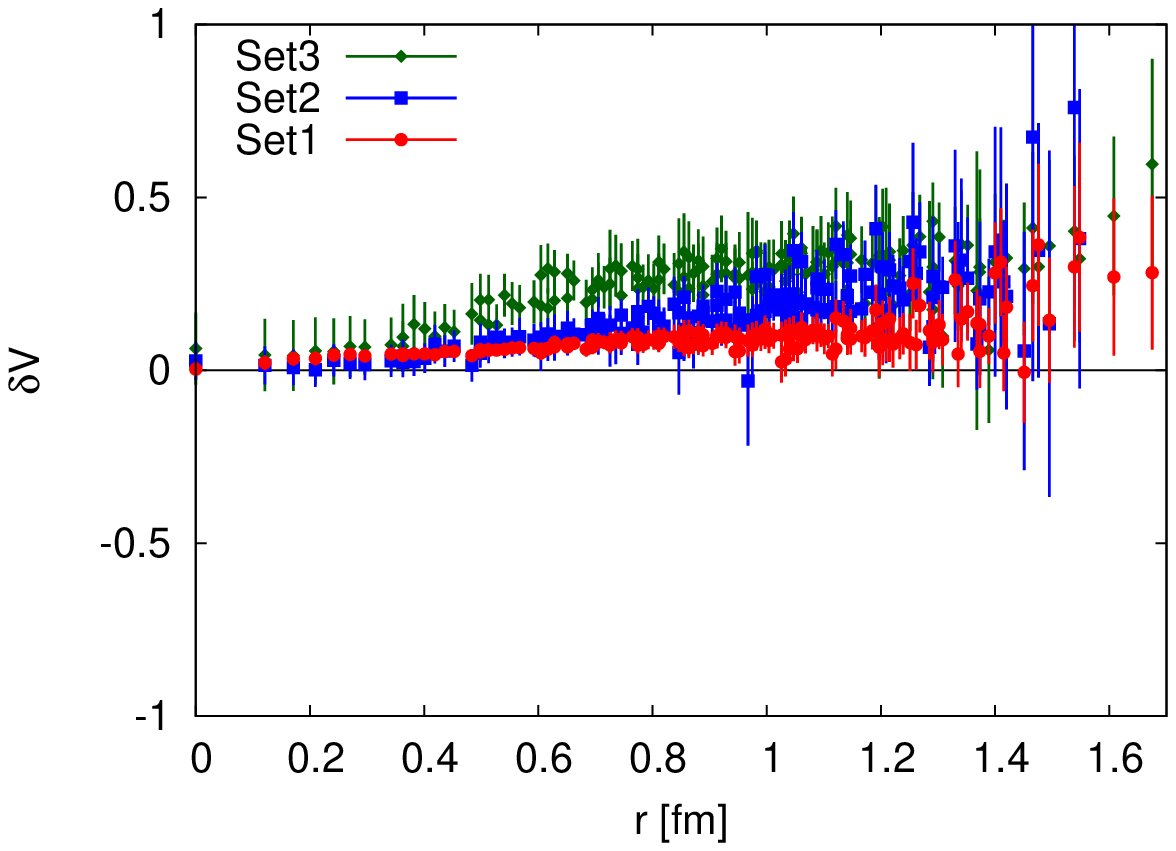} \hspace*{-1.5em}
& \includegraphics[scale=0.40]{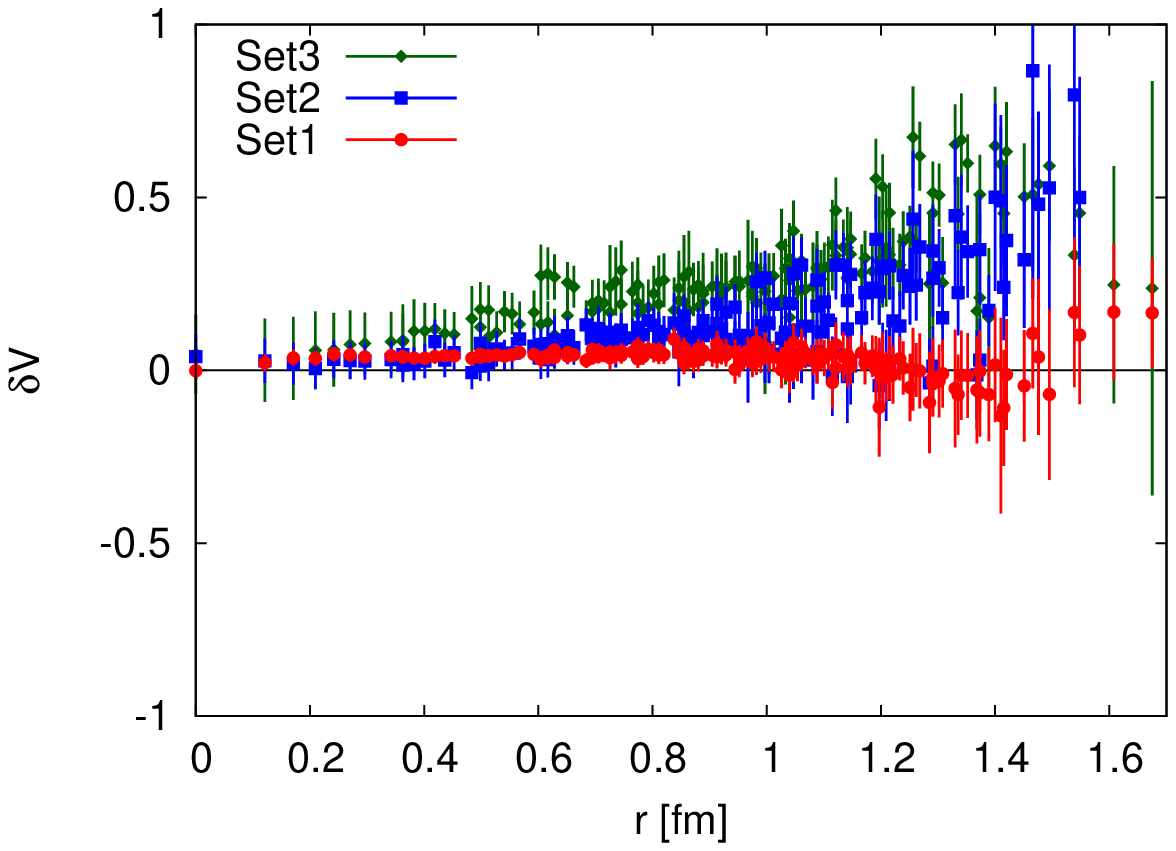} \hspace*{-1em}
\end{tabular}
\caption{Violation of Hermiticity  in the channel  ${^1S_0}$ ($I=0$): 
(left) $\delta V_{\Lambda\Lambda-N\Xi}$ (center) $\delta V_{\Lambda\Lambda-\Sigma\Sigma}$
(right) $\delta V_{N\Xi-\Sigma\Sigma}$ 
on Set 1 (red), Set 2 (blue) and Set 3 (green).}
\label{FIG:1S0I0-h}
\end{figure}

Fig.~\ref{FIG:3S1I1-h} shows $\delta V_{a-b}$ for $a,b = N\Xi, \Lambda\Sigma, \Sigma\Sigma$ in the $^3S_1$ ($I=1$) channel.
 Some violations of Hermiticity can be seen in $\delta V_{N\Xi-\Lambda\Sigma}$ and $\delta V_{\Lambda\Sigma-\Sigma\Sigma}$ at $r<0.5$~fm region.
Those for $a,b =  \Lambda\Lambda, N\Xi, \Sigma\Sigma$ in ${^1S_0}$ $(I=0)$ are given in Fig.~\ref{FIG:1S0I0-h}. Hermiticity is
more or less satisfied within the statistical errors.
It is our future problem to check whether possible Hermiticity breaking 
for small $r$ in Fig.6 and large $r$ in Fig.7 disappears or not by removing our assumption on the wave-function renormalization factor introduced after eq.~(\ref{EQ:NBSdef}).


\subsection{Potential matrices and their quark mass dependence} 
We here separately discuss properties of potentials in three cases, single channel, two channels and three channels.

\subsubsection{Single channel}
\begin{figure}
\begin{tabular}{cc}
  \includegraphics[scale=0.55]{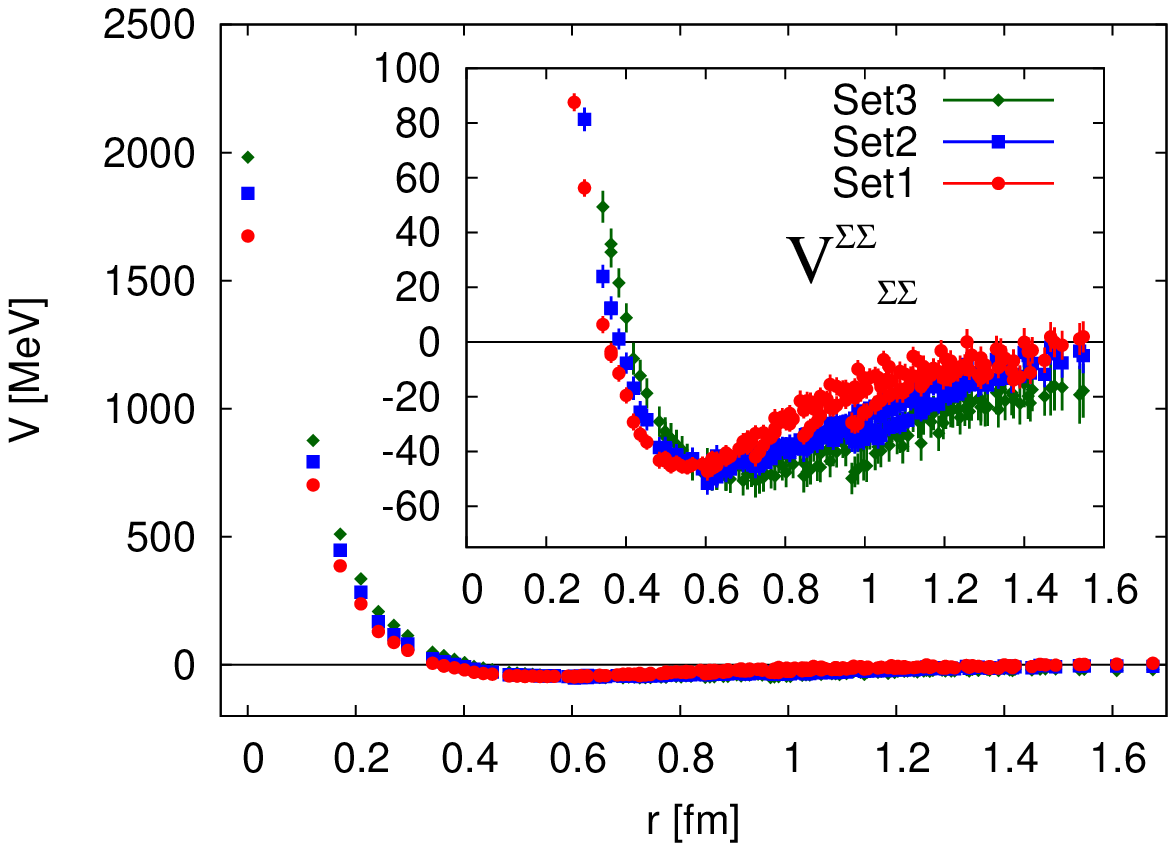} 
&
  \includegraphics[scale=0.55]{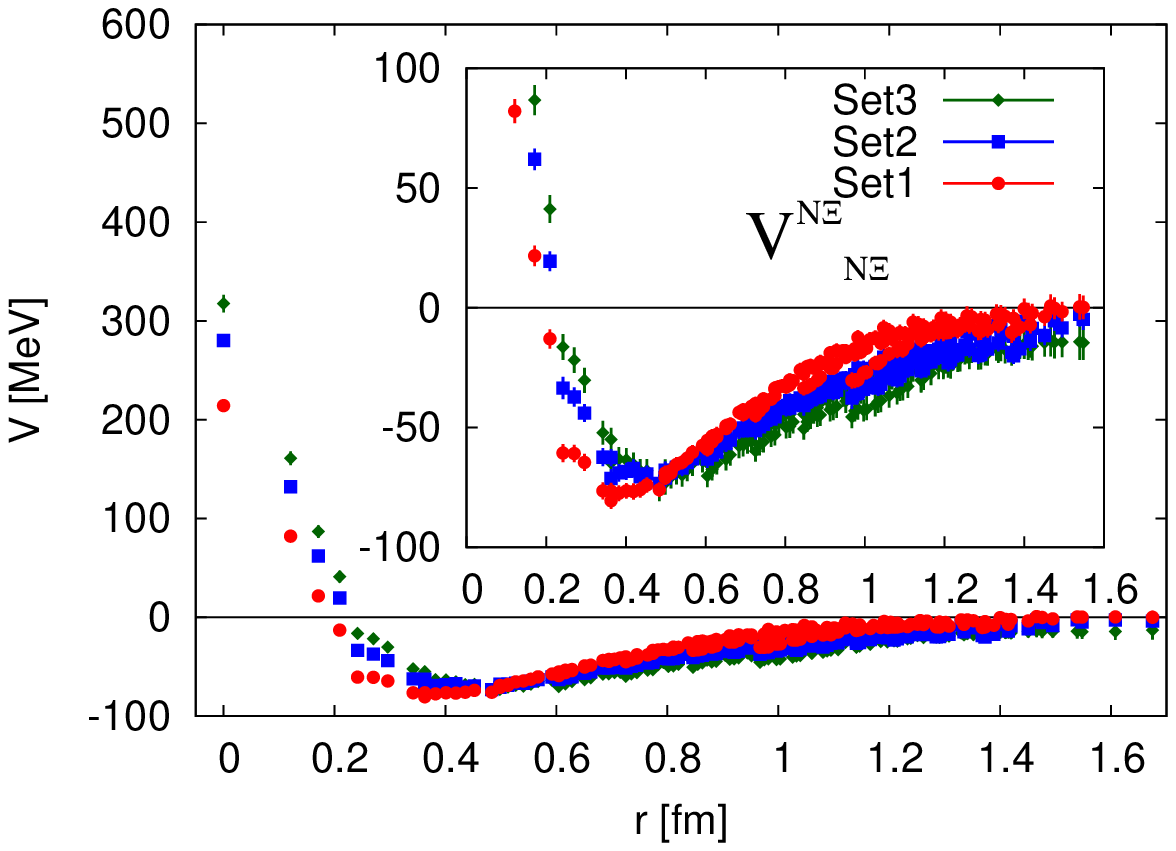} 
\end{tabular}
\caption{The $\Sigma\Sigma$ potential in the ${^1S_0} (I=2)$ channel (left) and the $N\Xi$ potential in the ${^3S_1} (I=0)$ channel  (right). 
Result from three gauge ensembles, Set 1 (red), Set 2 (blue) and Set 3 (green), are shown in one figure. 
Insets show the enlargement of the same plot.
}
\label{FIG:1S0I2}
\end{figure}
Fig.~\ref{FIG:1S0I2} shows quark mass dependences of the $\Sigma\Sigma$ potential in ${^1S_0} (I=2)$ channel (left) and the $N\Xi$ potential in the ${^3S_1} (I=0)$ channel (right).
We first notice non-smooth behaviors as a function of $r$ at large $r$ for both cases, which indicate that the spatial volume is not sufficiently large. In addition, non-smooth behavior at short distance observed in the $N\Xi$ ${^3S_1} (I=0)$ channel may be caused by the finite lattice spacing effect.
 With these systematics, we discuss only qualitative features of potentials in this report, and leaves
 quantitative analysis such as the extraction of scattering phase shifts  for future studies with larger and finer lattices.

The $\Sigma\Sigma$ potential in the ${^1S_0} (I=2)$ channel (left), which belongs to the $\boldmath{27}$-plet irreducible representation in the flavor SU(3), has repulsion at short distance and  attraction at long distance.
Also,  the magnitude of these two components increases as the light ($ud$) quark mass decreases,
as in the case of the $NN$ potential in the ${^1S_0}$ sector belonging to $\boldmath{27}$-plet. 
An increase of attraction at long distance, $r>0.8$~fm, may be related to the decrease of the mass of the pion
exchanged between two $\Sigma$'s.

Similarly, the $N\Xi$ potential in the  ${^3S_1} (I=0)$ channel (right) has
both  repulsion at short distance and  attraction at long distance.  The magnitude of these two components is enhanced as the light quark mass decreases. It should be remarked that the repulsion at short distance here is weaker than that of the $\Sigma\Sigma$ potential in the ${^1S_0} (I=2)$ channel. 
This difference may be related to the fact that the Pauli blocking in the quark level 
for $N\Xi$ in the  ${^3S_1} (I=0)$ channel  is  
 weaker than $\Sigma\Sigma$ in the ${^1S_0} (I=2)$ channel.

\subsubsection{Two channels}
\begin{figure}
\begin{tabular}{ccc}
\hspace*{-1em}
  \includegraphics[scale=0.40]{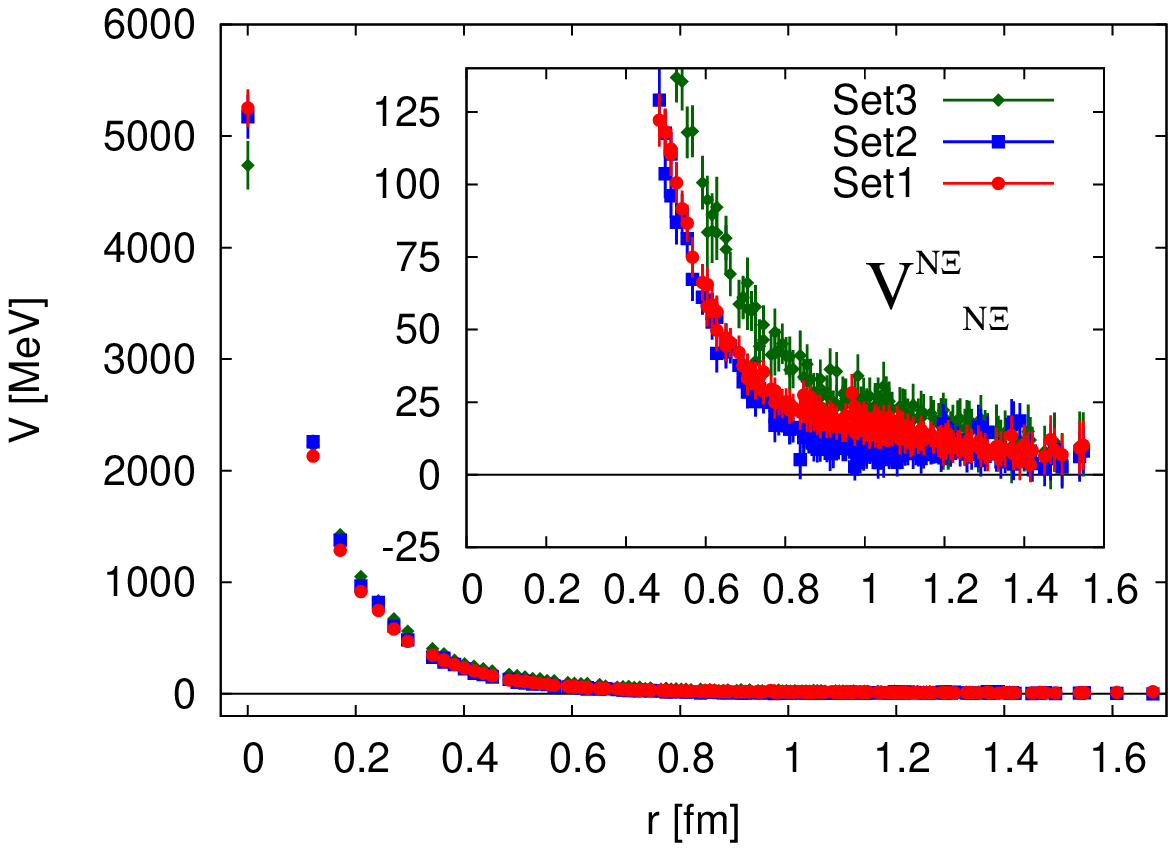} \hspace*{-1.5em}
& \includegraphics[scale=0.40]{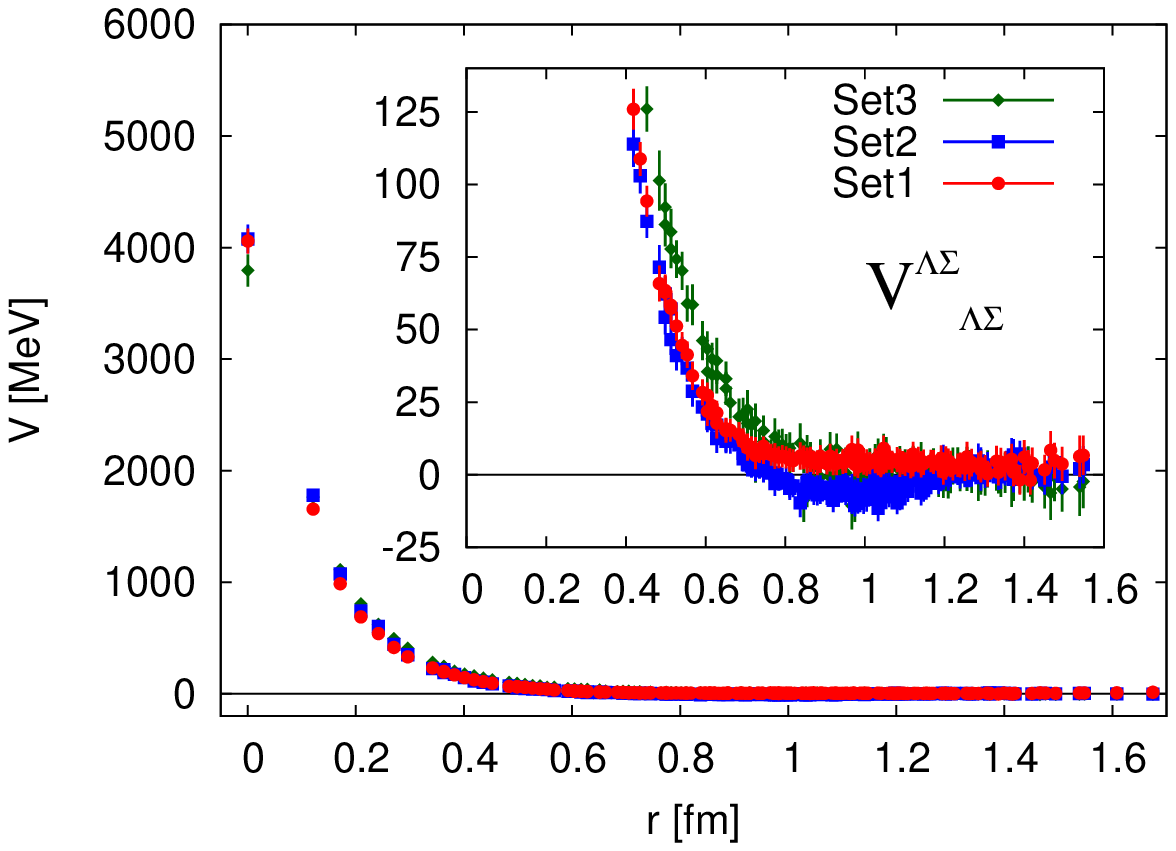} \hspace*{-1.5em}
& \includegraphics[scale=0.40]{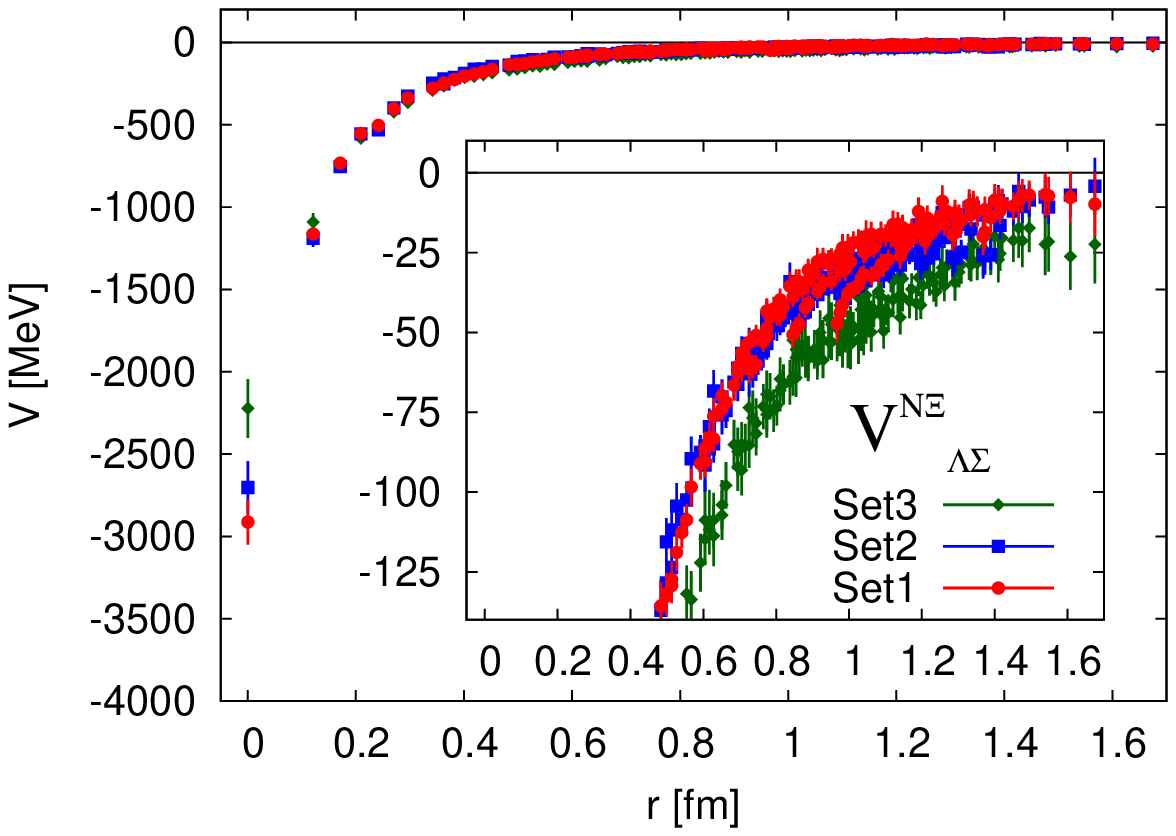} \hspace*{-1em}
\end{tabular}
\caption{Diagonal (left and center) and off-diagonal (right) elements of the potential matrix in the ${^1S_0} ( I=1)$ channel.
Result from three gauge ensembles, Set 1 (red), Set 2 (blue) and Set 3 (green), are shown in one figure. 
Insets show the enlargement of the same plot.
}
\label{FIG:1S0I1}
\end{figure}
A potential matrix in the ${^1S_0}$ ($ I=1$) channel, which has $N \Xi$ and $\Lambda \Sigma$ components, is given in Fig.~\ref{FIG:1S0I1}, which shows that diagonal elements of the potential matrix in this channel, ${V^{N \Xi}}_{N \Xi}$ and ${V^{\Lambda \Sigma}}_{\Lambda \Sigma}$,  are both strongly repulsive and the off-diagonal element, ${V^{N\Xi}}_{\Lambda\Sigma}$, is comparable to or even larger than the diagonal elements.
These features have been observed already in the flavor SU(3) symmetric limit~\cite{Inoue:2011ai}.

%
\subsubsection{Three channels}

\begin{figure}
\begin{tabular}{ccc}
  \includegraphics[scale=0.55]{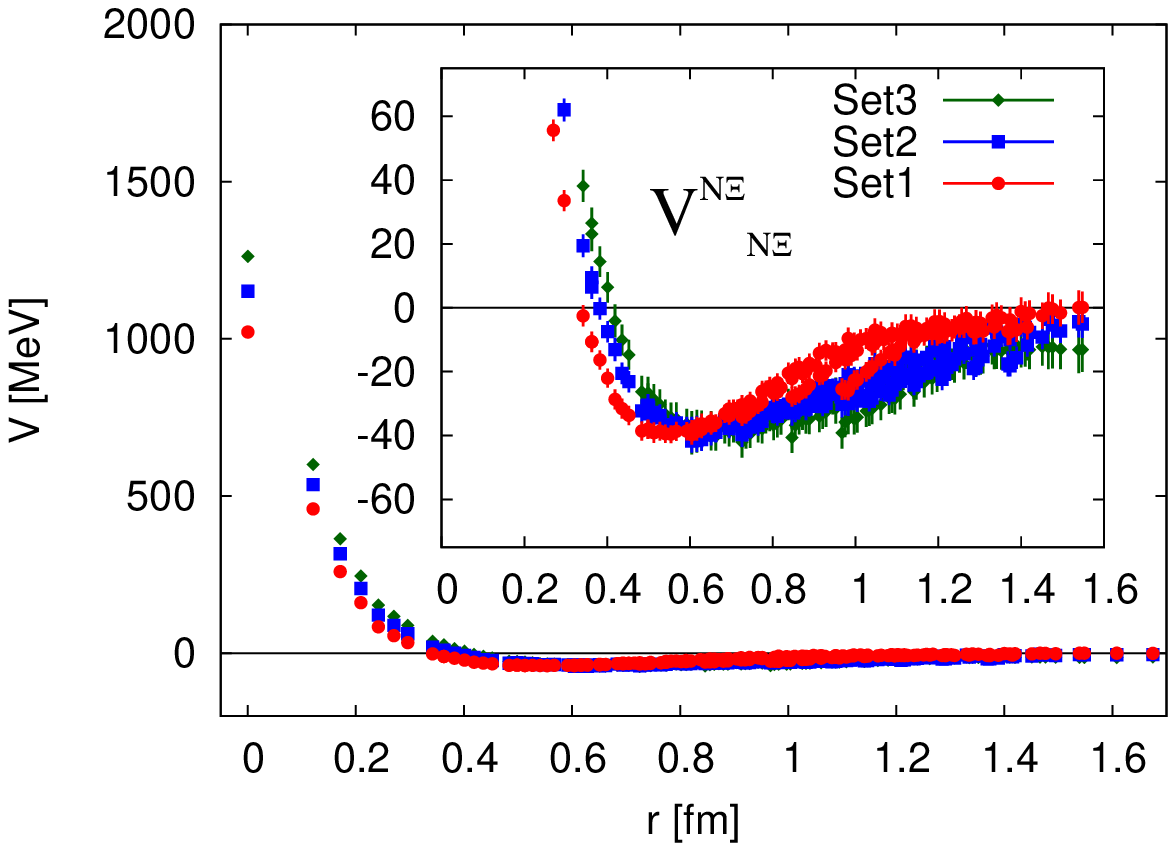} &
  \includegraphics[scale=0.55]{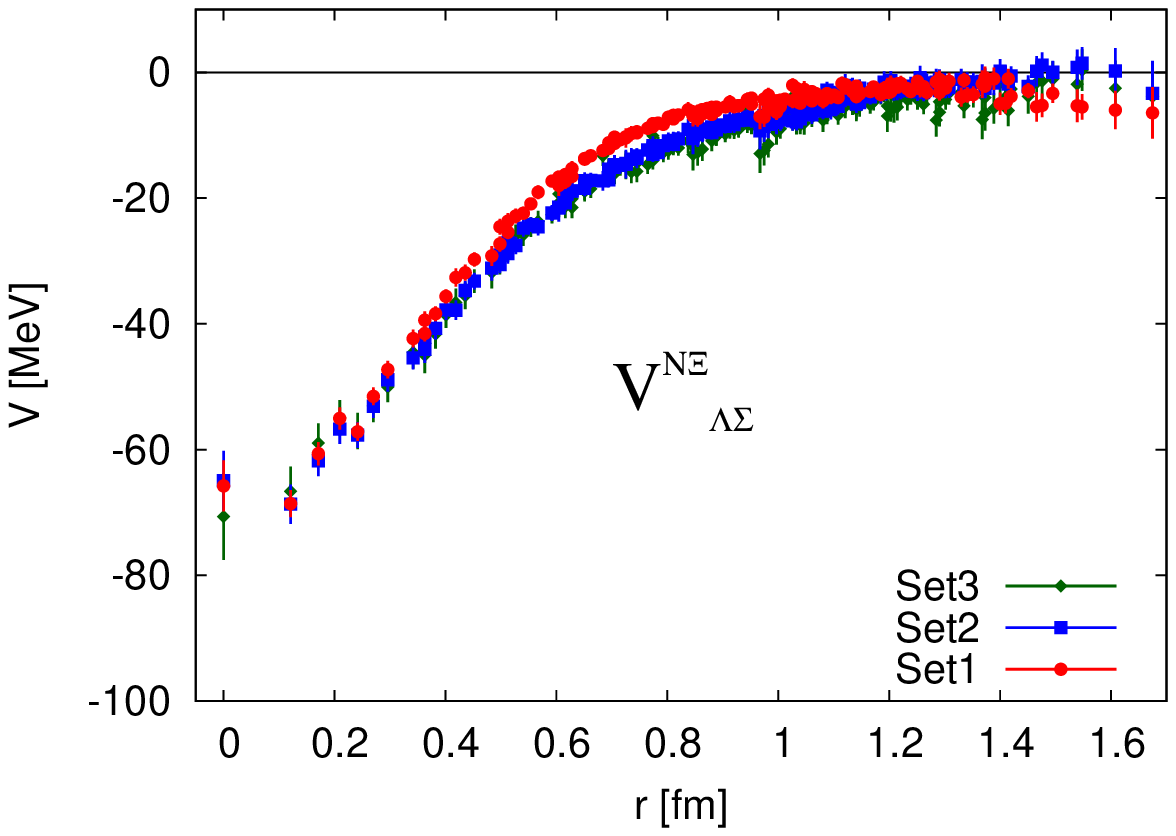} \\ 
  \includegraphics[scale=0.55]{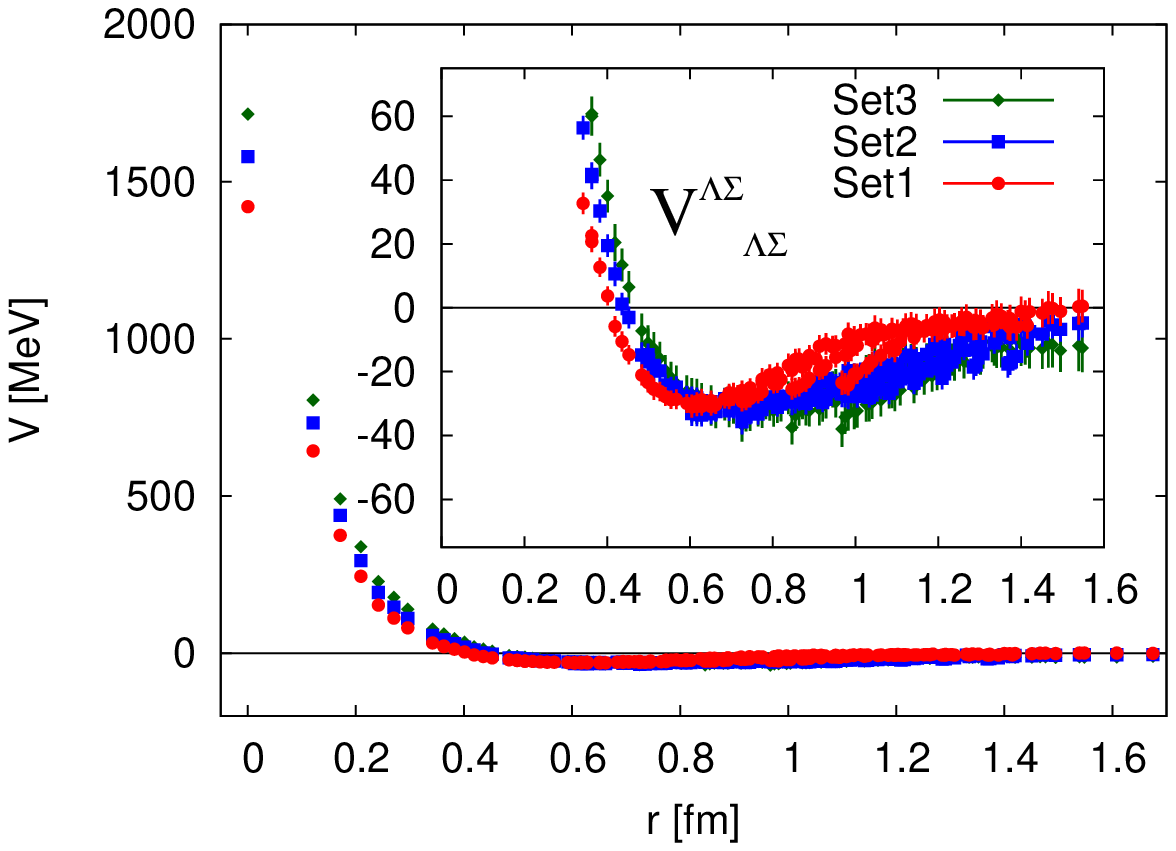} &
  \includegraphics[scale=0.55]{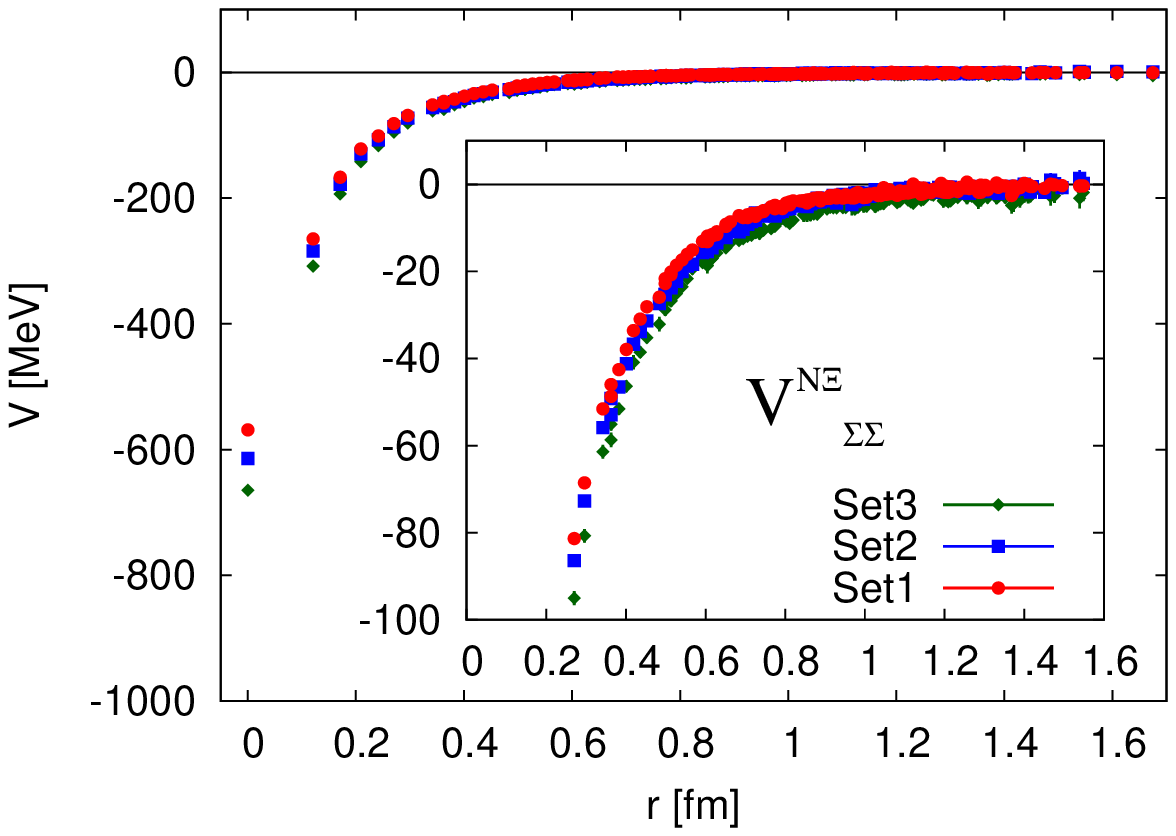} \\ 
  \includegraphics[scale=0.55]{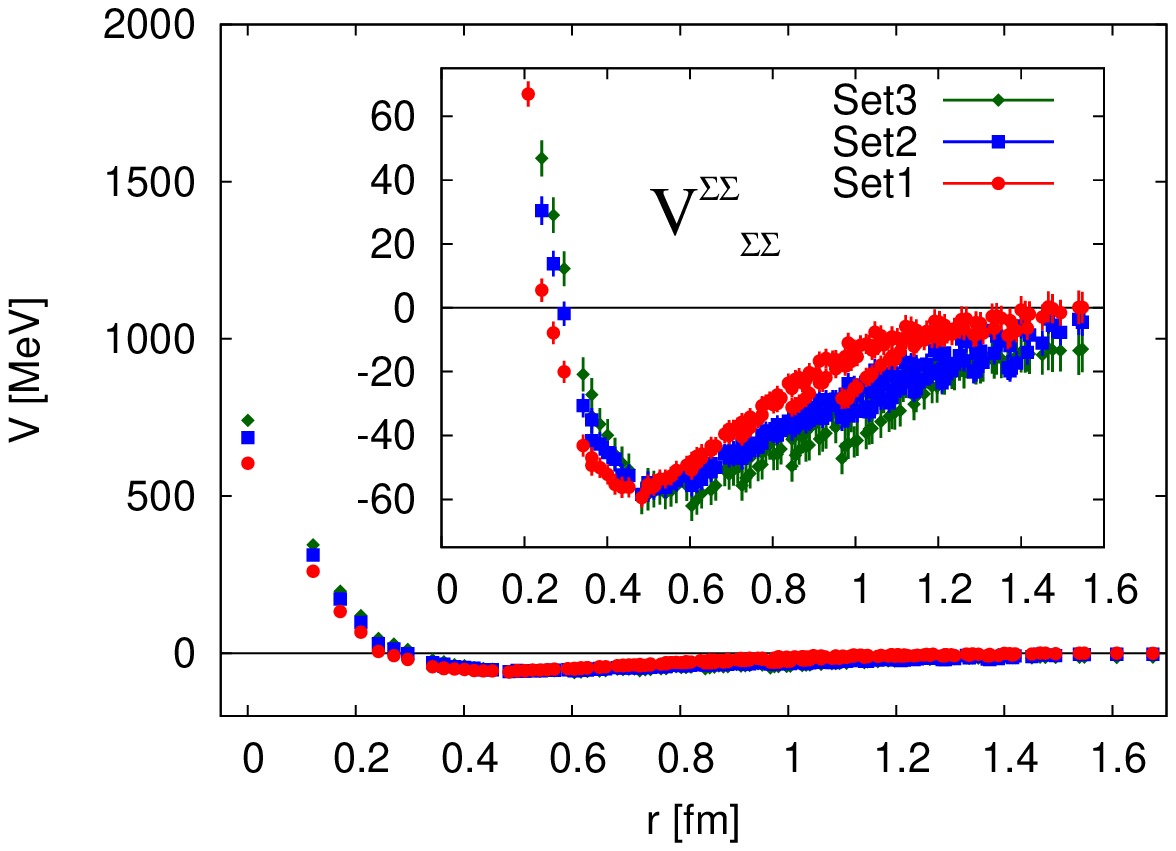} &
  \includegraphics[scale=0.55]{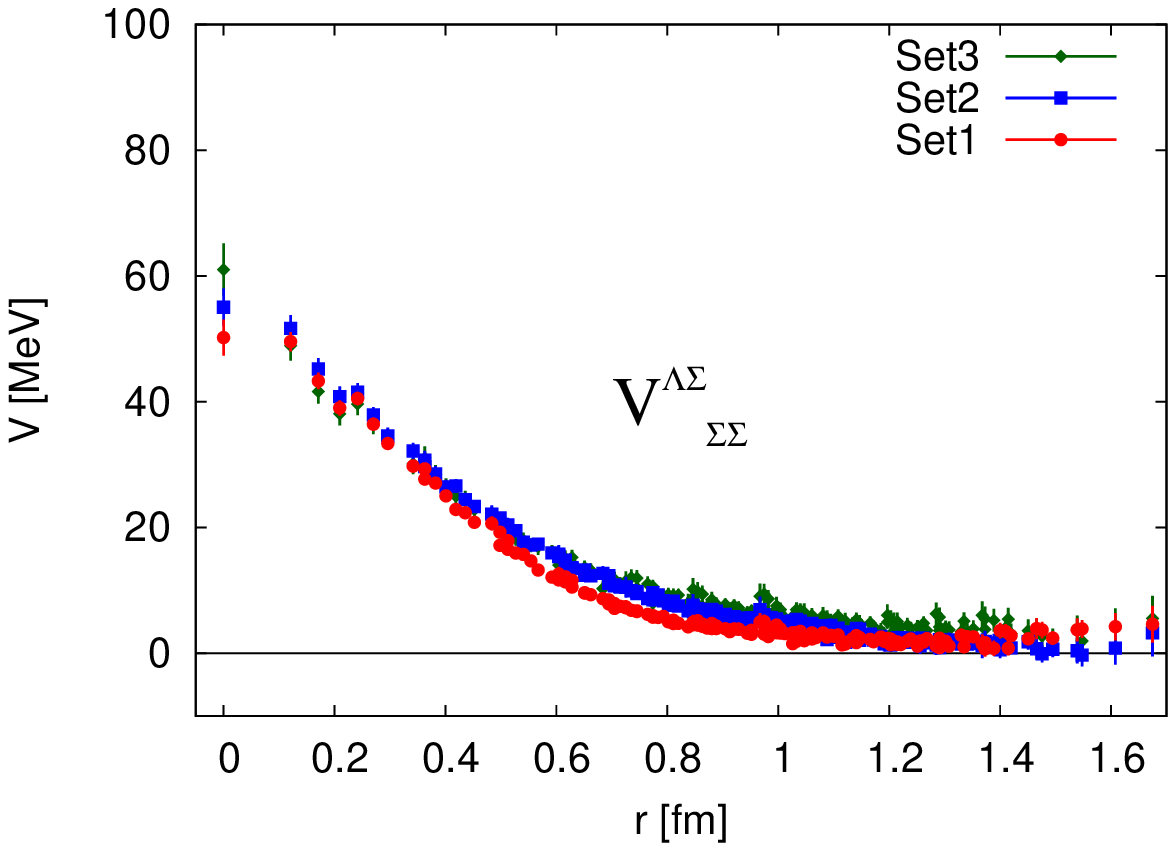} 
\end{tabular}
\caption{
Diagonal (left three panels) and off-diagonal (right three panels) elements of the potential matrix in the ${^3S_1} (I=1)$ channel.
Result from three gauge ensembles, Set~1 (red), Set~2 (blue) and Set~3 (green), are shown in one figure. 
Insets show the enlargement of the same plot.
}
\label{FIG:3S1I1}
\end{figure}

\begin{figure}
\begin{tabular}{ccc}
  \includegraphics[scale=0.55]{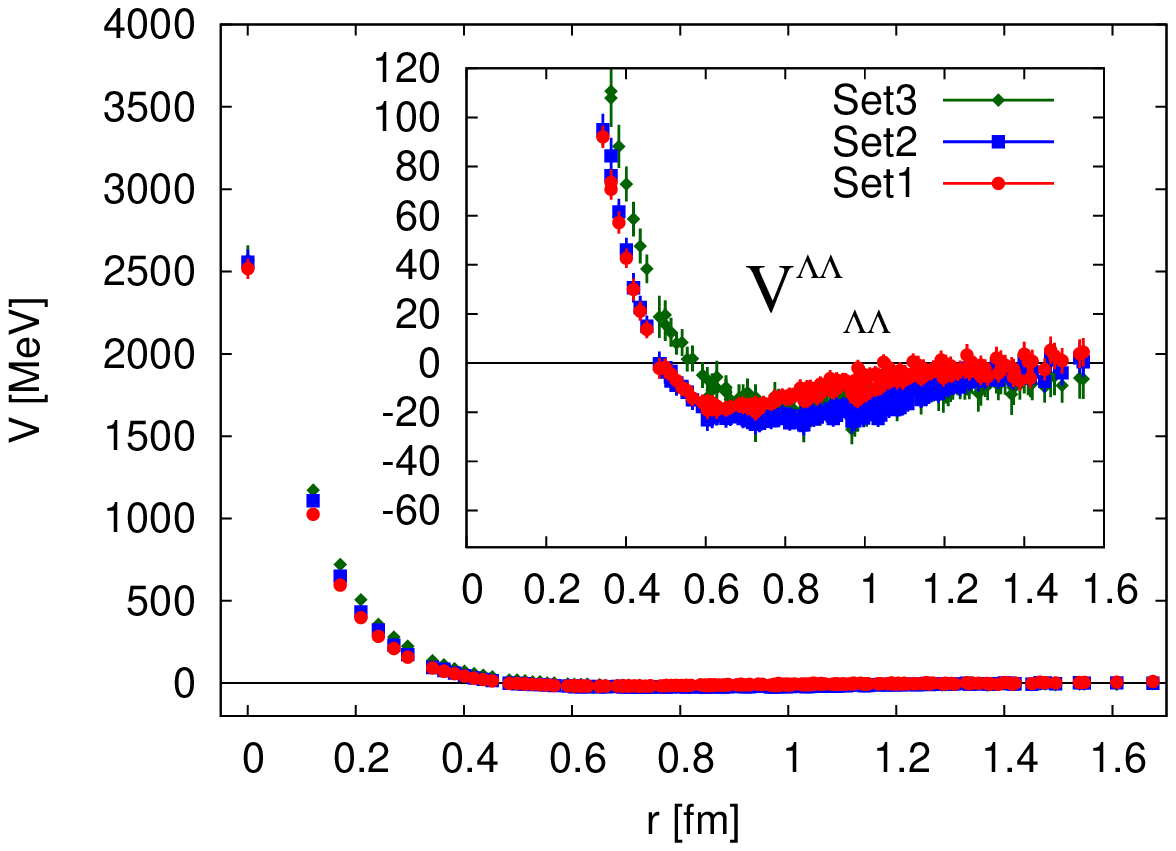} & 
  \includegraphics[scale=0.55]{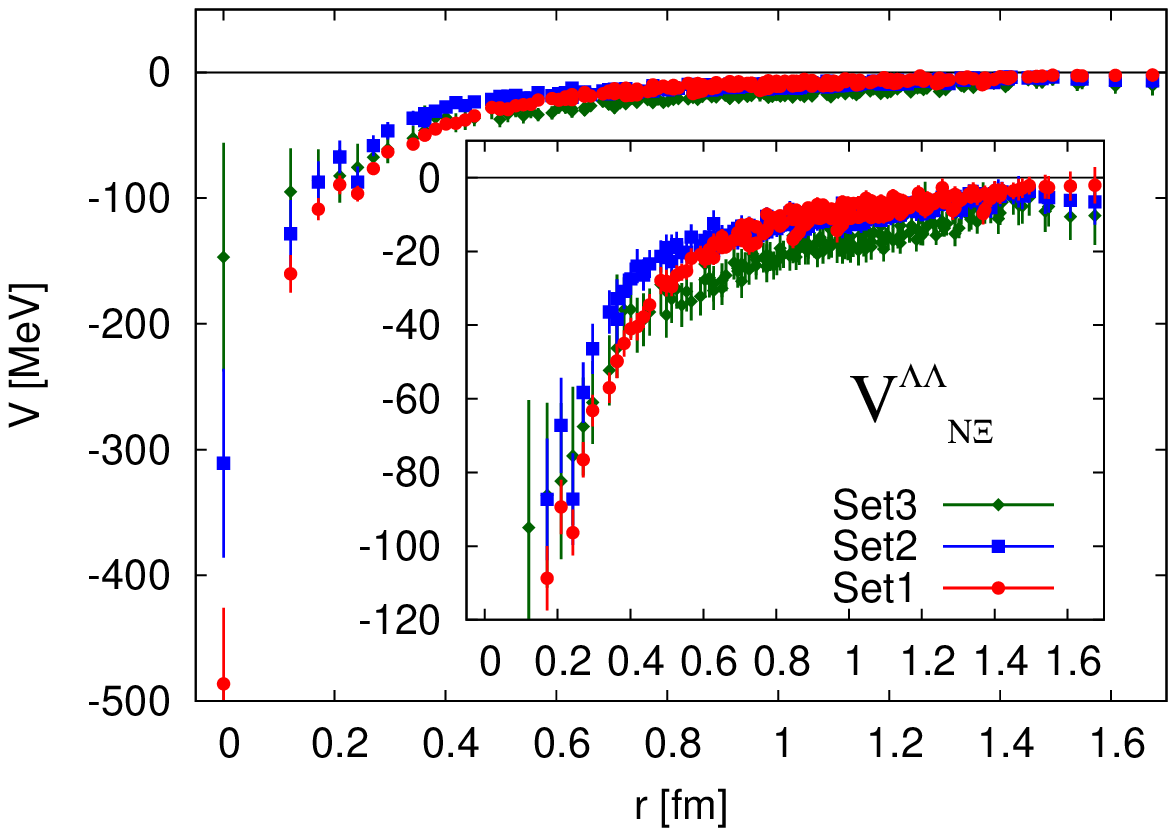} \\ 
  \includegraphics[scale=0.55]{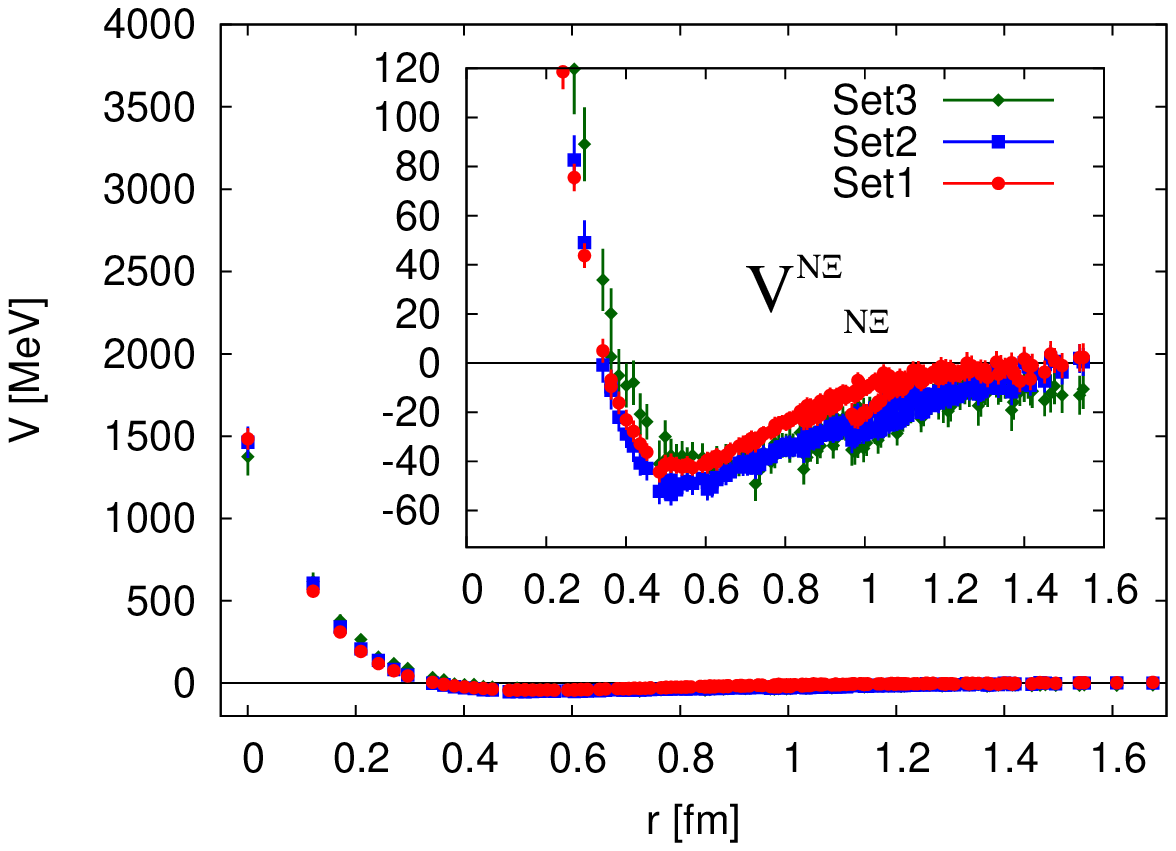} & 
  \includegraphics[scale=0.55]{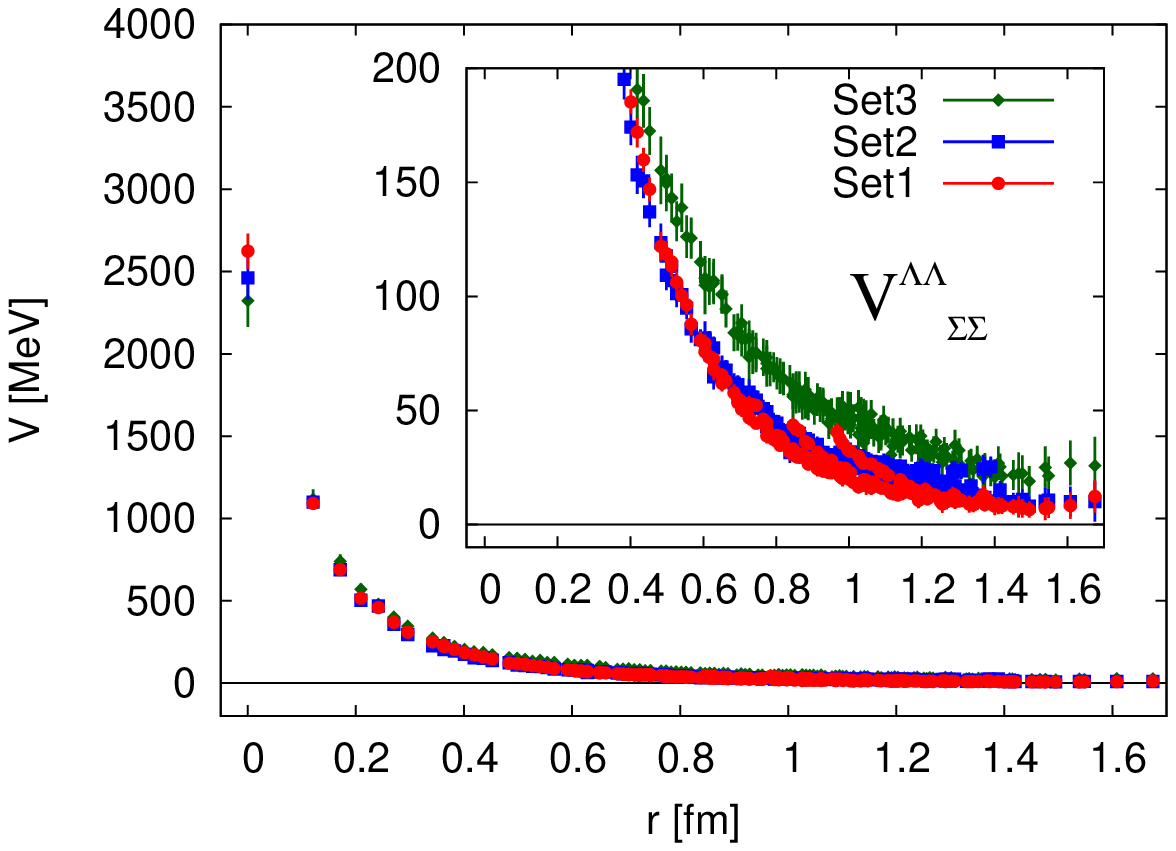} \\
  \includegraphics[scale=0.55]{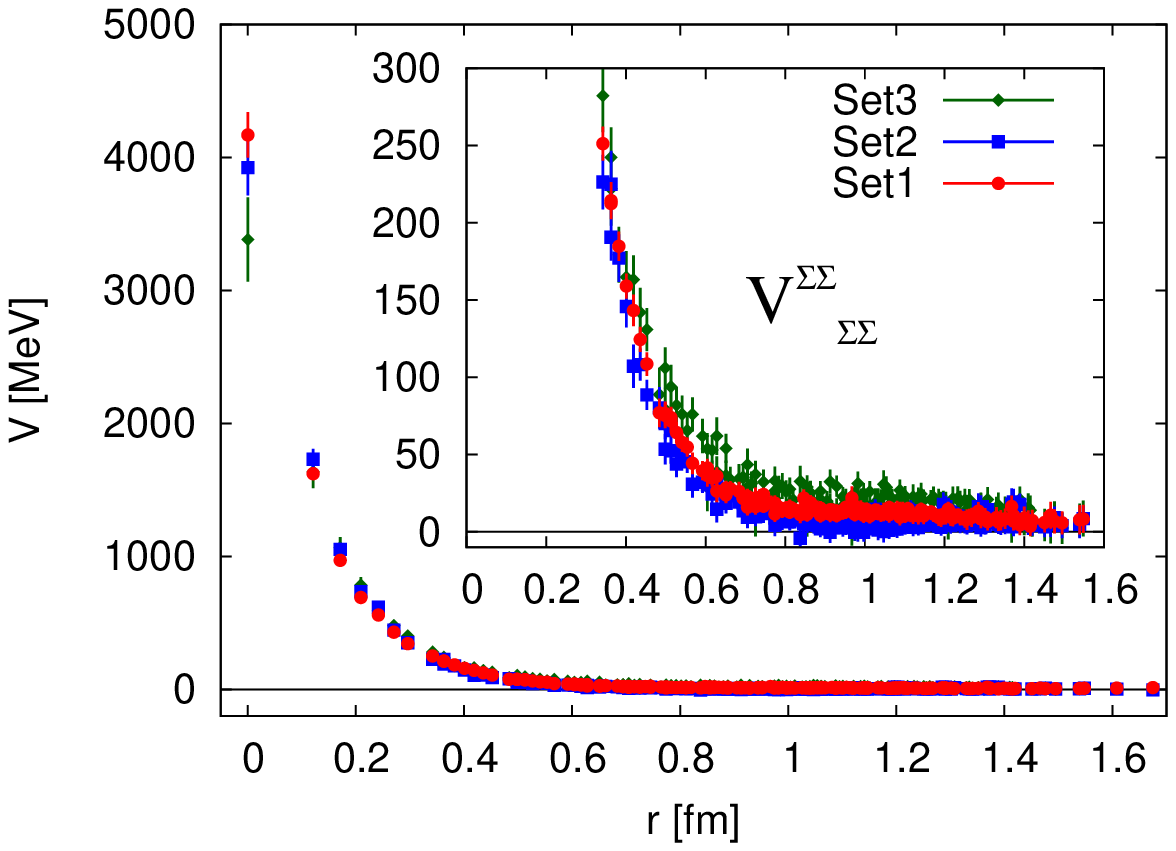} &
  \includegraphics[scale=0.55]{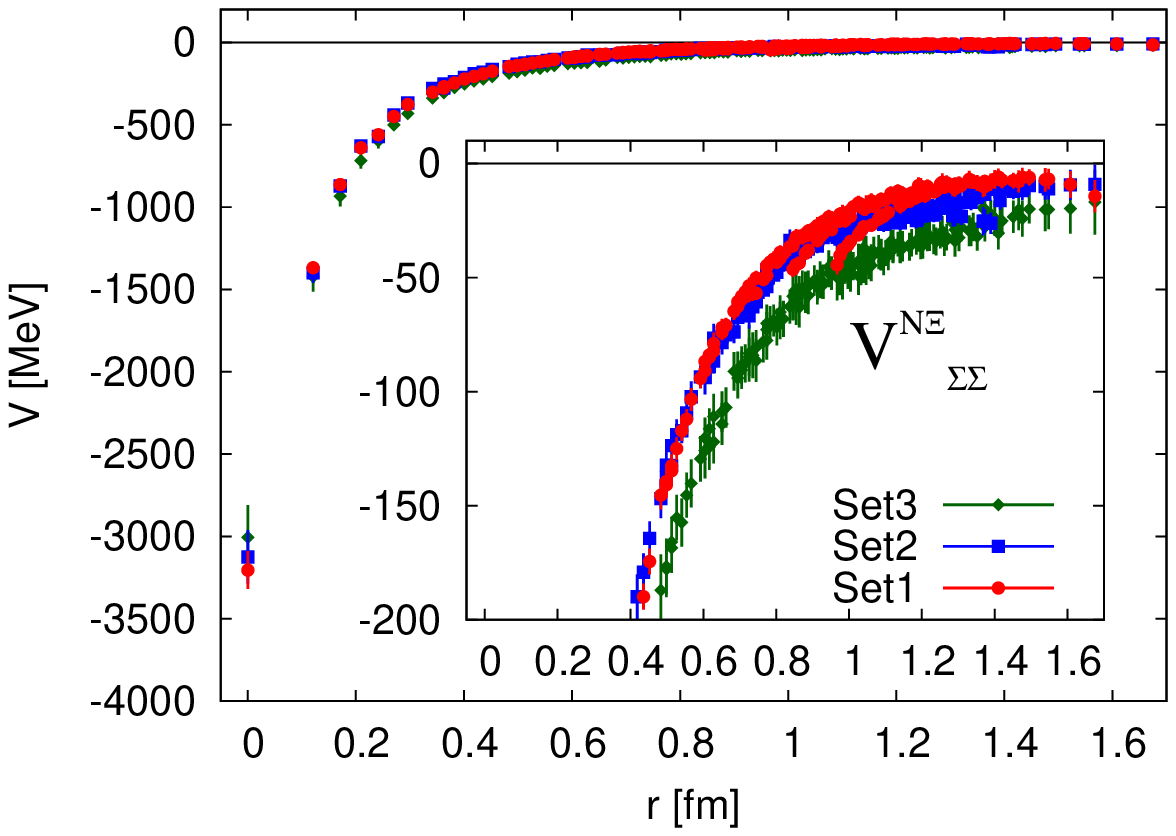} 
\end{tabular}
\caption{The potential matrix in the ${^1S_0} (I=0)$ channel, as in Fig.~\ref{FIG:3S1I1}. 
}
\label{FIG:1S0I0}
\end{figure}
Fig.~\ref{FIG:3S1I1} shows a potential matrix in  the ${^3S_1}$ ($I=1$) channel, which has $N \Xi$, $\Lambda \Sigma$ and $\Sigma \Sigma$ components.
All diagonal elements of the potential matrix, ${V^{N\Xi}}_{N\Xi}$ (left upper), ${V^{\Lambda\Sigma}}_{\Lambda\Sigma}$ (left middle) and ${V^{\Sigma\Sigma}}_{\Sigma\Sigma}$ (left lower), have an attraction at long distance and a repulsive core at short distance.
The largest attraction in this channel appears in ${V^{\Sigma \Sigma}}_{\Sigma\Sigma}$, whose maximum depth is about $-60$~MeV at around $r \sim 0.6$~fm.
All diagonal potentials have a tendency that magnitudes of both repulsion at short distance and attraction at long distance increase as the light quark masses decrease.

For the off-diagonal elements of potential matrix, ${V^{N\Xi}}_{\Lambda \Sigma}$ and ${V^{\Lambda \Sigma}}_{\Sigma \Sigma}$ are much smaller than ${V^{N \Xi}}_{\Sigma \Sigma}$.
These off-diagonal potentials,  ${V^{N \Xi}}_{\Lambda \Sigma}$ and ${V^{\Lambda \Sigma}}_{\Sigma \Sigma}$,  almost vanish  at $r>1.2$~fm and have a small quark mass dependence, while ${V^{N \Xi}}_{\Sigma \Sigma}$ increases as the light quark masses decrease.

Fig.~\ref{FIG:1S0I0} shows the potential matrix in the $^1S_0$ ($I=0$) channel, where the $H$ dibaryon state may appear. 
All diagonal elements of the potential matrix have a repulsive core at short distance, whose strength, however,  depends strongly on the state.
An attractive pocket, on the other hand,  appears only in two diagonal elements,  ${V^{\Lambda \Lambda}}_{\Lambda \Lambda}$ and ${V^{N \Xi}}_{N \Xi}$, where  
${V^{N \Xi}}_{N \Xi}$ has much deeper attractive pocket than ${V^{\Lambda \Lambda}}_{\Lambda \Lambda}$ does, while
${V^{\Sigma \Sigma}}_{\Sigma \Sigma}$ is totally repulsive in the whole range of $r$.

The off-diagonal element, ${V^{\Lambda \Lambda}}_{N \Xi}$, is smaller than other two, so that the decay rate from $N \Xi$ to $\Lambda \Lambda$ may be relatively suppressed. 
Diagonal elements of the potential matrix  generated with the configuration Set 2 are most attractive, while
off-diagonal potentials with the configuration Set 3 are strongest in magnitude for $r >0.5$fm.

\subsection{Potential matrix  in the SU(3) irreducible representation basis}
\begin{figure}[tb]
\begin{tabular}{cc}
  \includegraphics[scale=.55]{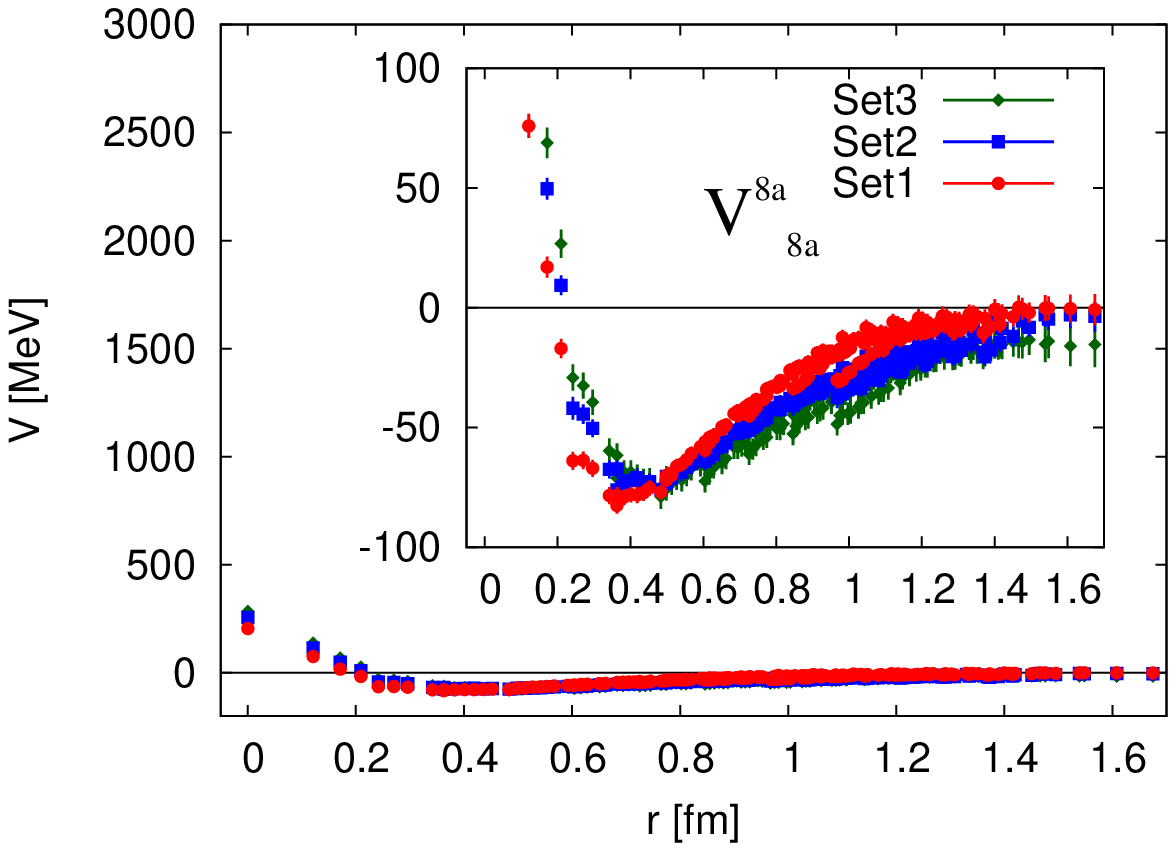} & 
  \includegraphics[scale=.55]{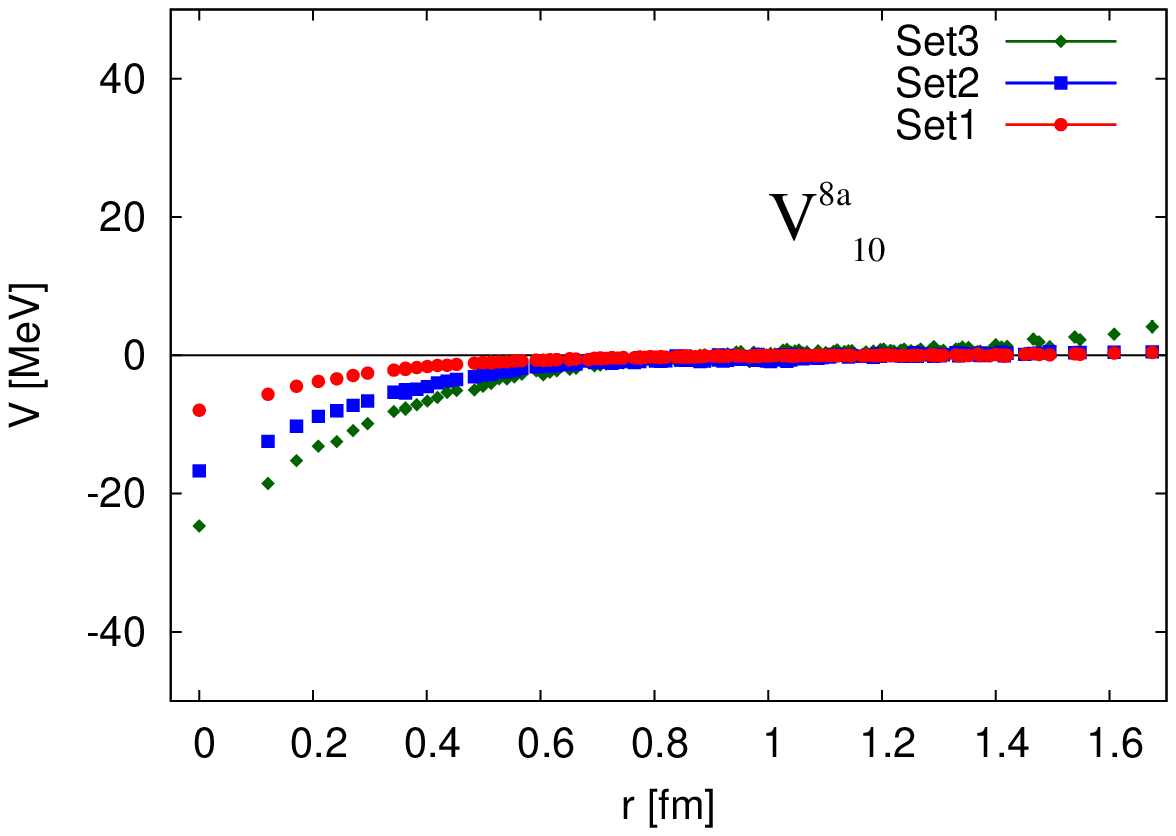} \\
  \includegraphics[scale=.55]{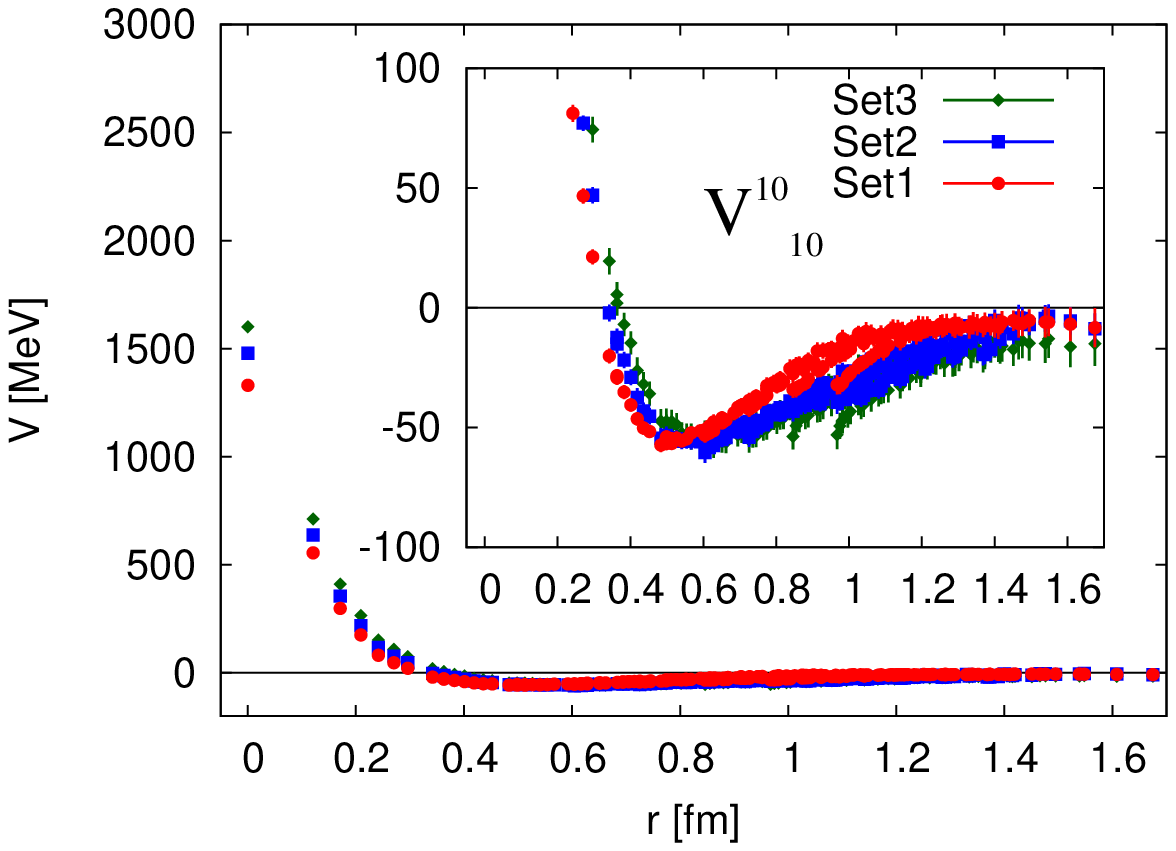} &
  \includegraphics[scale=.55]{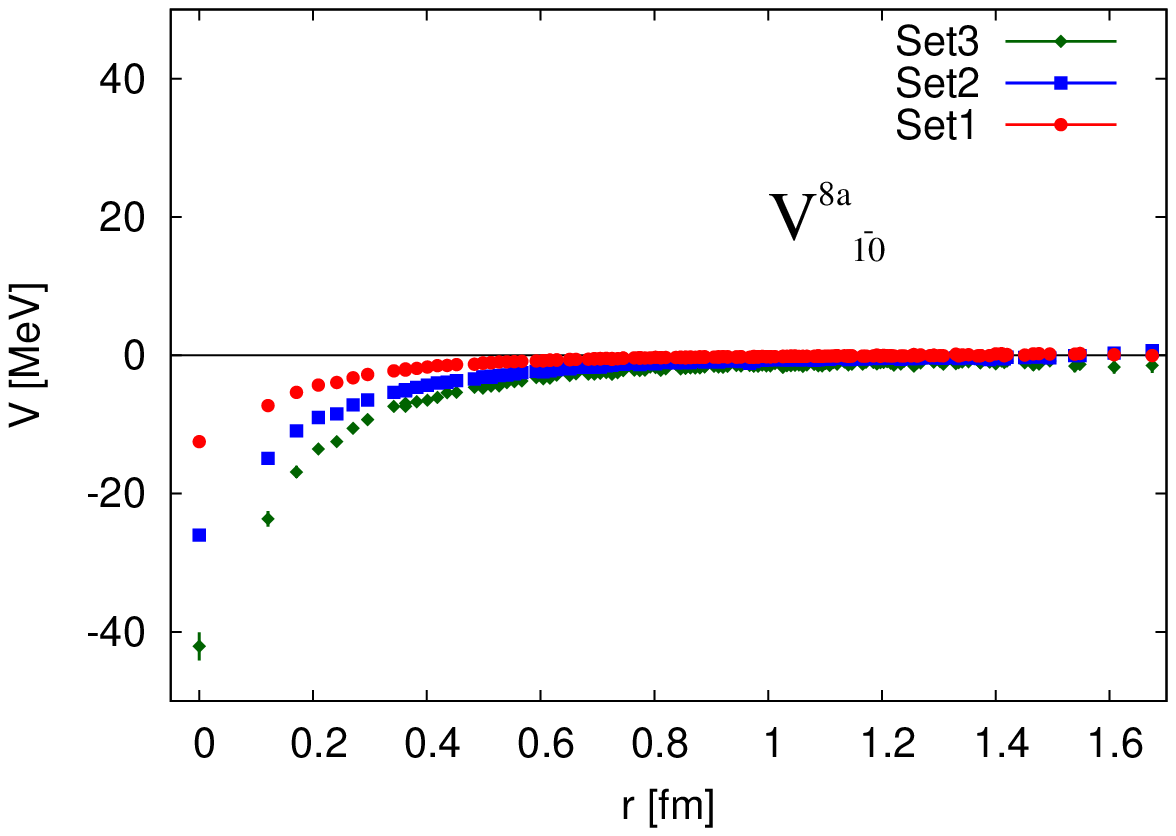} \\ 
  \includegraphics[scale=.55]{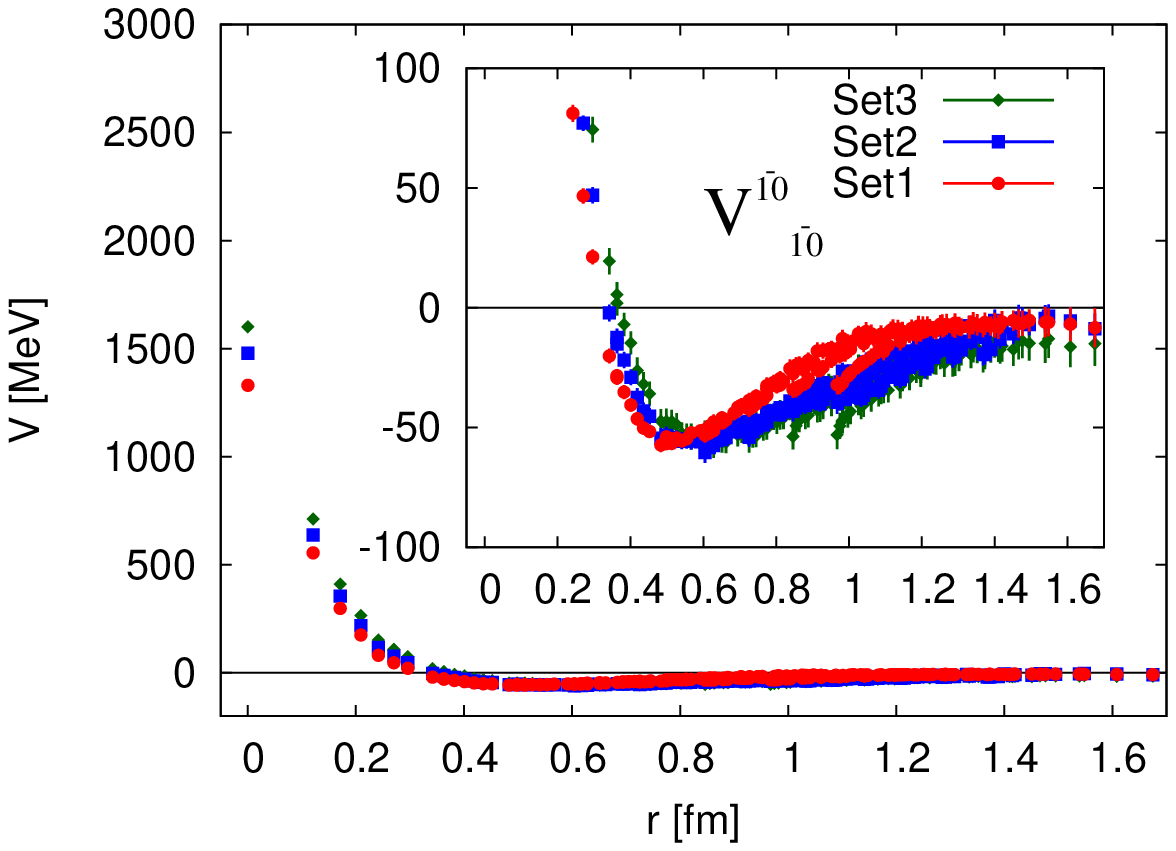} &
  \includegraphics[scale=.55]{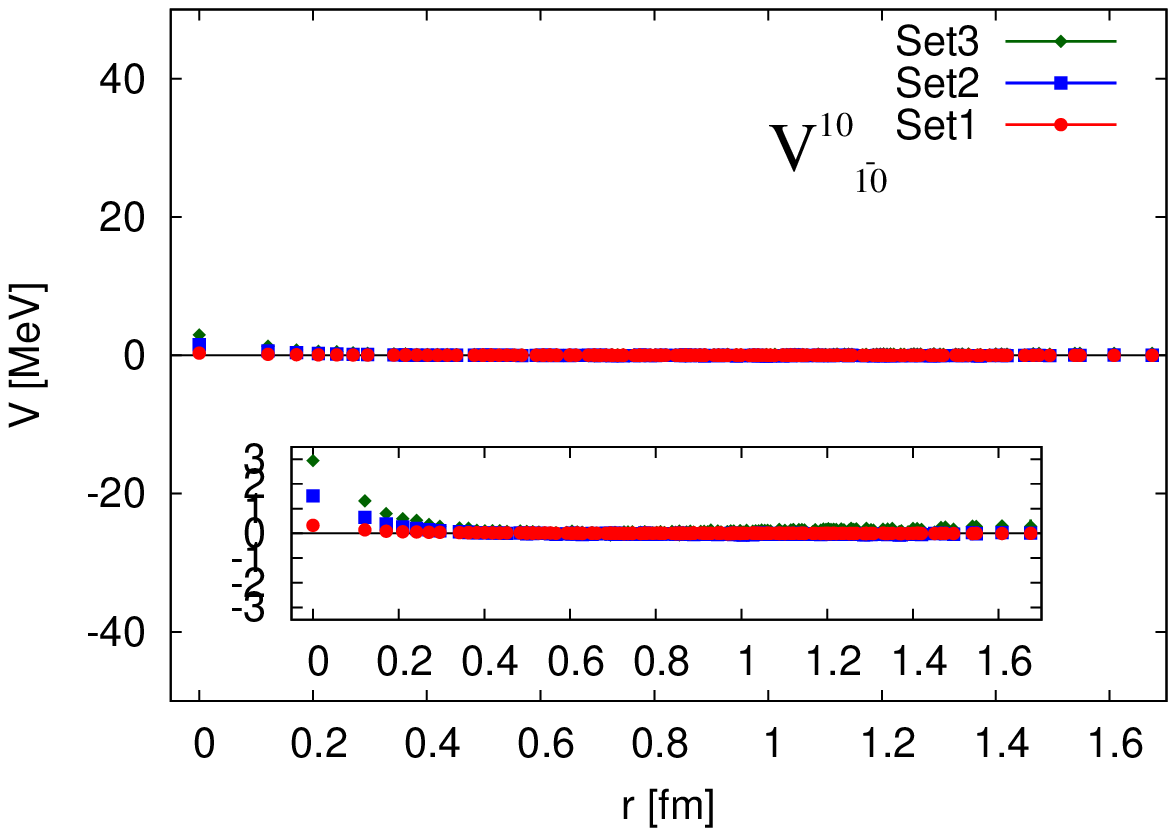}
\end{tabular}
\caption{Potential matrix in the SU(3) basis for $^3S_1$ with $S=-2$ and $I=1$.
Left three are diagonal elements, ${V^{8_a}}_{8_a}$ (upper), ${V^{\overline{10}}}_{\overline{10}}$ (midle) and ${V^{10}}_{10}$ (lower), while right three are off-diagonal ones, ${V^{8_a}}_{10}$ (upper), ${V^{8_a}}_{\overline{10}}$ (middle) and ${V^{10}}_{\overline{10}}$ (lower).
 Red, blue and green symbols stand for results with Set 1, 2 and 3, respectively.}
\label{FIG:3S1I1_IR} 
\end{figure}
We here present potential matrices in the SU(3)  irreducible representation  basis (SU(3) basis in short hereafter) such as
$1, 8_s, 8_a, 10, \overline{10}, 27$,  obtained from the particle basis  by using Clebsh-Gordan coefficients.
This makes us possible  to compare the results with those in the flavor SU(3) symmetric limit \cite{Inoue:2011ai}.

The transformation from particle basis to SU(3) basis is defined as
\begin{eqnarray}
\left| \begin{array}{c}
{\bf{1}} \\ {\bf{8}}_s \\ {\bf{27}}
\end{array} \right\rangle
=
\left( \begin{array}{ccc}
-\sqrt{\frac{1}{8}}   &  \sqrt{\frac{1}{2}}  &  \sqrt{\frac{3}{8}} \\
-\sqrt{\frac{1}{5}}   &  \sqrt{\frac{1}{5}}  & -\sqrt{\frac{3}{5}} \\
 \sqrt{\frac{27}{40}} &  \sqrt{\frac{3}{10}} & -\sqrt{\frac{1}{40}} \\
\end{array} \right)
\left| \begin{array}{c}
\Lambda \Lambda \\ N \Xi \\ \Sigma \Sigma
\end{array} \right\rangle
\end{eqnarray}
for ${^1S_0}(I=0)$ channel, and
\begin{eqnarray}
\left| \begin{array}{c}
{\bf{8}}_a \\ {\bf{10}} \\ \overline{\bf{10}}
\end{array} \right\rangle
=
\left( \begin{array}{ccc}
 \sqrt{\frac{1}{3}} & 0                  &  \sqrt{\frac{2}{3}} \\
-\sqrt{\frac{1}{3}} & \sqrt{\frac{1}{2}} &  \sqrt{\frac{1}{6}} \\
-\sqrt{\frac{1}{3}} &-\sqrt{\frac{1}{2}} &  \sqrt{\frac{1}{6}} \\
\end{array} \right)
\left| \begin{array}{c}
N \Xi \\ \Lambda \Sigma \\ \Sigma \Sigma
\end{array} \right\rangle
\end{eqnarray}
for ${^3S_1}(I=1)$ channel.

Fig.~\ref{FIG:3S1I1_IR} shows the potential matrix in the SU(3) basis for $^3S_1$($I=1$), which is composed of $8_a$, $10$ and $\overline{10}$.
While all diagonal elements of the potential matrix have a repulsive core,
the height of the repulsive core in ${V^{8_a}}_{8_a}$ is much lower than other two and its depth of the attractive pocket is the deepest among three.
On the other hand, ${V^{10}}_{10}$ is strongly repulsive and has a quite shallow attractive pocket at all quark masses, though the height of the repulsive core and a range of attractive pocket increase as the $ud$ quark mass decreases. 
As far as off-diagonal elements are concerned, they are very small.
In particular, ${V^{10}}_{\overline{10}}$ vanishes at all quark masses including Set 3, where the SU(3) breaking by the difference between $ud$ and $s$ quark masses is maximal in our calculation.
Other two off-diagonal elements, ${V^{8_a}}_{10}$ and ${V^{8_a}}_{\overline{10}}$, have small non-zero values at short distance region ($r<0.6$ fm), which gradually increase as the $ud$ quark masses decrease.

\begin{figure}[tb]
\begin{tabular}{cc}
  \includegraphics[scale=.55]{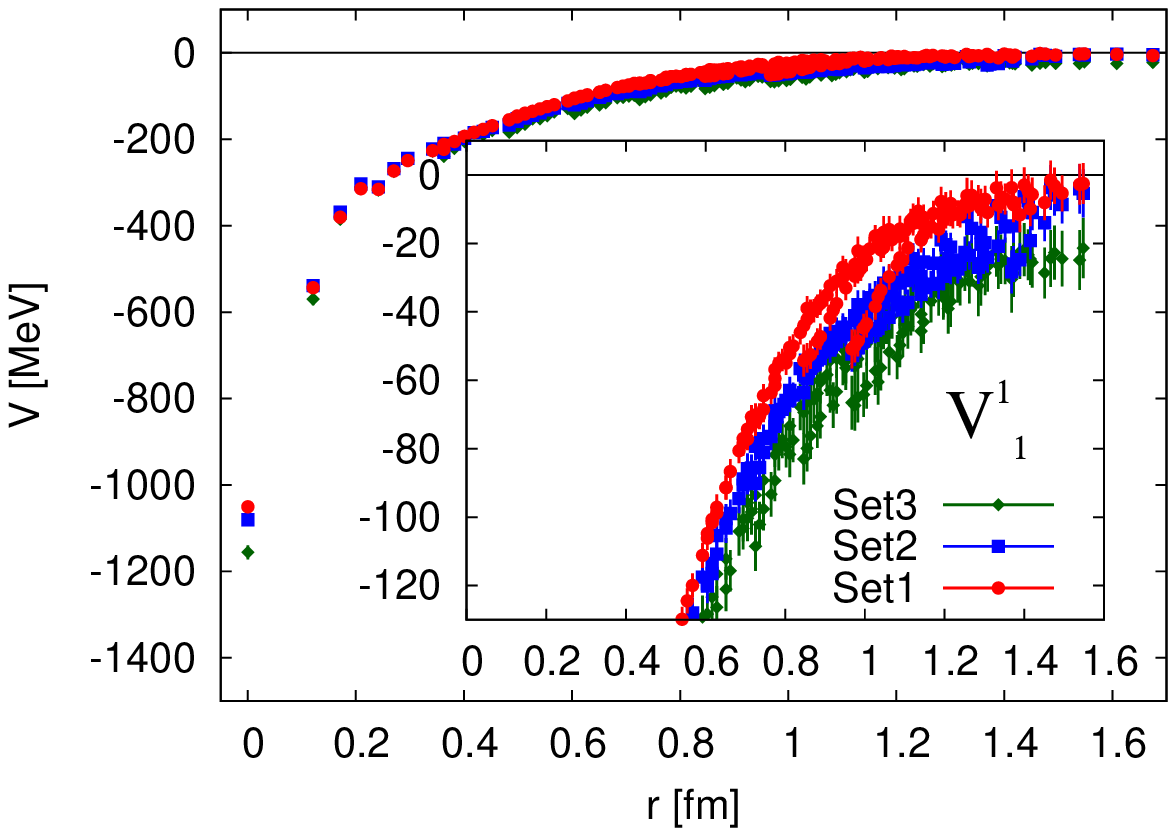} &
  \includegraphics[scale=.55]{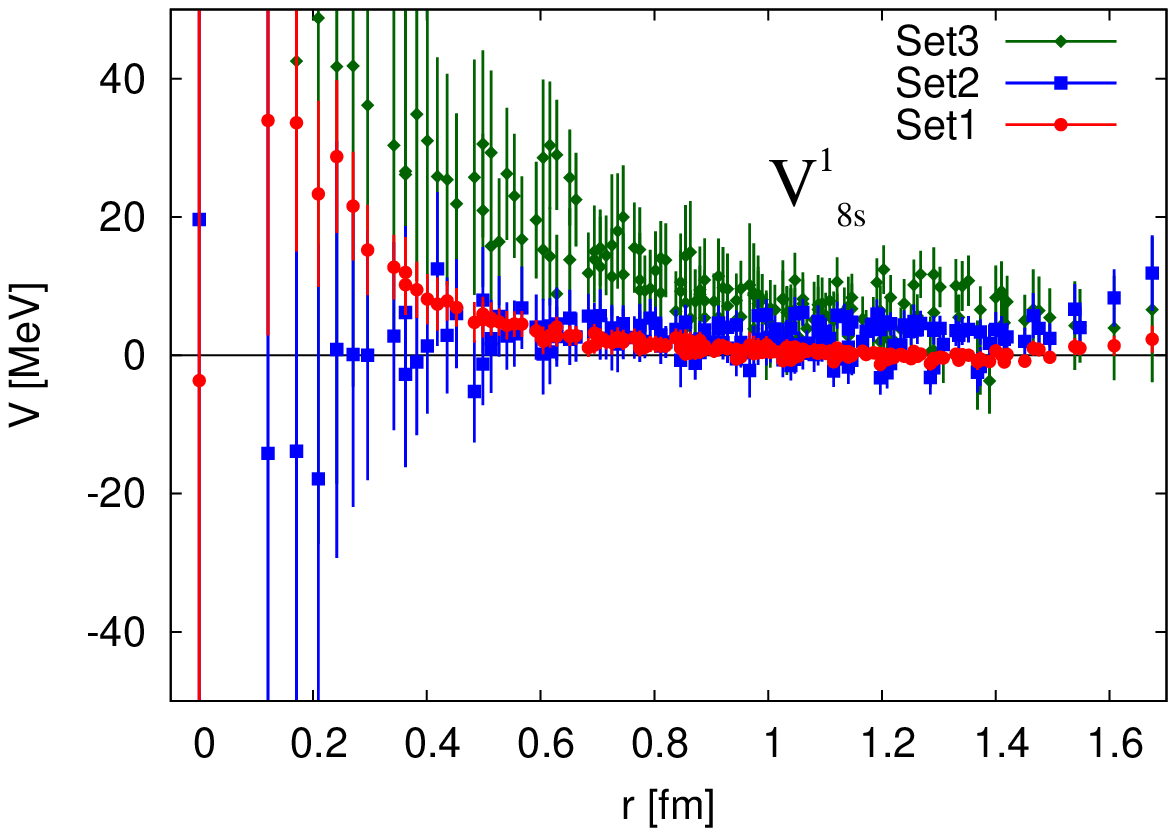} \\
  \includegraphics[scale=.55]{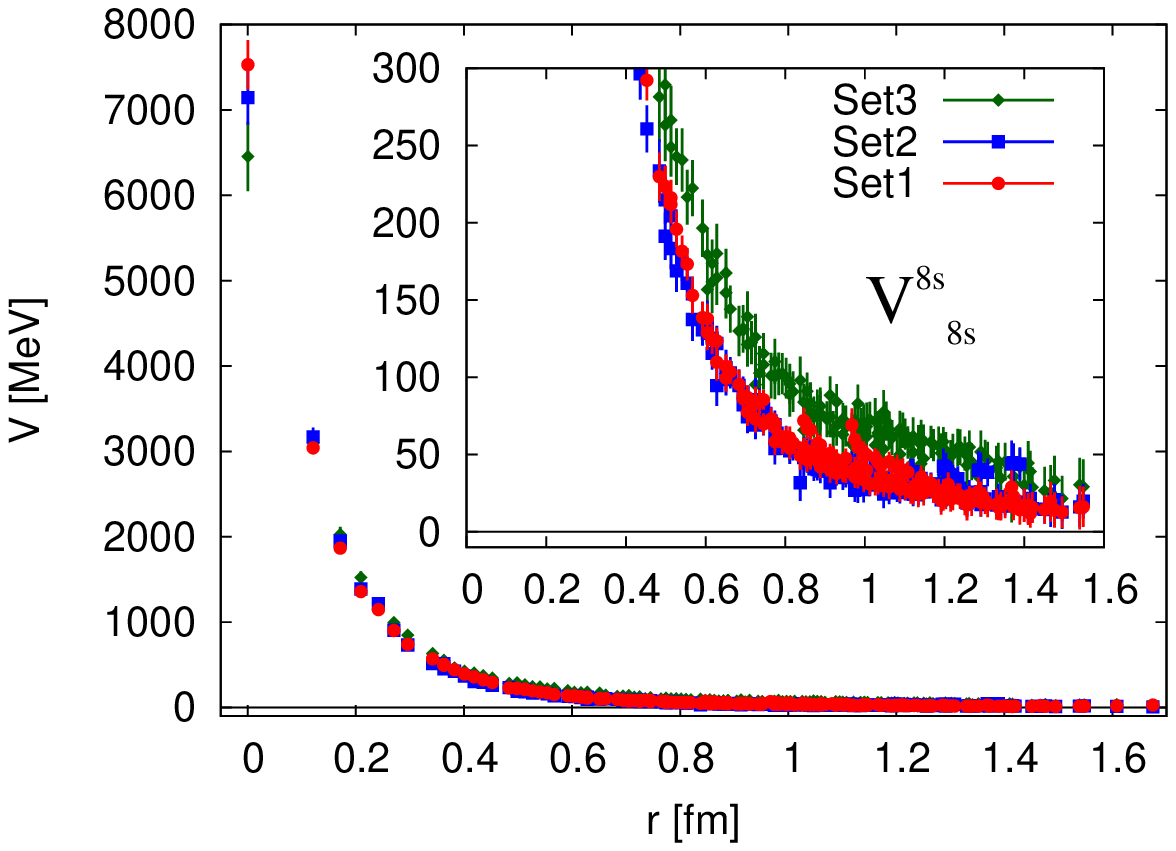} &
  \includegraphics[scale=.55]{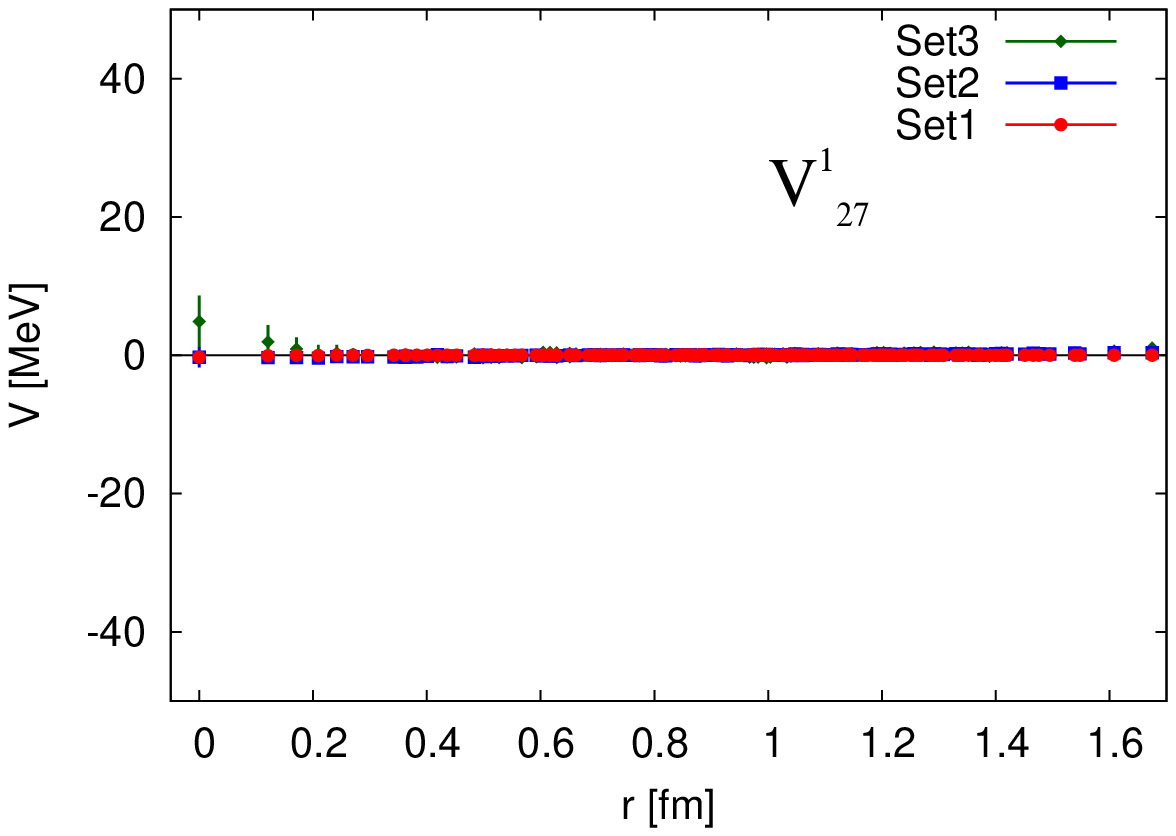} \\
  \includegraphics[scale=.55]{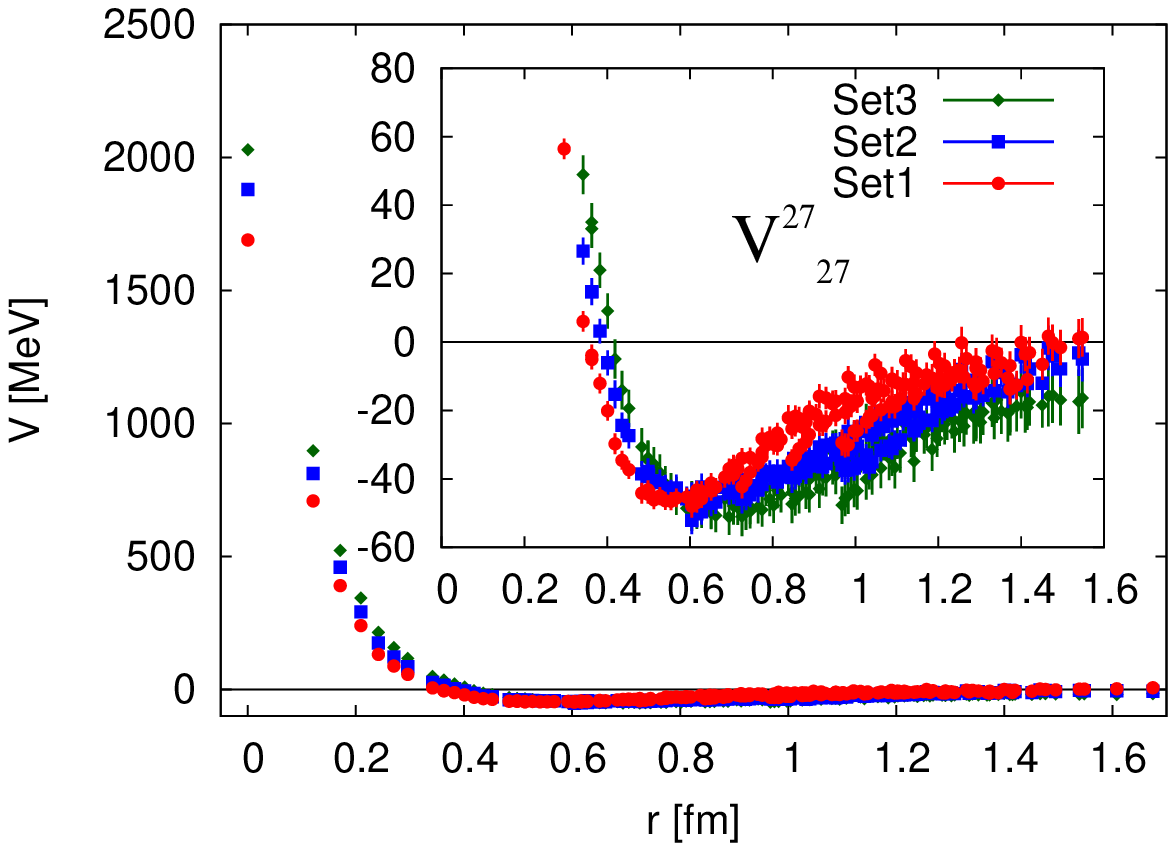} &
  \includegraphics[scale=.55]{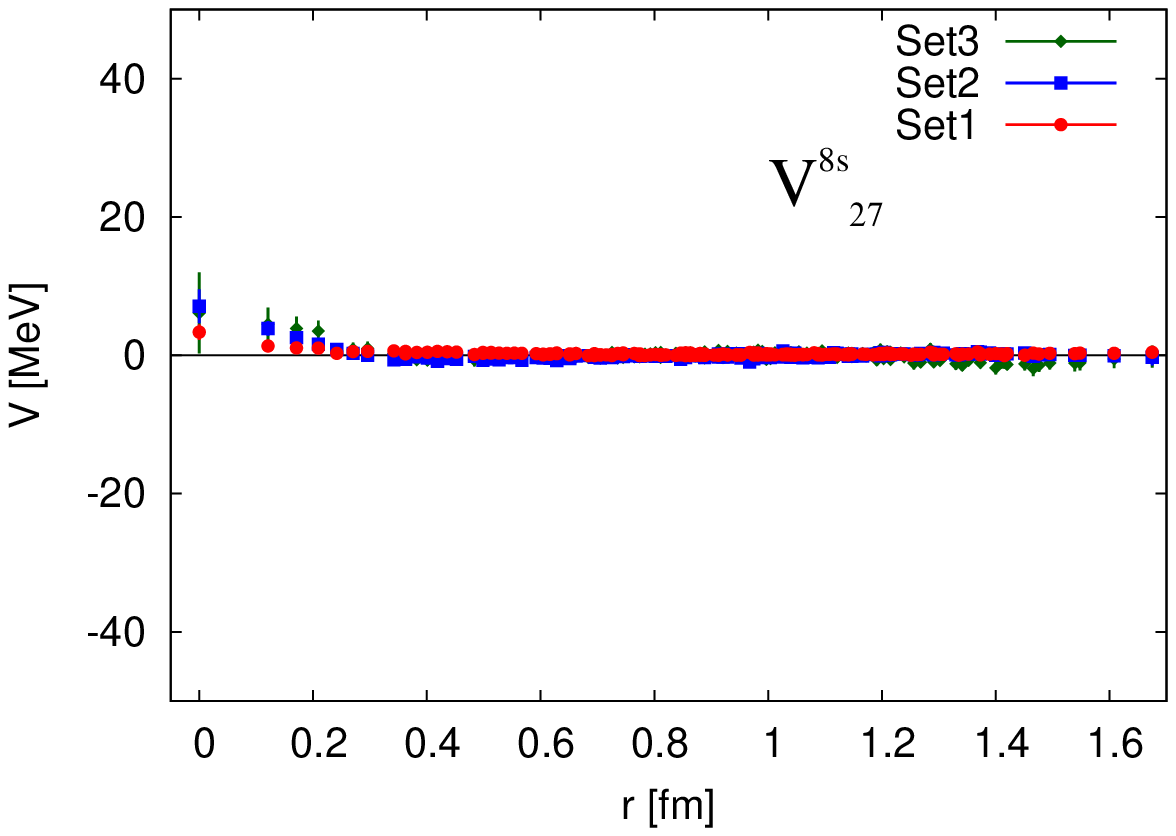} \\  
\end{tabular}
\caption{Potential matrix in the SU(3)  basis  for  $^1S_0$ with $S=-2$ and $I=0$.
Left three are diagonal elements, ${V^{1}}_{1}$ (upper), ${V^{8_s}}_{8_s}$ (middle) and ${V^{27}}_{27}$ (lower), while right three are off-diagonal ones, ${V^{1}}_{8_s}$ (upper), ${V^{1}}_{27}$ (middle) and ${V^{8_s}}_{27}$ (lower).
 Red, blue and green symbols stand for results with Set 1, 2 and 3, respectively.
}
\label{FIG:1S0I0_IR} 
\end{figure}
Fig.~\ref{FIG:1S0I0_IR} shows the potential matrix in the SU(3) basis for $^1S_0$ ($I=0$). 
As in the case of the SU(3) limit \cite{Inoue:2011ai},  the diagonal element for the flavor singlet state, ${V^{1}}_{1}$, is strongly attractive, while  ${V^{8_s}}_{8_s}$ is repulsive,  at all distances.
The absence of repulsive core in ${V^{1}}_{1}$ is consistent with the absence of the quark Pauli blocking effect.
A shape of ${V^{27}}_{27}$ is similar to the ${^1S_0}$ nuclear force, which also belongs to 27-plet. 

Quark mass dependences of diagonal potentials can be seen clearly in the flavor basis.
As the light quark mass decreases, the attraction in ${V^{1}}_{1}$ gradually increases, while both repulsive core and  attraction at long distance in ${V^{27}}_{27}$ are enhanced.

Off-diagonal elements of potential matrix in the SU(3) basis are presented in right three panels in Fig.~\ref{FIG:1S0I0_IR}, which give effective measures of the flavor SU(3) breaking effects since they are absent in the flavor SU(3) symmetric limit.
Fig.~\ref{FIG:1S0I0_IR} shows that ${V^{1}}_{8_s}$ (upper) is small but non-zero while ${V^{1}}_{27}$ (middle) and ${V^{8_s}}_{27}$ (lower) are consistent with zero except for very short distance, $r<0.2$~fm, where cutoff effects could be sizeable.
These results tell us that flavor SU(3) breaking effects in the off-diagonal parts is much smaller than that in the diagonal part within the quark masses adopted in this paper.
  The 27-plet state is almost uncoupled even if $ud$ quark mass is different from the $s$ quark mass.

\begin{figure}
\begin{tabular}{ccc}
\hspace*{-1em}
  \includegraphics[scale=0.40]{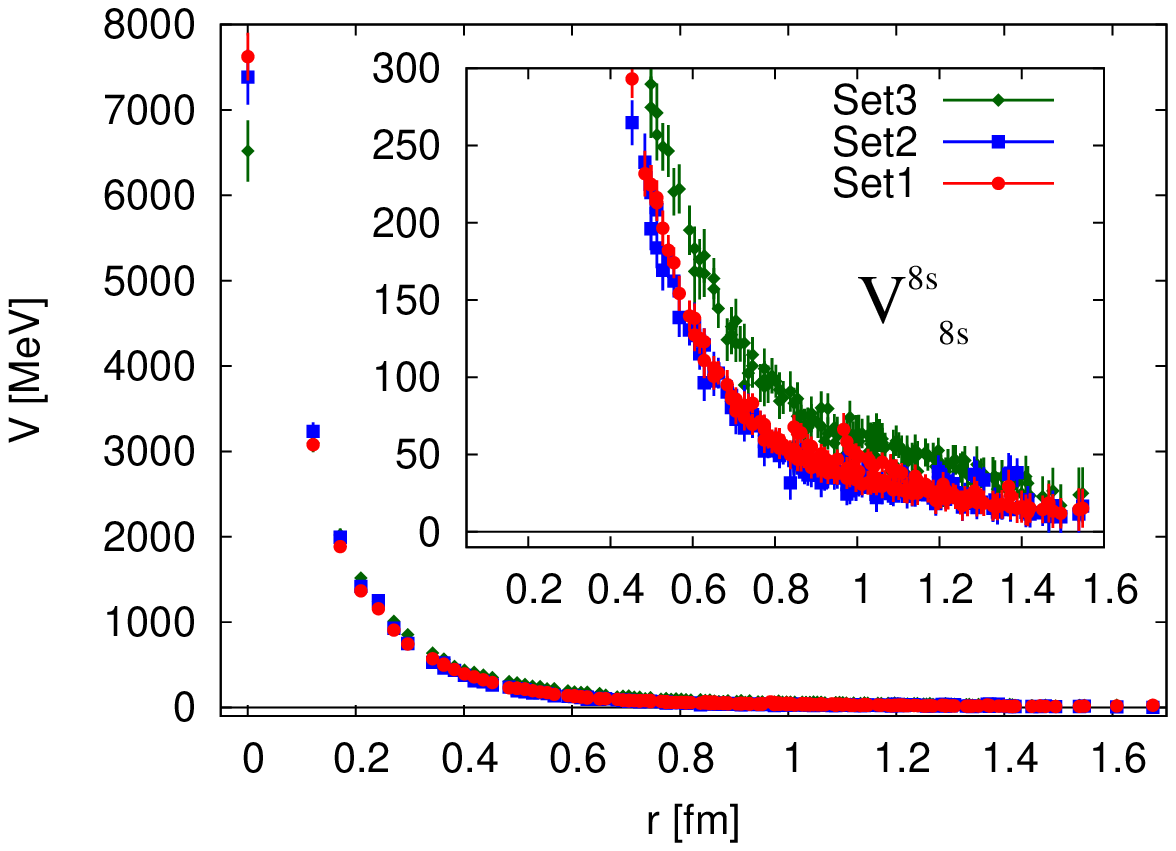} \hspace*{-1.5em}
& \includegraphics[scale=0.40]{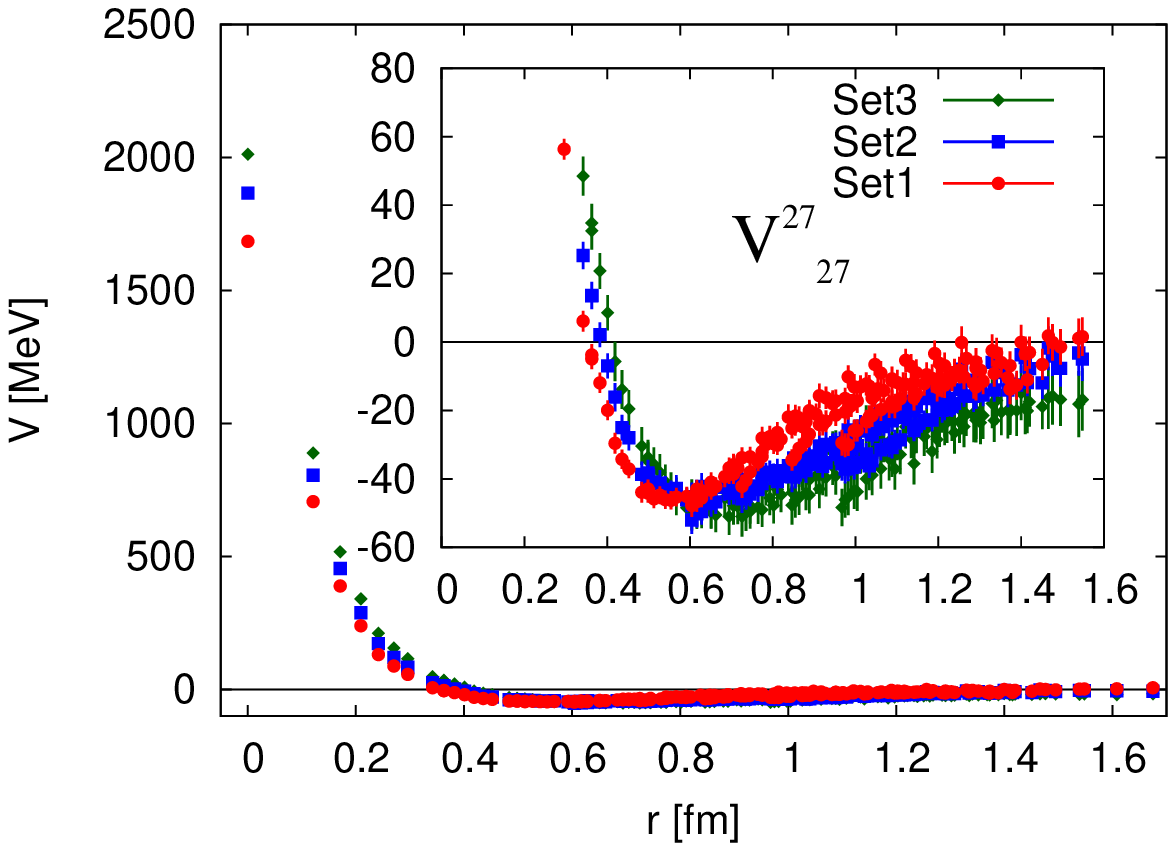} \hspace*{-1.5em}
& \includegraphics[scale=0.40]{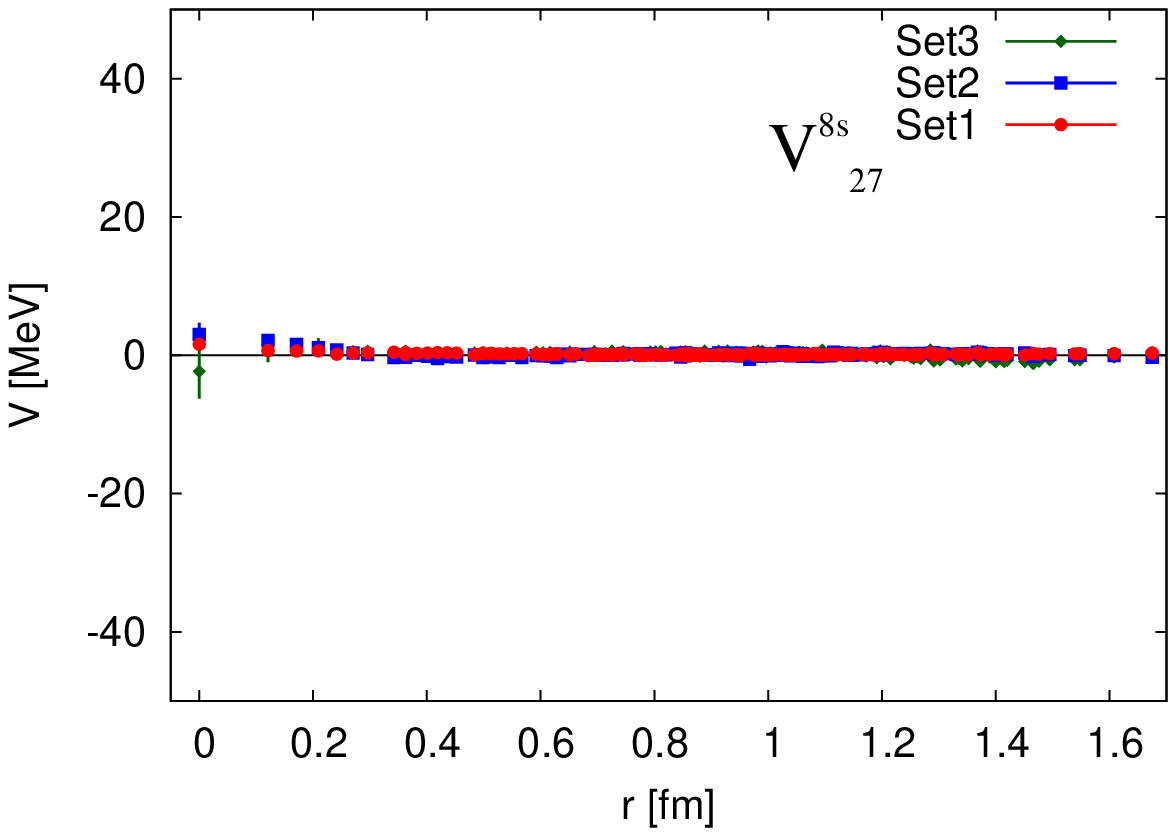} \hspace*{-1em}
\end{tabular}
\caption{
Diagonal (left and center) and off-diagonal (right) elements of the potential matrix in the SU(3) basis for the ${^1S_0} (I=1)$ channel.
Left and center panels are diagonal elements, ${V^{8_s}}_{8_s}$ (Left), ${V^{27}}_{27}$ (center), while right panel is off-diagonal ones, ${V^{8_s}}_{27}$.
 Red, blue and green symbols stand for results with Set 1, 2 and 3, respectively.}
\label{FIG:1S0I1_IR}
\end{figure}
For ${^1S_0} (I=1)$, the potential matrix in the SU(3) basis is obtained by
\begin{eqnarray}
\left| \begin{array}{c}
8_s \\ 27
\end{array} \right\rangle
=
\left( \begin{array}{cc}
- \sqrt{\frac{3}{5}} & \sqrt{\frac{2}{5}} \\
\sqrt{\frac{2}{5}} & \sqrt{\frac{3}{5}}
\end{array} \right)
\left| \begin{array}{c}
N \Xi \\ \Lambda \Sigma
\end{array} \right\rangle.
\end{eqnarray}
Fig.~\ref{FIG:1S0I1_IR} shows diagonal and off-diagonal parts of potential matrix in the SU(3) basis for ${^1S_0} (I=1)$ channel.
We find that diagonal elements, ${V^{8_s}}_{8_s}$ and ${V^{27}}_{27}$, have similar behaviors to ones obtained from ${^1S_0} (I=0)$, and the transition potential between $8_s$-plet and $27$-plet is quite small for all Sets.

\begin{figure}
\begin{tabular}{cc}
  \includegraphics[scale=0.55]{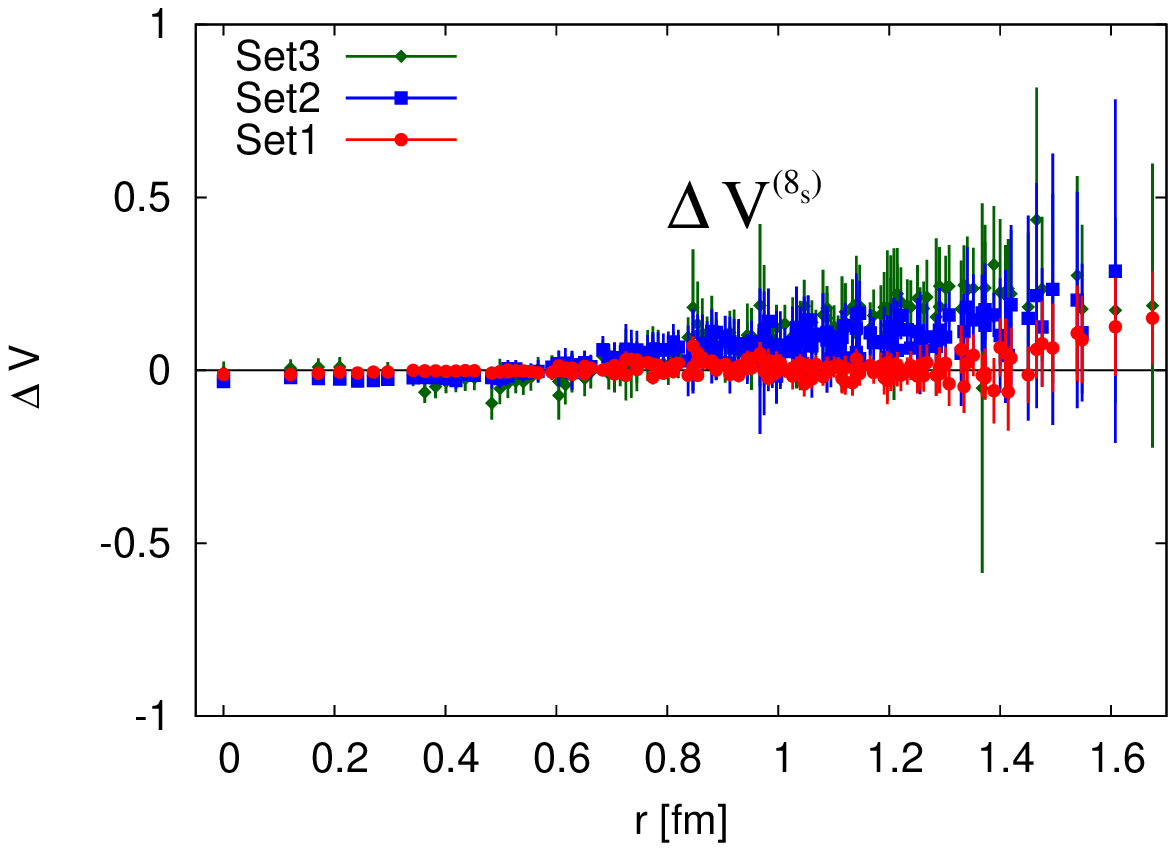} &  
  \includegraphics[scale=0.55]{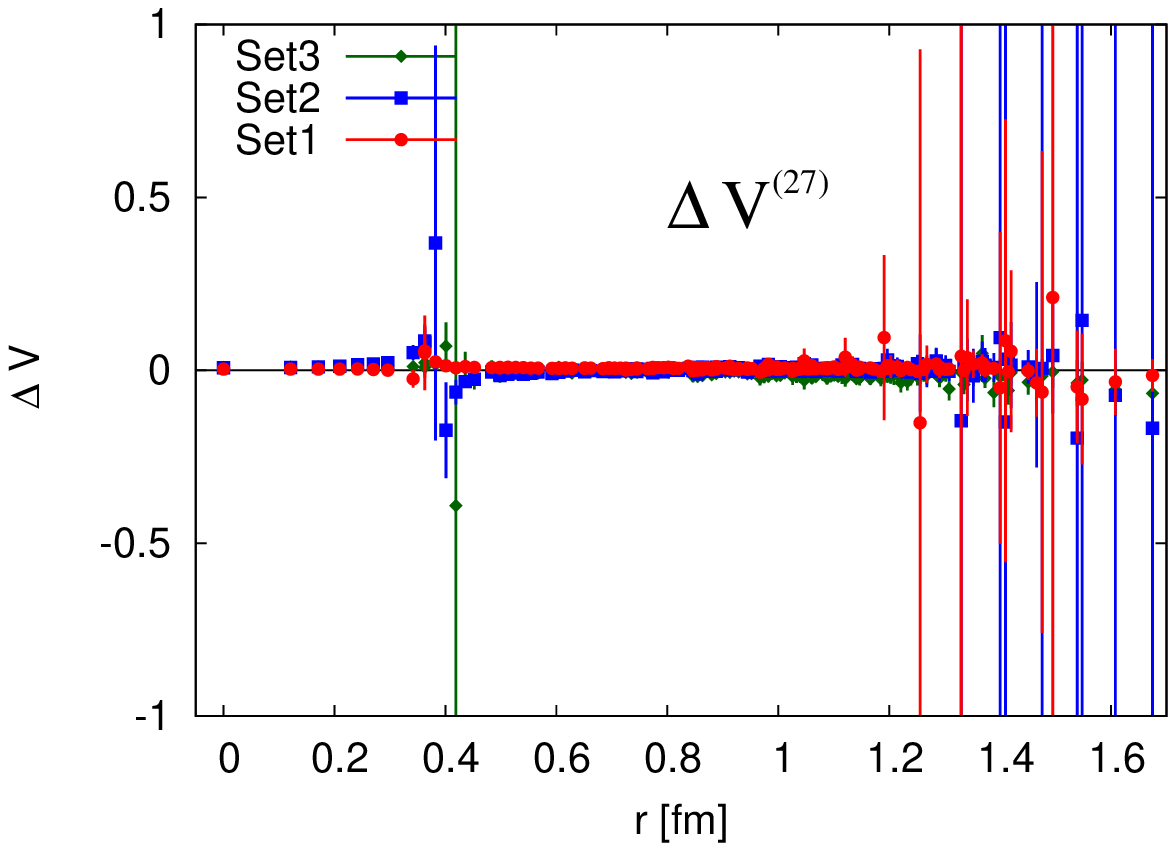} 
\end{tabular}
\caption{ 
Measures of SU(3) breaking effects: $\Delta V^{(8_s)}$ (left) and $\delta V^{(27)}$ (right) as a function of $r$.
 Red, blue and green symbols stand for results with Set 1, 2 and 3, respectively.
}
\label{FIG:pot_comp_IR_basis}
\end{figure}
To see the effects of SU(3) breakings, it is interesting to compare the potentials in the SU(3) basis extracted from two different channels, ${^1S_0} (I=0)$ and ${^1S_0} (I=1)$.
In Fig.~\ref{FIG:pot_comp_IR_basis}, we show the measure of SU(3) breaking defined as 
$\Delta V^{(c)} \equiv 2(V^{c}_{I=0} - V^{c}_{I=1})/(V^{c}_{I=0} + V^{c}_{I=1})$.
We find that, for $\Delta V^{(8_s)}$, there are no significant deviations from zero within statistical errors for all Sets.
A similar behavior to $\Delta V^{(8_s)}$ case can be seen again for $\delta V^{(27)}$ case except for a singular behavior at $r \simeq 0.4$ fm where the 27-plet potentials themselves almost vanish.

\section{Summary and conclusions}
In this paper, we have investigated the $S = -2$ $BB$ potentials from $2+1$ flavor lattice QCD by using the
 HAL QCD method extended to coupled channel systems in Ref.~\cite{Aoki:2011gt}.
Combining the coupled channel formalism with the time-dependent Schr\"odinger equation \cite{HALQCD:2012aa}, we could extract {\em potential matrices} for the first time without the ground state saturation and without the diagonalization of the source operators.

By considering two baryon systems with $S=-2$,  $\Lambda \Lambda$, $N \Xi$, $\Sigma \Sigma$ and $\Lambda \Sigma$ which are mutually coupled,
we successfully extracted potential matrices. 
They are approximately  hermitician within the 
statistical errors, which is not guaranteed  from the definition. 
A small violation of Hermiticity may be removed at least partly by the 
proper treatment of renormalization factors, which is left for future studies.

We discussed properties of potential matrices for all $S=-2$ two baryon systems.
We found that all diagonal elements of the potential matrix have a repulsive core, while
their heights largely depend on their flavor structure.
Our previous works show that decreasing $ud$ quark mass leads to the enhancement of
 the short ranged repulsion and the long-ranged attraction.
Although such quark mass dependence was seen clearly in single channel case, 
it becomes less pronounced in two and three channel cases.

The potentials in the SU(3) basis are also investigated, where
 we could see clear quark mass dependence.
We found a strongly attractive potential for ${V^{1}}_{1}$ whose strength increases as the quark mass decreases. 
 The off-diagonal potentials in the SU(3) basis is a proper measure of the
 SU(3) breaking. In the $^3S_1 (I=1)$ channel, except for ${V^{\overline{10}}}_{10}$, only 
 a small transition potential between irreducible representations could be seen at short distances.
 In the $^1S_0 (I=0)$ channel, we found a clear mixture of the flavor singlet state and the octet state.
The other off-diagonal potentials have only a small magnitude at short distances.
Note, however, that the SU(3) breaking introduced in this paper ($m_{\pi}/m_K=0.96, 0.90, 0.86$)
 is still small compared to the realistic magnitude of the breaking ($m_{\pi}/m_K=0.27$).
  Nevertheless, the present paper provides a first theoretical
and numerical step toward the realistic  $BB$ potential matrix at the physical quark masses.

\section*{Acknowledgements}
We greatly appreciate the USQCD for their offer of computer resources to overcome the electricity crisis due to the Catastrophic Earthquake in East of Japan on Mar. 11th.
We are grateful for authors and maintainers of CPS++~\cite{CPS}, whose modified version is used for our simulations. 
We also thank CP-PACS/JLQCD Collaborations and ILDG/JLDG~\cite{JLDGILDG} for providing gauge configurations.
This work is supported in part by the  Strategic program for Innovative Research (SPIRE) Field 5,
the Large Scale Simulation Program of High Energy Accelerator Research Organization (KEK), 
Grant-in-Aid of the Ministry of Education, Science and Technology, Sports and Culture (Nos. 19540261, 20340047, 22540268, 24740144, 24740146, 25287046) and the Grant-in-Aid for Scientific Research on Innovative Areas (Nos. 20105001, 20105003).
T.H. was partially supported by RIKEN iTHES Project.

%

\end{document}